\documentclass[prd,aps,nofootinbib,noshowpacs,10pt]{revtex4}
\usepackage{amsmath,graphicx,epsfig,amssymb,dsfont,mathtools,}
\usepackage[usenames]{color}
\usepackage{ulem} 
\usepackage{bigstrut}
\usepackage{slashed}
\usepackage{multirow}
\usepackage{makecell}
\usepackage{verbatim}
\usepackage{makecell}

\usepackage[colorlinks=true,
            linkcolor=red,
            urlcolor=blue,
            citecolor=blue]{hyperref}
\newcommand{\red}{\textcolor{red}}

\allowdisplaybreaks[4]

\begin{document}
\title{Unification  of Flavor  SU(3) Analyses  of Heavy Hadron Weak Decays}
\author{Xiao-Gang He$^{1,2,3,4}$~\footnote{Email:hexg@sjtu.edu.cn}, 
Yu-Ji  Shi$^{1}$~\footnote{Email:shiyuji@sjtu.edu.cn}
and  Wei Wang $^{1}$~\footnote{Email:wei.wang@sjtu.edu.cn}
}
\affiliation{
 $^1$ INPAC,  SKLPPC,  MOE KLPPC, School of Physics and Astronomy, \\ 
Shanghai Jiao Tong University, Shanghai  200240 \\
$^2$ T.-D. Lee Institute, Shanghai Jiao Tong University,  Shanghai  200240\\
$^3$ Department of Physics, National Taiwan University, Taipei 106 \\
$^4$ National Center for Theoretical Sciences,  Hsinchu 300}

\begin{abstract}
Analyses of heavy mesons and baryons hadronic charmless decays using the flavor SU(3) symemtry can be formulated  in two  different forms. One is to construct the SU(3) irreducible representation amplitude (IRA) by decomposing effective Hamiltonian, and the other is to draw the topological diagrams (TDA).  In the flavor SU(3) limit, we study   various $B/D\to PP,VP,VV$, $B_c\to DP/DV$ decays, and  two-body  nonleptonic decays of beauty/charm baryons,   and demonstrate  that when all terms are included these two ways of analyzing the decay amplitudes are completely equivalent.    Furthermore we clarify some confusions in drawing topological diagrams using different ways of describing beauty/charm baryons. 
\end{abstract}
\maketitle

\section{Introduction}
Weak decays of heavy mesons and baryons carrying a bottom and/or a charm quark are of great interests and   have been studied extensively on both experimental and theoretical sides.  These decays   provide useful information about the strong and electroweak interactions in the standard model (SM). Rare decays   are ideal  to look for  new physics effects beyond SM, and recent measurements of lepton flavor universality  have shown notable  deviations from the standard model (see Ref.~\cite{Li:2018lxi} for a recent brief review on the anomalies in $B$ decays).  Quite a number of physical  observables like  branching fractions, CP asymmetries and polarizations have been precisely measured  by experiments~\cite{Amhis:2016xyh,Patrignani:2016xqp,Tanabashi:2018oca}.    On the other hand, due to our limited  understanding of QCD at low energy regions,  theoretical calculations of  decay amplitudes are not well understood. Most of the current calculations rely on the factorization methods. 
Among them, many  available studies are conducted at leading power in $1/m_b$, while recent analyses of semileptonic and radiative processes have indicated the importance of next-to-leading power corrections~\cite{Shen:2016hyv,Lu:2018cfc}.

Apart from factorization approaches, the flavor SU(3) symmetry is a powerful tools frequently  used in two-body and three-body heavy meson decays~\cite{Zeppenfeld:1980ex,Savage:1989ub,Deshpande:1994ii,He:1998rq,He:2000ys,Hsiao:2015iiu,Chau:1986du,Chau:1987tk,Chau:1990ay,Gronau:1994rj,Gronau:1995hm,Cheng:2014rfa,Zhou:2016jkv,Muller:2015lua}.  
Although  flavor SU(3) symmetry is  approximate, yet it still provides very useful information about the decays.
Since the SU(3) invariant amplitudes can be determined by fitting the data,  the SU(3) analysis bridges experimental data and the dynamical approaches. 

Among different realizations of carrying out $SU(3)$ analysis for decay amplitudes there are two popular methods. One of them is topological diagram amplitude (TDA) method, where decay amplitudes are represented by connecting quark lines flows in different ways and then relate them by $SU(3)$ symmetry, and another way is to construct the SU(3) irreducible representation amplitude (IRA) by decomposing effective Hamiltonian. The TDA approach gives a better understanding of decay dynamics especially in the bottom hadron decays, where the power expansion of $1/m_b$ can be properly conducted so that each TDA  amplitude can be related with matrix elements of SCET operators \cite{Bauer:2004ck}. On the other hand, the IRA approach shows a convenient connection with the SU(3) symmetry. These two methods looks very different in formulations, one may wonder whether they will obtain the same results.
This equivalence for some considered decays has been discussed in Refs.~\cite{Zeppenfeld:1980ex,Gronau:1994rj,Muller:2015lua}, in particular for charmed meson decays, Ref.~\cite{Muller:2015lua} has shown the equivalence  in the presence of   SU(3) symmetry breaking effects. 
In Ref.~\cite{He:2018php}, two of us have explored   two-body $B/D$ meson decays, $B\to PP$ and $D \to PP$ and pointed out that in the exact SU(3) symmetry limit, the two methods are  consistent when all contributions are  included. However  this equivalence is nontrivial: a few amplitudes  are suppressed and thus were not included in some TDA analysis; among the known diagrammatic amplitudes, one of them is not SU(3) independent, and should be absorbed into other amplitudes. Actually, in charm meson decays, the fact that one of the known amplitudes, $T,A,C,E$, is not independent has already been noticed  in Ref.~\cite{Muller:2015lua}.

In this work, we extend our analysis to several other types of two body decays of $B/D$ and $B_c$ mesons and also beauty/charm baryons to show the equivalence of the TDA and IRA methods. We will also work in the exact flavor SU(3) limit throughout this work. For two-body decays of beauty/charm baryons, we clarify some subtleties  including the description of baryon representation, one or two indices for $\bar 3$, in relation to TDA.  As we will show, it is easy to determine the independent amplitudes in IRA while TDA gives some redundancy.  
A few amplitudes are not independent and therefore should be absorbed into other amplitudes. Despite this disadvantage, the topological nature of TDA is still helpful for understanding the internal dynamics underlying in b decays in a more intuitive way.


The rest of this paper is arranged as  follows. In Section II, we briefly summarize the SU(3) properties of various inputs. In Section III, we give  the results for $B \to PP$ in the TDA and IRA methods to set up the notation. Then we provide results for $B \to PV, VV$ and discuss some points specialized to these decays. In section IV, we give the results for $D \to PP, VV, PV$ in the TDA and IRA methods. In section V, we carry out a similar analysis for $B_c \to DP, DV$.  In section VI and VII, we respectively discuss beauty and charm baryon decays into an octet baryon and an octet pseudo-scalar meson, while for beauty baryon decays we also consider the final states containing a decuplet baryon. The expanded  amplitudes and relations given these sections are useful for a global  analysis when enough data is available in the future.  In section VIII we summarize our results. In the Appendix, we give the relations for different parametrizations in TDA and IRA methods for bottom and charmed baryon decays.

\section{SU(3) Properties of Hamiltonian and Hadron States}
\label{sec:multiplet}

\subsection{Hadron Multiplets}

Several classes of heavy hadron, containing at heavy quark $b$ or $c$, will be considered in this work. The involved  processes include decays of heavy
SU(3) triplet mesons $B$ and $D$  into  $PP,\;PV,\;VV$, and the $ B_c$ meson into $DP,\;DV$. For heavy baryons, the decay processes include a heavy 
anti-triplet $T_{c\bar 3}$ or a  $T_{b\bar 3}$ decays into a  baryon in the decuplet $T_{10}$ plus a light meson,  and  decay into a baryon in the octet $T_{8}$ plus a light meson. We display the hadron $SU(3)$ properties and their component fields in this subsection.

The $B_c$ meson contains no light quark and it is a singlet. Heavy mesons containing one heavy quark
\begin{eqnarray}
(B_i) = (B^-(b \bar u), \overline B^0(b  \bar d), \overline B_s^0(b \bar s))\;, \;\;
(D_i)=  (D^0(c \bar u), D^+ (c \bar d), D^+_s (c \bar s))\;,
\end{eqnarray}
are flavor SU(3) anti-triplets.  

Light pseudoscalar $P$ and vector $V$ mesons are mixtures of octets and singlets so that each of them contain nine hadrons: 
\begin{eqnarray}
 P=\begin{pmatrix}
 \frac{\pi^0}{\sqrt{2}}+\frac{\eta_8}{\sqrt{6}}+\frac{\eta_1}{\sqrt3}  &\pi^+ & K^+\\
 \pi^-&\frac{\eta_8}{\sqrt{6}}-\frac{\pi^0}{\sqrt{2}}+\frac{\eta_1}{\sqrt3}&{K^0}\\
 K^-&\overline K^0 &\frac{\eta_1}{\sqrt3}-2\frac{\eta_8}{\sqrt{6}}
 \end{pmatrix},\;\;\;  V=\begin{pmatrix}
 \frac{\rho^0+\omega}{\sqrt{2}}  &\rho^+ & K^{*+}\\
 \rho^-& \frac{-\rho^0+\omega}{\sqrt{2}}&{K^{*0}}\\
 K^{*-}&\overline K^{*0} & \phi
 \end{pmatrix}, 
\end{eqnarray}
where $\omega$ and $\phi$  mix in an ideal form.  The $\eta$ and $\eta'$ are mixtures of $\eta_8$ and $\eta_1$ with the mixing angle $\theta$:
\begin{eqnarray}
\eta&=& \cos\theta\eta_8+ \sin\theta \eta_1, \nonumber\\
\eta'&=&-\sin\theta\eta_8+ \cos\theta \eta_1. 
\end{eqnarray}
Since $\eta_8$ and $\eta_1$ are   not physical states, optionally one can choose the $\eta_q$ and $\eta_s$ basis for the $\eta$ mixing, which are defined  so that the pseudoscalar octets $P$ has the same form of parametrization as vector octets $V$: 
\begin{eqnarray}
P=\begin{pmatrix}
 \frac{\pi^0+\eta_q}{\sqrt{2}}  &\pi^+ & K^{+}\\
 \pi^-& \frac{-\pi^0+\eta_q}{\sqrt{2}}&{K^{0}}\\
 K^{-}&\overline K^{0} & \eta_s
 \end{pmatrix}.  \label{eq:new_pseudo_scalar}
\end{eqnarray}
with 
\begin{eqnarray}
\eta_8=\frac{1}{\sqrt{3}}\eta_q- \sqrt{\frac{2}{3}}\eta_s,\;\;\; \eta_1= \sqrt{\frac{2}{3}}\eta_q+ \frac{1}{\sqrt{3}}\eta_s
\end{eqnarray} 
An advantage of the  parametrization in Eq.~\eqref{eq:new_pseudo_scalar} is that  there is  a one-to-one correspondence between the decay amplitudes of channels with vector final state  and that of channels with pseudoscalar final state  in the SU(3) limit.

A charmed or a bottom baryons with two light quarks  can form an anti-triplet or sextet. Most  members of the sextet  can decay through strong interaction or electromagnetic interactions. The only exceptions are $\Omega_{b}$ and $\Omega_{c}$ \cite{Wang:2017vnc}. We will concentrate on anti-triplet weak decays. For the anti-triplet bottom and charmed baryons, we have the following matrix expressions:
\begin{eqnarray}
 (T_{\bf{c\bar 3}}^{ij})= \left(\begin{array}{ccc} 0 & \Lambda_c^+  &  \Xi_c^+  \\ -\Lambda_c^+ & 0 & \Xi_c^0 \\ -\Xi_c^+   &  -\Xi_c^0  & 0
  \end{array} \right), \;\;\;  (T_{\bf{b\bar 3}}^{ij})= \left(\begin{array}{ccc} 0 & \Lambda_b^0  &  \Xi_b^0  \\ -\Lambda_b^0 & 0 & \Xi_b^- \\ -\Xi_b^0   &  -\Xi_b^-  & 0
  \end{array} \right). 
\end{eqnarray}
One can also contract the above matrix with the anti-symmetric tensor $\epsilon_{ijk}$ ($\epsilon_{123} = +1$) to have $T_{\bar 3,i} = \epsilon_{ijk} T^{jk}_{\bar 3}$ with
\begin{eqnarray}
( (T_{\bf{c\bar 3}})_i)=\left(\begin{array}{ccc} \Xi_c^0  &-\Xi_c^+ & \Lambda_c^+  \end{array} \right), \;\;\; 
 ((T_{\bf{b\bar 3}})_i)=\left(\begin{array}{ccc} \Xi_b^-  &-\Xi_b^0 & \Lambda_b^0  \end{array} \right). 
\end{eqnarray}

The lowest-lying  baryon octet is given by:
\begin{eqnarray}
((T_8)^i_j )= \left(\begin{array}{ccc} \frac{1}{\sqrt{2}}\Sigma^0+\frac{1}{\sqrt{6}}\Lambda^0 & \Sigma^+  &  p  \\ \Sigma^-  &  -\frac{1}{\sqrt{2}}\Sigma^0+\frac{1}{\sqrt{6}}\Lambda^0 & n \\ \Xi^-   & \Xi^0  & -\sqrt{\frac{2}{3}}\Lambda^0
  \end{array} \right).
\end{eqnarray}
One can also contract the above with $\epsilon_{ijk}$ to have $(T_8)_{ijk}\equiv \epsilon_{ijn} (T_8)^n_k$.

The light baryon decuplet is given as:
\begin{eqnarray}
T_{10}^{111} &=&  \Delta^{++},\;\;\; T_{10}^{112}= T_{10})^{121}=T_{10}^{211}= \frac{1}{\sqrt3} \Delta^+,
T_{10}^{222} =  \Delta^{-},\;\;\; T_{10}^{122}= T_{10}^{212}=T_{10}^{221}= \frac{1}{\sqrt3} \Delta^0, \nonumber\\
T_{10}^{113} &=& T_{10}^{131}=T_{10}^{311}= \frac{1}{\sqrt3} \Sigma^{\prime+},\;\;T_{10}^{223} = T_{10}^{232}=T_{10}^{322}= \frac{1}{\sqrt3} \Sigma^{\prime-},\nonumber\\
T_{10}^{123} &=& T_{10}^{132}=T_{10}^{213}=T_{10}^{231}=T_{10}^{312}=T_{10}^{321}= \frac{1}{\sqrt6} \Sigma^{\prime0},\nonumber\\
T_{10}^{133} &=& T_{10}^{313}=T_{10}^{331}= \frac{1}{\sqrt3} \Xi^{\prime0},\;\;T_{10}^{233} = T_{10}^{323}=T_{10}^{332}= \frac{1}{\sqrt3}  \Xi^{\prime-},\;\;
T_{10}^{333} =\Omega^-.
\end{eqnarray}

\subsection{ $SU(3)$ properties of effective Hamiltonian }

\noindent
{\bf Effective Hamiltonian for charmless $b$ decays}

In the SM weak decays of  charmless $b$ decays are induced by the following electroweak effective Hamiltonian~\cite{Buchalla:1995vs,Ciuchini:1993vr,Deshpande:1994pc}:
 \begin{eqnarray}
 {\cal H}^b_{eff}  = \frac{G_{F}}{\sqrt{2}}
     \big\{ V_{ub} V_{uq}^{*}  [
     C_{1}  O_{1}
  +  C_{2}  O_{2}]
 - V_{tb} V_{tq}^{*}   {\sum\limits_{i=3}^{10}} C_{i}  O_{i} \big\}+ \mbox{h.c.}. 
 \label{eq:hamiltonian}
\end{eqnarray}
Here $G_F$ is the Fermi constant, and the $V_{uq}$ and $V_{tq}$ are   CKM matrix elements. The  $O_{i}$ is a four-quark operator with $C_i$ as its Wilson coefficient.    The explicit forms of $O_i$s are given as follows: 
\begin{eqnarray}
 O_1 = (\bar q^i u^j)_{V-A} (\bar u^j b^i)_{V-A}, && O_2 = (\bar q u)_{V-A} (\bar u b)_{V-A}, \nonumber\\
 O_3= (\bar q b)_{V-A} \sum_{q'} (\bar q'q')_{V-A}, && O_4= (\bar q^i b^j)_{V-A} \sum_{q'} (\bar q^{\prime j}q^{\prime i})_{V-A}, \nonumber\\
 O_5= (\bar q b)_{V-A} \sum_{q'} (\bar q'q')_{V+A}, && O_6= (\bar q^i b^j)_{V-A} \sum_{q'} (\bar q^{\prime j}q^{\prime i})_{V+A}, \nonumber\\
 O_7=\frac{3}{2} (\bar q b)_{V-A} \sum_{q'} e_{q'}(\bar q'q')_{V+A}, && O_8= \frac{3}{2}(\bar q^i b^j)_{V-A} \sum_{q'}e_{q'} (\bar q^{\prime j}q^{\prime i})_{V+A}, \nonumber\\
 O_9=\frac{3}{2} (\bar q b)_{V-A} \sum_{q'}e_{q'}  (\bar q'q')_{V-A}, &&  O_{10}= \frac{3}{2}(\bar q^i b^j)_{V-A} \sum_{q'}e_{q'}  (\bar q^{\prime j}q^{\prime i})_{V-A}. \label{operators}
\end{eqnarray}
$q=d,s$ and $q'=u,d,s$. 
Here  the $V-A$ and $V+A$ corresponds a left-handed $\gamma_\mu (1-\gamma_5)$ and a right-handed current $\gamma_\mu (1+\gamma_5)$ respectively.  

In the  SU(3) group for light flavors, tree operators $O_{1,2}$ and electroweak penguin  operators $O_{7-10}$ can
be decomposed in terms of a vector $H_{\bf \bar 3}^i$, a traceless
tensor antisymmetric in upper indices, $(H_{\bf6})^{[ij]}_{k}$, and a
traceless tensor symmetric in   upper indices,
$(H_{\bf{\overline{15}}})^{\{ij\}}_{k}$.  
For the $\Delta S=0 (b\to d)$ decays, the non-zero components of the effective Hamiltonian are~\cite{Savage:1989ub,He:2000ys,Hsiao:2015iiu}:
\begin{eqnarray}
 (H_{\bf \bar3})^2=1,\;\;\; (H_{6})^{12}_1=-(H_{6})^{21}_1=(H_{6})^{23}_3=-(H_{6})^{32}_3=1,\nonumber\\
 2(H_{\overline{15}})^{12}_1= 2(H_{\overline{15}})^{21}_1=-3(H_{\overline{15}})^{22}_2 = 
 -6(H_{\overline{15}})^{23}_3=-6(H_{\overline{15}})^{32}_3=6.\label{eq:H3615_bb}
\end{eqnarray}
For the $\Delta S=-1(b\to s)$
decays the nonzero entries in the $H_{\bf{\bar 3}}$, $H_{\bf 6}$,
$H_{\bf{\overline{15}}}$ can be  obtained from Eq.~\eqref{eq:H3615_bb}
with the exchange  $2\leftrightarrow 3$ corresponding to the $d \leftrightarrow s$ exchange.  

QCD penguin operators $O_{3-6}$  behave as the ${\bf  \bar 3}$ representation.
For the magnetic moment operators, the color magnetic moment  operator $O_{8g} = (g_sm_b/4\pi) \bar s \sigma^{\mu\nu}T^a G^a_{\mu\nu} (1+\gamma_5) b$ is an SU(3) triplet, while the electromagnetic moment operator $O_{7\gamma} = \frac{em_b}{4\pi} \bar s \sigma^{\mu\nu}F_{\mu\nu} (1+\gamma_5) b$ can be effectively incorporated into the $O_{7-10}$. Thus both of them are not included in Eq.~\eqref{eq:hamiltonian} and the above decomposition is complete. 

The irreducible representation amplitude (IRA) method of describing related decays is  to decompose effective Hamiltonian according to the above mentioned representations and   construct the amplitudes accordingly. On the other hand the the topological diagrams (TDA) method
is to take the effective Hamiltonian with two light anti-quarks and a light quark $H^{ij}_k$ to represent  $\bar q u \bar u b$ with $i = \bar u$, $k=u$ and $j=\bar q$ (omitting the Lorentz indicies), and then contract the indices with initial and final hadron states. In this way the decays are represented by diagrams following the quark line flows. Note that in the TDA method, the indices
$i$ and $j$ ordering matters which are neither symmetry nor anti-symmetric. They are not traceless neither.

The effective Hamiltonian have both tree and loop contributions. When strong penguin and electroweak penguin are all included
the tree and loop contributions have $\bar 3$, $6$ and $\overline {15}$ representations. 
The independent amplitudes have the same numbers, except that one can make one of the tree or penguin amplitude real and the rest all in principle complex. Using the unitarity property of the CKM matrix $V_{ub}V^*_{uq}+ V_{cb}V^*_{cq}+V_{tb}V_{tq}^* =0$, one can also rewrite $c$ loop induced penguin contributions into amplitude proportional to $V_{ub}V^*_{uq}$ and $V_{tb}V_{tq}^*$,
\begin{eqnarray}
{\cal A} = V_{ub}V^*_{uq}{\cal A}_u + V_{tb}V_{tq}^* {\cal A}_t\;.
\end{eqnarray}
For simplicity, we refer to ${\cal A}_u$ as ``tree" amplitude since it is dominated by tree contributions with modifications from $u$ and $c$ loop contributions.
${\cal A}_t$ is ``penguin" amplitude with $c$ and $t$ loop contributions. It is necessary to stress that not all contributions in ${\cal A}_u$ are tree diagrams in topology, and the same for ${\cal A}_t$.

In Ref.~\cite{He:2018php}, two of us   have shown that both ${\cal A}_u$ and ${\cal A}_t$ have similar form of amplitudes with $SU(3)$ representations. Thus in our later discussions we will concentrate on the ${\cal A}_u$ amplitudes. One can easily obtain the ${\cal A}_t$ amplitudes by just changing the amplitude labels.
\\


\noindent
{\bf Effective electroweak Hamiltonian for $c$ hadronic decays}

For  hadronic decays of charmed  hadrons, the effective Hamiltonian with $\Delta C= 1$ is given as:
\begin{eqnarray}
{\cal H}^c_{eff} &=& \frac{G_F}{\sqrt2}\big\{V_{cs} V_{ud}^* [C_1O_1^{sd}+C_2O_2^{sd}]+V_{cd} V_{ud}^* [C_1O_1^{dd}+C_2O_2^{dd}] \nonumber\\
&&+ V_{cs} V_{us}^* [C_1O_1^{ss}+C_2O_2^{ss}]+V_{cd} V_{us}^* [C_1O_1^{ds}+C_2O_2^{ds}]  \big\}, 
\end{eqnarray}
where we have neglected the highly suppressed penguin contributions, and 
\begin{eqnarray}
O_1^{sd} &=& [\bar s^i  \gamma_{\mu}(1-\gamma_5) c^j ][\bar u^i \gamma^{\mu}(1-\gamma_5) d^j], \;\; 
O_2^{sd} = [\bar s  \gamma_{\mu}(1-\gamma_5) c][\bar u \gamma^{\mu}(1-\gamma_5) d], 
\end{eqnarray}
while other operators can be obtained by replacing the $d,s$ quark fields. 
Tree operators  transform
under the flavor SU(3) symmetry as ${\bf \bar  3}\otimes {\bf  3}\otimes {\bf  \bar
3}={\bf   \bar 3}\oplus {\bf   \bar3}\oplus {\bf  6}\oplus {\bf   {\overline {15}}}$. 

For the Cabibbo allowed $c\to s  u \bar d$ transition, we have amplitudes proportional to $V_{cs}V^*_{ud}$ and the Hamiltonians are:
\begin{eqnarray}
(H_{  6})^{31}_2=-(H_{  6})^{13}_2=1,\;\;\;
 (H_{\overline {15}})^{31}_2= (H_{\overline {15}})^{13}_2=1.\label{eq:H3615_c_allowed}
\end{eqnarray}
For the  doubly Cabbibo suppressed $c\to d  u \bar s$ transition, we have amplitudes to be proportional to $V_{cd}V^*_{us}$ and the Hamiltonians are:
\begin{eqnarray}
(H_{  6})^{21}_3=-(H_{  6})^{12}_3=1,\;\;
 (H_{\overline {15}})^{21}_3= (H_{\overline {15}})^{12}_3=1. \label{eq:H3615_c_doubly_suprressed}
\end{eqnarray}
For decays proportional to $V_{cs}V_{us}^*$, we have:
\begin{eqnarray}
(H_{  6})^{31}_3 =-(H_{  6})^{13}_3 =1,\;\;\;
 (H_{\overline {15}})^{31}_3= (H_{\overline  {15}})^{13}_3 = 1, \label{eq:H3615_cc_singly_suppressed}
\end{eqnarray}
and for decays proportional to $V_{cd}V_{ud}^*$, we have:
\begin{eqnarray}
(H_{  6})^{12}_2 =-(H_{  6})^{21}_2 =1,\;\;\;
(H_{\overline  {15}})^{12}_2=(H_{\overline  {15}})^{21}_2= - 1. \label{eq:H3615_cc_singly_suppressed}
\end{eqnarray}

For singly Cabbibo suppressed decays,  $c\to u \bar dd$ and $c\to u \bar ss$ transitions have approximately equal magnitudes but opposite signs: $V_{cd}V_{ud}^* = - V_{cs}V_{us}^* -V_{cb}V_{ub}^* \approx - V_{cs}V_{us}^*$ (with $10^{-3}$ deviation). As a result, the contributions from the $\bar 3$ representation vanish, and   one has the nonzero components contributed only by 6 and $\bar{15}$ representations.

For the singly Cabibbo-suppressed transition,  there are also loop contributions proportional to $V_{cb}V^*_{ub}$. Such loop contributions are small so that we will concentrate on the dominant amplitude proportional to $V_{cs}V^*_{us}$.  However, one can include these contributions by adding a $3$ representation in the Hamiltonian.

Again, we use the above $SU(3)$ decompositions for IRA analysis and use the effective Hamiltonian $H^{ij}_k$ with $i = \bar s$, $j = \bar u$ and $k = q$ for TDA analysis to trace the quark line flows. 



\section{Charmless two-body $B$ decays } 

\subsection{$B \to PP$ decays}

Let us start with  the $B \to PP$ decays. The generic amplitude   is  decomposed according to   CKM matrix elements: 
\begin{eqnarray}
{\cal A}&=& V_{ub}V_{uq}^* {\cal A}^{IRA}_u  + V_{tb}V^*_{tq} {\cal A}^{IRA}_t \nonumber\\
&=& V_{ub}V_{uq}^* {\cal A}^{TDA}_u  + V_{tb}V^*_{tq} {\cal A}^{TDA}_t\;,
\end{eqnarray}
where the amplitudes expressed by IRA and TDA  should be equivalent. Although in fact this equivalence is concrete and obvious, as we argued in~\cite{He:2018php}, it has not been thoroughly discussed in the literature.

To obtain IRA, one takes   various representations in Eq.~\eqref{eq:H3615_bb} and contracts all indices in   $B_i$ and light meson $P^i_j$ with various combinations: 
\begin{eqnarray} 
 {\cal A}^{IRA}_u &=&A_3^T B_i (H_{\bar 3})^i P_k^jP_j^k +C_3^T B_i (H_{\bar 3})^kP^i_j P^j_k +B_3^T   B_i (H_3)^i P_k^kP_j^j +D_3^T B_i  (H_{\bar 3})^j P^i_j P^k_k\nonumber\\
  &&\;\;\;     
  +A_6^T B_i (H_{ 6})^{[ij]}_k P_j^lP_l^k 
  +C_6^T B_i (H_{ 6})^{[jl]}_k P^i_j P_l^k +B_6^T B_i (H_{ 6})^{[ij]}_k P_j^kP_l^l \nonumber\\
  && \;\;\;
  +A_{15}^T B_i (H_{\overline{15}})^{\{ij\}}_k P_j^lP_l^k 
  +C_{15}^T B_i (H_{\overline{15}})^{\{jk\}}_l P^i_j P_k^l +B_{15}^T B_i (H_{\overline{15}})^{\{ij\}}_k P_j^kP_l^l. \label{eq:B2PP_IRA_tree}
\end{eqnarray}

\begin{figure}
\begin{center}
\includegraphics[scale=0.5]{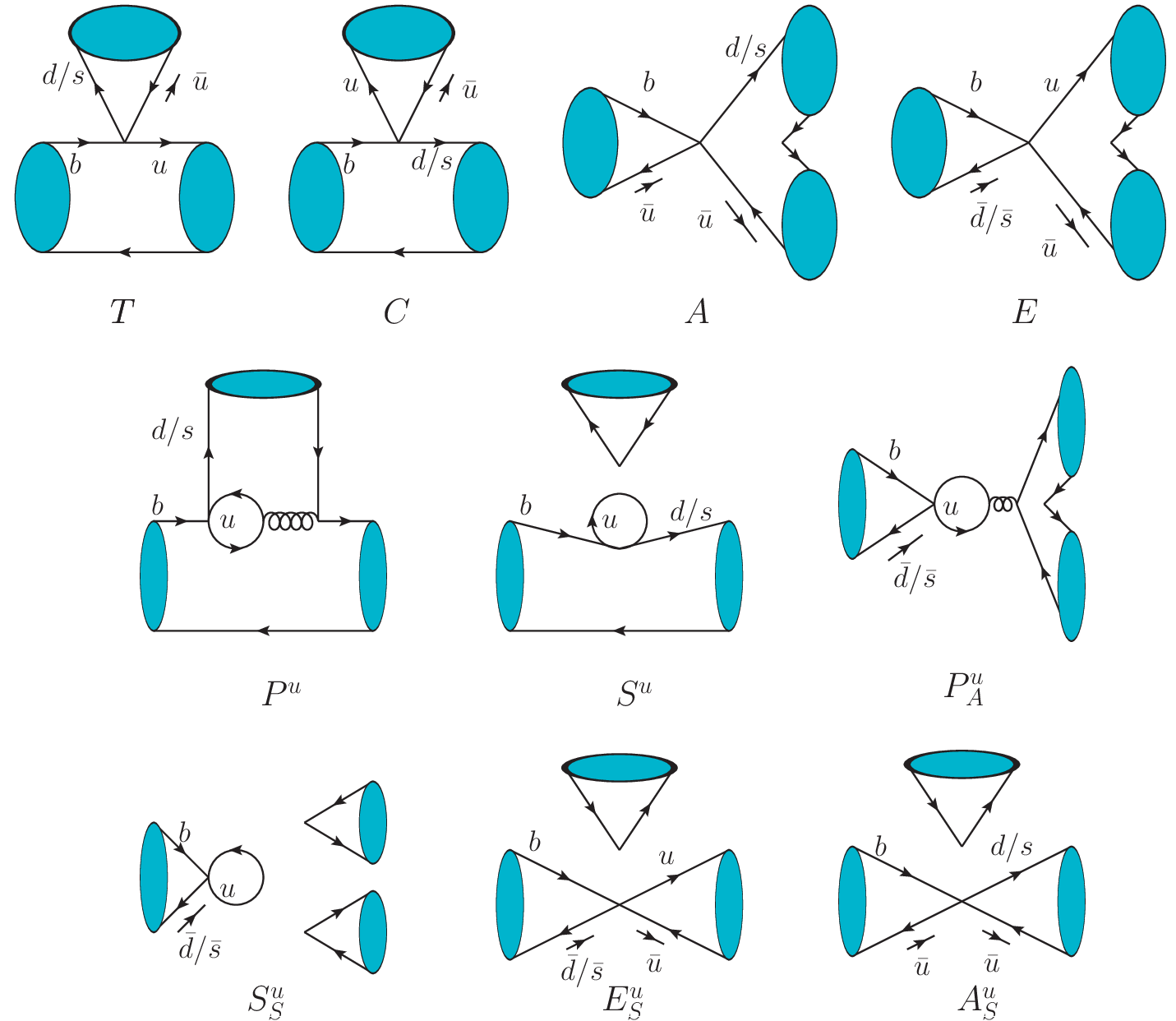}
\end{center}
\caption{Topological diagrams for the amplitudes with CKM factor   $V_{ub}V_{uq}^*$ in $B\to PP$ and $B\to VV$ decays. \red{We work in the effective field theory at $m_b$ scale, and thus there exists four-quark interaction operators as shown in the above.  The quark flavors corresponding to these operators are explicitly given. In these diagrams, when the  $\bar u u$ annihilate,  two or more gluons are needed to create one pair of quarks with flavor $u, d, s$, which are denoted by the unspecified lines. }  }\label{fig:Feynman_B_PP}
\end{figure}

For the TDA decomposition, one can classify the different topologies of diagram as \cite{Cheng:2014rfa}:
\begin{itemize}
\item[(i)] $T$, denoting the color-allowed tree amplitude with $W$ emission;
\item [(ii)] $C$, denoting the color-suppressed tree diagram;
\item [(iii)] $E$ denoting the $W$-exchange diagram;
\item [(iv)] $P$, corresponding to the QCD penguin contributions;
\item[(v)] $S$, being the flavor singlet QCD penguin;
\item[(vi)] $A$,  annihilation diagrams. 
\end{itemize}
With each coefficient denoted by the above topologies, one has the TDA decomposition as:
\begin{eqnarray}
{\cal A}^{TDA}_u &=&  T   B_i H^{jl}_k P^{i}_j   P^k_l   +C   B_i H^{lj}_k P^{i}_j  P^k_l + A   B_i H^{il}_j   P^j_k P^{k}_l  + E   B_i  H^{li}_j P^j_k P^{k}_l\nonumber\\
&& + S^{u} B_i  H^{lj}_{l} P^{i}_j  P^k_k +P^{u} B_i H^{lk}_{l} P^{i}_j   P^j_k + P_{A}^{u} B_i H^{li}_{l}  P^j_k P^{k}_j   + S_{S}^{u} B_i H^{li}_{l}  P^j_j P^{k}_{k}   \nonumber\\
&& +E_{S}^{u} B_i H^{ji}_{l}  P^{l}_j  P^k_k+A_{S}^{u} B_i  H^{ij}_{l}   P^{l}_j    P^k_k.  \label{eq:B2PP_TDA_tree}
\end{eqnarray}
  According to this decomposition, topological diagrams for $B\to PP$ decays can be found  in Fig.~\ref{fig:Feynman_B_PP}.  \red{ Since we work in the effective field theory at $m_b$ scale, as shown in this figure there exists four-quark interaction operators.  In these diagrams, the quark flavors corresponding to these operators are explicitly given. When the $\bar u u$ annihilate,  two or more gluons are needed to create one pair of quarks with flavor $u, d, s$, which are denoted by the unspecified lines.}  Apart from the ordinary $T, C, A, E$, we have also included the other SU(3) irreducible amplitudes, most of which come from   loop diagrams, and/or   the flavor singlet diagram.

Expanding Eqs.~(\ref{eq:B2PP_IRA_tree},\ref{eq:B2PP_TDA_tree}), one obtains $B\to PP$ amplitudes in Table~\ref{tab:Two_body_PP}, where the IRA amplitudes are consistent with Ref.~\cite{Hsiao:2015iiu} and an earlier work \cite{Grossman:2003qp}. 
Since we have decomposed the effective Hamiltonian into irreducible representations, one may  expect   that there are 10 independent  amplitudes for ${\cal A}_{u}$ and similarly 10 amplitudes for ${\cal A}_t$.   A careful examination  shows that the   $A_{6}^T$ can be absorbed into $B_{6}^T$ and $C_6^T$ with a redefinition:
\begin{eqnarray}
C_{6}^{T\prime}= C_{6}^T-A_{6}^T, \;\; B_{6}^{T\prime}= B_{6}^T+A_{6}^T\;.
\end{eqnarray}
This  combination  can also be found   explicitly from Table~\ref{tab:Two_body_PP}. 
After eliminating the redundant amplitude, 
there are actually only 18 (${\cal A}_u$ and ${\cal A}_t$ contribute 9 each) SU(3) independent amplitudes. 
Since one overall phase can be chosen free,   there are   35 independent real independent parameters.
If one consider $\eta_1-\eta_8$ mixing, the mixing angle $\theta$ should also be introduced.

\begin{table}
\renewcommand\arraystretch{1.3}
\caption{Decay amplitudes for two-body $B\to PP$ decays. Only the 
amplitudes  with   CKM factor $V_{ub}V_{uq}^{*}$  are shown in
this table and the following ones.   ``Penguin" amplitudes with $V_{tb}V_{tq}^{*}$ can be obtained with the replacement Eq.~\eqref{replaceforPenguinIRA} and Eq.~\eqref{replaceforPenguinTDA}. }
\label{tab:Two_body_PP}%
\begin{tabular}{cccccc}
\hline 
\hline 
{$b\to d$}  & {IRA}  & {TDA}  &  &  & \tabularnewline
\hline 
$B^{-}\to\pi^{0}\pi^{-}$  & $4\sqrt{2}C_{15}^{T}$ & $(\text{C}+\text{T})/{\sqrt{2}}$ &  &  & \tabularnewline
\hline 
$B^{-}\to\pi^{-}\eta_{q}$  & $\sqrt{2}\left(A_{6}^{T}+3A_{15}^{T}+B_{6}^{T}+3B_{15}^{T}+C_{3}^{T}+2C_{15}^{T}+D_{3}^{T}\right)$ & $(2A_{s}^{u}+2\text{A}+\text{C}+2P^{u}+2S^{u}+\text{T})/{\sqrt{2}}$ &  &  & \tabularnewline
\hline 
$B^{-}\to\pi^{-}\eta_{s}$  & $B_{6}^{T}+3B_{15}^{T}+C_{6}^{T}-C_{15}^{T}+D_{3}^{T}$ & $A_{s}^{u}+S^{u}$ &  &  & \tabularnewline
\hline 
$B^{-}\to K^{0}K^{-}$  & $A_{6}^{T}+3A_{15}^{T}+C_{3}^{T}-C_{6}^{T}-C_{15}^{T}$ & $\text{A}+P^{u}$ &  &  & \tabularnewline
\hline 
$\overline{B}^{0}\to\pi^{+}\pi^{-}$  & $2A_{3}^{T}-A_{6}^{T}+A_{15}^{T}+C_{3}^{T}+C_{6}^{T}+3C_{15}^{T}$ & $2P_{A}^{u}+\text{E}+P^{u}+\text{T}$ &  &  & \tabularnewline
\hline 
$\overline{B}^{0}\to\pi^{0}\pi^{0}$  & $2A_{3}^{T}-A_{6}^{T}+A_{15}^{T}+C_{3}^{T}+C_{6}^{T}-5C_{15}^{T}$ & $2P_{A}^{u}-\text{C}+\text{E}+P^{u}$ &  &  & \tabularnewline
\hline 
$\overline{B}^{0}\to\pi^{0}\eta_{q}$  & $-A_{6}^{T}+5A_{15}^{T}-B_{6}^{T}+5B_{15}^{T}-C_{3}^{T}+2C_{15}^{T}-D_{3}^{T}$ & $\text{E}_{S}^{u}+\text{E}-P^{u}-S^{u}$ &  &  & \tabularnewline
\hline 
$\overline{B}^{0}\to\pi^{0}\eta_{s}$  & $-(B_{6}^{T}-5B_{15}^{T}+C_{6}^{T}-C_{15}^{T}+D_{3}^{T})/{\sqrt{2}}$ & $(\text{E}_{S}^{u}-S^{u})/{\sqrt{2}}$ &  &  & \tabularnewline
\hline 
$\overline{B}^{0}\to K^{+}K^{-}$  & $2\left(A_{3}^{T}+A_{15}^{T}\right)$ & $2P_{A}^{u}+\text{E}$ &  &  & \tabularnewline
\hline 
$\overline{B}^{0}\to K^{0}\overline{K}^{0}$  & $2A_{3}^{T}+A_{6}^{T}-3A_{15}^{T}+C_{3}^{T}-C_{6}^{T}-C_{15}^{T}$ & $2P_{A}^{u}+P^{u}$ &  &  & \tabularnewline
\hline 
$\overline{B}^{0}\to\eta_{q}\eta_{q}$  & $2A_{3}^{T}-A_{6}^{T}+A_{15}^{T}+4B_{3}^{T}-2B_{6}^{T}+2B_{15}^{T}+C_{3}^{T}-C_{6}^{T}+C_{15}^{T}+2D_{3}^{T}$ & $2P_{A}^{u}+\text{C}+2\text{E}_{S}^{u}+\text{E}+P^{u}+2S^{u}+4S_{S}^{u}$ &  &  & \tabularnewline
\hline 
$\overline{B}^{0}\to\eta_{q}\eta_{s}$  & $(4B_{3}^{T}+B_{6}^{T}-B_{15}^{T}+C_{6}^{T}-C_{15}^{T}+D_{3}^{T})/{\sqrt{2}}$ & $(\text{E}_{S}^{u}+S^{u}+4S_{S}^{u})/{\sqrt{2}}$ &  &  & \tabularnewline
\hline 
$\overline{B}^{0}\to\eta_{s}\eta_{s}$  & $2(A_{3}^{T}+A_{6}^{T}-A_{15}^{T}+B_{3}^{T}+B_{6}^{T}-B_{15}^{T})$ & $2(P_{A}^{u}+S_{S}^{u})$ &  &  & \tabularnewline
\hline 
$\overline{B}_{s}^{0}\to\pi^{0}K^{0}$  & $(A_{6}^{T}+A_{15}^{T}-C_{3}^{T}-C_{6}^{T}+5C_{15}^{T})/{\sqrt{2}}$ & $(\text{C}-P^{u})/{\sqrt{2}}$ &  &  & \tabularnewline
\hline 
$\overline{B}_{s}^{0}\to\pi^{-}K^{+}$  & $-A_{6}^{T}-A_{15}^{T}+C_{3}^{T}+C_{6}^{T}+3C_{15}^{T}$ & $P^{u}+\text{T}$ &  &  & \tabularnewline
\hline 
$\overline{B}_{s}^{0}\to K^{0}\eta_{q}$  & $-(A_{6}^{T}+A_{15}^{T}+2B_{6}^{T}+2B_{15}^{T}-C_{3}^{T}+C_{6}^{T}-C_{15}^{T}-2D_{3}^{T})/{\sqrt{2}}$ & $(\text{C}+P^{u}+2S^{u})/{\sqrt{2}}$ &  &  & \tabularnewline
\hline 
$\overline{B}_{s}^{0}\to K^{0}\eta_{s}$  & $-A_{6}^{T}-A_{15}^{T}-B_{6}^{T}-B_{15}^{T}+C_{3}^{T}-2C_{15}^{T}+D_{3}^{T}$ & $P^{u}+S^{u}$ &  &  & \tabularnewline
\hline 
\hline 
{$b\to s$}  & {IRA}  & {TDA}  &  &  & \tabularnewline
\hline 
$B^{-}\to\pi^{0}K^{-}$  & $(A_{6}^{T}+3A_{15}^{T}+C_{3}^{T}-C_{6}^{T}+7C_{15}^{T})/{\sqrt{2}}$ & $(\text{A}+\text{C}+P^{u}+\text{T})/{\sqrt{2}}$ &  &  & \tabularnewline
\hline 
$B^{-}\to\pi^{-}\overline{K}^{0}$  & $A_{6}^{T}+3A_{15}^{T}+C_{3}^{T}-C_{6}^{T}-C_{15}^{T}$ & $\text{A}+P^{u}$ &  &  & \tabularnewline
\hline 
$B^{-}\to K^{-}\eta_{q}$  & $(A_{6}^{T}+3A_{15}^{T}+2B_{6}^{T}+6B_{15}^{T}+C_{3}^{T}+C_{6}^{T}+5C_{15}^{T}+2D_{3}^{T})/{\sqrt{2}}$ & $(2A_{s}^{u}+\text{A}+\text{C}+P^{u}+2S^{u}+\text{T})/{\sqrt{2}}$ &  &  & \tabularnewline
\hline 
$B^{-}\to K^{-}\eta_{s}$  & $A_{6}^{T}+3A_{15}^{T}+B_{6}^{T}+3B_{15}^{T}+C_{3}^{T}-2C_{15}^{T}+D_{3}^{T}$ & $A_{s}^{u}+\text{A}+P^{u}+S^{u}$ &  &  & \tabularnewline
\hline 
$\overline{B}^{0}\to\pi^{+}K^{-}$  & $-A_{6}^{T}-A_{15}^{T}+C_{3}^{T}+C_{6}^{T}+3C_{15}^{T}$ & $P^{u}+\text{T}$ &  &  & \tabularnewline
\hline 
$\overline{B}^{0}\to\pi^{0}\overline{K}^{0}$  & $(A_{6}^{T}+A_{15}^{T}-C_{3}^{T}-C_{6}^{T}+5C_{15}^{T})/{\sqrt{2}}$ & $(\text{C}-P^{u})/{\sqrt{2}}$ &  &  & \tabularnewline
\hline 
$\overline{B}^{0}\to\overline{K}^{0}\eta_{q}$  & $-(A_{6}^{T}+A_{15}^{T}+2B_{6}^{T}+2B_{15}^{T}-C_{3}^{T}+C_{6}^{T}-C_{15}^{T}-2D_{3}^{T})/{\sqrt{2}}$ & $(\text{C}+P^{u}+2S^{u})/{\sqrt{2}}$ &  &  & \tabularnewline
\hline 
$\overline{B}^{0}\to\overline{K}^{0}\eta_{s}$  & $-A_{6}^{T}-A_{15}^{T}-B_{6}^{T}-B_{15}^{T}+C_{3}^{T}-2C_{15}^{T}+D_{3}^{T}$ & $P^{u}+S^{u}$ &  &  & \tabularnewline
\hline 
$\overline{B}_{s}^{0}\to\pi^{+}\pi^{-}$  & $2\left(A_{3}^{T}+A_{15}^{T}\right)$ & $2P_{A}^{u}+\text{E}$ &  &  & \tabularnewline
\hline 
$\overline{B}_{s}^{0}\to\pi^{0}\pi^{0}$  & $2(A_{3}^{T}+A_{15}^{T})$ & $2(P_{A}^{u}+(\text{E})/{2})$ &  &  & \tabularnewline
\hline 
$\overline{B}_{s}^{0}\to\pi^{0}\eta_{q}$  & $-2\left(A_{6}^{T}-2A_{15}^{T}+B_{6}^{T}-2B_{15}^{T}\right)$ & $\text{E}_{S}^{u}+\text{E}$ &  &  & \tabularnewline
\hline 
$\overline{B}_{s}^{0}\to\pi^{0}\eta_{s}$  & $-\sqrt{2}\left(B_{6}^{T}-2B_{15}^{T}+C_{6}^{T}-2C_{15}^{T}\right)$ & $(\text{C}+\text{E}_{S}^{u})/{\sqrt{2}}$ &  &  & \tabularnewline
\hline 
$\overline{B}_{s}^{0}\to K^{+}K^{-}$  & $2A_{3}^{T}-A_{6}^{T}+A_{15}^{T}+C_{3}^{T}+C_{6}^{T}+3C_{15}^{T}$ & $2P_{A}^{u}+\text{E}+P^{u}+\text{T}$ &  &  & \tabularnewline
\hline 
$\overline{B}_{s}^{0}\to K^{0}\overline{K}^{0}$  & $2A_{3}^{T}+A_{6}^{T}-3A_{15}^{T}+C_{3}^{T}-C_{6}^{T}-C_{15}^{T}$ & $2P_{A}^{u}+P^{u}$ &  &  & \tabularnewline
\hline 
$\overline{B}_{s}^{0}\to\eta_{q}\eta_{q}$  & $2(A_{3}^{T}+A_{15}^{T}+2\left(B_{3}^{T}+B_{15}^{T}\right))$ & $2(P_{A}^{u}+\text{E}_{S}^{u}+(\text{E})/{2}+2S_{S}^{u})$ &  &  & \tabularnewline
\hline 
$\overline{B}_{s}^{0}\to\eta_{q}\eta_{s}$  & $\sqrt{2}\left(2B_{3}^{T}-B_{15}^{T}+C_{15}^{T}+D_{3}^{T}\right)$ & $(\text{C}+\text{E}_{S}^{u}+2S^{u}+4S_{S}^{u})/{\sqrt{2}}$ &  &  & \tabularnewline
\hline 
$\overline{B}_{s}^{0}\to\eta_{s}\eta_{s}$  & $A_{3}^{T}-2A_{15}^{T}+B_{3}^{T}-2B_{15}^{T}+C_{3}^{T}-2C_{15}^{T}+D_{3}^{T}$ & $P_{A}^{u}+P^{u}+S^{u}+S_{S}^{u}$ &  &  & \tabularnewline
\hline 
\hline 
\end{tabular}
\end{table}

We list all  TDA amplitudes in Table~\ref{tab:Two_body_PP}.
It is necessary to point out that the last 6 diagrams in Fig.~\ref{fig:Feynman_B_PP} are   omitted in some SU(3) TDA analysis~\cite{Gronau:1994rj,Cheng:2014rfa}, while other  amplitudes are consistent with Refs.~\cite{Gronau:1994rj,Cheng:2014rfa}. However  only by including them the complete equivalence of IRA and TDA can be established.  One of the 10 TDA amplitudes must be redundant.  Such redundancy can be understood through the following relations between the IRA and TDA amplitudes: 
\begin{eqnarray}
&&T+E = 4A_{15}^T +2C_{6}^{\prime T} +4C_{15}^T, \hspace{1.3cm} C-E=-4A_{15}^T -2C_{6}^{\prime T} +4C_{15}^T, \nonumber\\
&&A+E = 8A_{15}^T , \hspace{3.7cm} P^{u}-E= -5A_{15}^T +C_{3}^T-C_{6}^{\prime T} -C_{15}^T, \nonumber\\
&&P_{A}^{u}+ \frac{E}{2}= A_{3}^T+A_{15}^T , \hspace{2.5cm} E_{S}^{u}+E = 4A_{15}^T -2B_{6}^{\prime T} +4B_{15}^T, \nonumber\\
&&A_{S}^{u}-E = -4A_{15}^T +2B_{6}^{\prime T} +4B_{15}^T, \;\; \;\;\;\;\;S_{S}^{u}- \frac{E}{2} = -2A_{15}^T +B_{3}^T +B_{6}^{\prime T} -B_{15}^T, \nonumber\\
&&S^{u}+E = 4A_{15}^T -B_{6}^{\prime T} -B_{15}^T +C_{6}^{\prime T} -C_{15}^T +D_{3}^T. 
\end{eqnarray}
We have adopted the choice in which  $E$ is always in companion with another amplitude.
It is also possible to replace the role of $E$ by one of the amplitudes  $A$, $C$ or even $T$. 
One can also reversely obtain:
\begin{eqnarray}
&&A_3^T= -\frac{A}{8} + \frac{3E}{8}+P_{A}^{u}, \;\hspace{2.3cm}
B_3^T=  S_{S}^{u} +\frac{3E_{S}^{u}-A_{S}^{u}}{8},\;\;\;\;\;\;\;\;\;
C_3^T=  \frac{1}{8} ({3A-C-E+3T})+P^u,  \nonumber\\
&&D_3^T=  S^{u} +\frac{1}{8} (3C-E_{S}^{u}+3A_{S}^{u}-T),\;\; B_6^{\prime T}=  \frac{1}{4}(A-E+A_{S}^{u}-E_{S}^{u}), \;\; C_6^{\prime T}=  \frac{1}{4}(-A-C+E+T),\nonumber\\
&&A_{15}^T=  \frac{A+E}{8}, \hspace{4cm}
B_{15}^T= \frac{A_{S}^{u}+E_{S}^{u}}{8}, \hspace{2cm} 
C_{15}^T=  \frac{C+T}{8}. \label{eq:relation_TDA2IRA}
\end{eqnarray}

Similar analysis for the ${\cal A}_t$  contributions  can be obtained with the replacement for the IRA: 
 \begin{eqnarray}
  A_i^T \to A_i^P,\ \   B_i^T \to B_i^P,\ \   C_i^T \to C_i^P,\ \   D_i^T \to D_i^P. \label{replaceforPenguinIRA}
 \end{eqnarray}
while for   TDA, we have: 
 \begin{eqnarray}
 T\to P_{T}, \;\; C\to P_{C}, \;\; A\to P_{TA}, \;\; P^{u} \to P,\;\; E\to P_{TE},  \nonumber\\
 P_{A}^{u}\to P_A, \;\; E_{S}^{u}\to P_{AS}, \;\; A_{S}^{u}\to P_{ES}, \;\; S_{S}^{u}\to P_{SS}, \;\; S^u\to S.  \label{replaceforPenguinTDA}
 \end{eqnarray}
 \red{ It should be pointed out that the amplitudes generated $Q_{7,8,9,10}$ operators  have the form $(2/3)\bar u u - (1/3) (\bar d d +\bar s s)$ and these amplitudes can be  can again be as a sum of  a tree-like operator $\bar u u$ and a  penguin-like operator $-(1/3) (\bar u u + \bar d d +\bar s s)$.  Penguin-like contributions have been absorbed into $P,P_A,S$, $P_{SS}$,  while tree-like amplitudes are denoted as  $P_T,\;, P_C,\;P_{TA}, \;P_{TE},\; P_{ES}$.  It should be noticed that the tree-like amplitudes  $P_T$ and $P_C$   are also denoted as $P_{EW}, P_{EW,C}$ in some references.  }

\subsubsection{Impact of the new TDA amplitudes}

The  new TDA amplitudes in Fig.~\ref{fig:Feynman_B_PP} may play an important role in understanding CP violation  (CPV) phenomena. 
Without the new TDA amplitudes, some decays only have terms proportional to $V_{tq}^*V_{tb}$, such as $\overline B^0 \to K^0 \bar K^0$ and $\overline B^0_s \to K^0 \bar K^0$. For instance,  in Ref.~\cite{Cheng:2014rfa}, the amplitudes for $\overline B^0 \to K^0 \bar K^0$ read: 
\begin{eqnarray}
{\cal A}(\overline B^0\to K^0\bar K^0)=   V_{tb}V_{td}^* \left(P-\frac{1}{2}P_{EW}^C+2P_A\right). 
\end{eqnarray}
This would imply  the  CP  violating   asymmetry    is {\it identically zero}. However,  as we have shown, these two decays receive contributions from the   $P^{u} + 2 P_{A}^{u}$   multiplied by $V_{uq}^*V_{ub}$:
\begin{eqnarray}
{\cal A}(\overline B^0\to K^0\bar K^0)= V_{ub}V_{ud}^* (P^{u} + 2 P_{A}^{u})+ V_{tb}V_{td}^* (P+2P_{A}^{u}). 
\end{eqnarray}
Therefore a non-vanishing direct CP asymmetry is obtained, as noticed in many references for instance Ref.~\cite{Fleischer:2016ofb}, ~\cite{DescotesGenon:2006wc} and ~\cite{Fleischer:2004vu}. This would certainly affect the search for new physics in a precise  CP violation measurement.

Most new TDA amplitudes in Fig.~\ref{fig:Feynman_B_PP} arise from higher order loop corrections, and thus they are likely small in magnitude. However, sometimes  they can not be completely neglected.  In Ref.~\cite{Hsiao:2015iiu}, the authors have  performed a  fit of $B\to PP$ decays in the IRA framework. Depending on different choices of   data, four cases were considered in their analysis~\cite{Hsiao:2015iiu}. Here for illustration, we give  their results in  case 4: 
\begin{eqnarray}
|C_{\bar 3}^T|= -0.211\pm 0.027,\;\; \delta_{\bar 3}^T= (-140\pm6)^\circ,  \;\; |B_{\overline {15}}^T| = -0.038\pm 0.016, \;\; \delta_{B_{\overline {15}}^T}= (78\pm 48)^\circ,  \label{ea:IRA_fit}
\end{eqnarray}
where the magnitudes and  strong phases are defined    relative to the amplitude $C_{\bar 3}^P$. From Eq.~\eqref{eq:relation_TDA2IRA}, one can find  that  the $C_{\bar 3}^T$ is a mixture  of  $T$,   $C$ and others, while the $B_{\overline{15}}^T$ equals $(E_{S}^{u}+ A_{S}^{u})/8$.  The fitted results in Eq.~\eqref{ea:IRA_fit} indicate,  compared to $C_{\bar 3}^T$, the $B_{\overline {15}}^T$ could  reach $20\%$ in magnitude, and more notably, the strong phases are sizably different.  The fact that the $B_{\overline {15}}^T$, namely  $E_{S}^{u}$ and $A_{S}^{u}$,  have non-negligible contributions supports  our call for a complete analysis.

\subsubsection{Comparison with QCDF amplitudes}
 
The topological amplitudes in $B\to PP$ decays can be compared to the QCDF amplitude in Ref.~\cite{Beneke:2003zv}. Such a comparison requests two remarks. 
Firstly, in our decomposition, we adopt the CKM matrix elements $V_{ub}V_{uq}^*$ and  $V_{tb}V_{tq}^*$, while Ref.~\cite{Beneke:2003zv} used $V_{ub}V_{uq}^*$ and  $V_{cb}V_{cq}^*$. The unitarity of CKM matrix guarantees the equivalence of the two approaches.  So we will directly compare the ``tree" ${\cal A}_u$ and ``penguin" ${\cal A}_t$ amplitudes, though some of them might be recombined in order to have the same CKM factors. 
Secondly, we have decomposed one part of the electroweak penguin into the QCD penguin as shown in Sec.~\ref{sec:multiplet}, and we will do so for QCDF amplitudes too.

We have the following correspondence between the SU(3) TDA amplitudes and the QCDF amplitudes  for ``tree" amplitudes: 
\begin{eqnarray}
T\to \alpha_1, \;\;\; P^u\to \alpha_4^u+ \beta_3^u, \;\;\; C\to \alpha_2, \;\;\; S^u\to \alpha_3^u+ \beta_{S3}^u, \;\;\; A\to \beta_2, \nonumber\\
E\to \beta_1,\;\;\; P_{A}^{u}\to \beta_4^u, \;\;\; A_{S}^{u}\to \beta_{S2},\;\;\; E_{S}^{u}\to \beta_{S1},\;\;\; S_{S}^{u}\to \beta_{S4}^u.
\end{eqnarray}
where the notations $\alpha_i$ and $\beta_i$ are from Ref.~\cite{Beneke:2003zv}. 
The correspondence for   ``penguin" ones is given as: 
\begin{eqnarray}
P_T\to \alpha_{4,EW}^c, \;\;\; P\to   \alpha_4^c+ \beta_3^c, \;\;\; P_C\to \alpha_{3,EW}^c, \;\;\; S\to \alpha_3^c+ \beta_{S3}^c,\;\;\; P_{TA}\to \beta_{3,EW}^c, \nonumber\\
P_{TE}\to  \beta_{4,EW}^c,\;\;\; P_A\to \beta_4^c,\;\;\; P_{ES}\to \beta_{S3,EW}^c, \;\;\; P_{AS}\to \beta_{S4,EW}^c, \;\;\; P_{SS}\to \beta_{S4}^c. 
\end{eqnarray}

\subsubsection{U-Spin relations}
Some  decay channels shown in Table~\ref{tab:Two_body_PP} with $\Delta S=0$ and $\Delta S=1$ are related by U-spin, the $d\leftrightarrow s$ exchange symmetry. The relations will be discussed explicitly in the following.  These pairs of channels include:\\
$B^{-}\to K^{0}K^{-}$ and $B^{-}\to\pi^{-}\overline{K}^{0}$; 
$\overline{B}^{0}\to\pi^{+}\pi^{-}$ and $\overline{B}_{s}^{0}\to K^{+}K^{-}$;   
$\overline{B}^{0}\to K^{0}\overline{K}^{0}$ and $\overline{B}_{s}^{0}\to K^{0}\overline{K}^{0}$;
$\overline{B}^{0}\to K^{-}K^{+}$ and $\overline{B}_{s}^{0}\to\pi^{+}\pi^{-}$;
$\overline{B}_{s}^{0}\to\pi^{0}K^{0}$ and $\overline{B}^{0}\to\pi^{0}\overline{K}^{0}$;
$\overline{B}_{s}^{0}\to\pi^{-}K^{+}$ and $\overline{B}^{0}\to\pi^{+}K^{-}$;
$\overline{B}_{s}^{0}\to K^{0}\pi^{0}$ and $\overline{B}^{0}\to\pi^{0}\overline{K}^{0}$.

In the past years, there have been extensive examinations on the U-spin symmetry. One of the interesting features of these U-spin pairs is that there are CP violating relation among them. 
Here we consider two $U$-spin related decays with the same ``tree" ${\cal A}_u$ and and ``penguin" ${\cal A}_t$~\footnote{For   decay modes to be discussed in the following, there are non-trivial Clebsch-Gordon coefficients, such that  Eq.~\eqref{eq:U-spin-amplitudes} is modified as: \begin{eqnarray}
&&A(\Delta S = 0) =  r(V_{ub}V_{ud}^*{\cal A}_u + V_{tb}V^*_{td}{\cal A}_t)\;,\nonumber\\
&&A(\Delta S = 1) = V_{ub}V_{us}^*{\cal A}_u + V_{tb}V^*_{ts} {\cal A}_t\;.  \nonumber
\end{eqnarray}
The  relation in Eq.~\eqref{CPR} is changed to: 
\begin{eqnarray}
{A^i_{CP}(\Delta S=0)\over A^j _{CP}(\Delta S = 1)}= - r^2{\tau_j{\cal B}(\Delta S = 1)\over \tau_i {\cal B}(\Delta S =0)}\;.\nonumber
\end{eqnarray}
}:
\begin{eqnarray}
&&A(B_i \to PP, \Delta S = 0) =  V_{ub}V_{ud}^*{\cal A}_u + V_{tb}V^*_{td}{\cal A}_t\;,\nonumber\\
&&A(B_i\to PP, \Delta S = 1) = V_{ub}V_{us}^*{\cal A}_u + V_{tb}V^*_{ts} {\cal A}_t\;. \label{eq:U-spin-amplitudes}
\end{eqnarray} 

Through the relation
${\rm Im}(V_{ub}V_{ud}^*V^*_{tb}V_{td}) = -  {\rm Im}(V_{ub}V_{us}^*V^*_{tb}V_{ts})$, one can obtain 
the CP violating rate difference $\Delta (B_i \to PP, \Delta S ) = \Gamma(\Delta S) - \overline{\Gamma}(\Delta S)$~\cite{Deshpande:1994ii,He:1998rq,Gronau:2000zy}
\begin{eqnarray}
 \Delta(B_i \to PP, \Delta S=0) = -\Delta (B_j\to PP, \Delta S = 1)\;.
\end{eqnarray}
This leads to a relation between branching ratio and CP asymmetry $A^i_{CP} (\Delta S ) =  \Delta (B_i \to PP, \Delta S) /{\cal B}(B_i \to PP)$:
\begin{eqnarray}
{A^i_{CP}(\Delta S=0)\over A^j _{CP}(\Delta S = 1)}= - {\tau_j{\cal B}(\Delta S = 1)\over \tau_i {\cal B}(\Delta S =0)}\;.\label{CPR}
\end{eqnarray}
Here ${\cal B}(B_i \to PP)$ is the branching ratio of $B_i \to PP$ and $\tau_i$ is the lifetime of $B_i$.

One of the most prominent example is the case of the U-spin pair $\overline{B}_{s}^{0}\to\pi^{-}K^{+}$ and $\overline{B}^{0}\to\pi^{+}K^{-}$.  Their CP asymmetry have been studied in Ref.~\cite{Grossman:2013lya}. Here we will comment on the experimental situation for this case and introduce a parameter $r_c$ to account for  the deviation from SU(3) symmetry.
\begin{eqnarray}
\frac{A_{CP}(\overline{B}^{0}\to\pi^{+}K^{-})} {A_{CP}(\overline{B}_{s}^{0}\to\pi^{-}K^{+})}+ r_c \frac{\tau_B {\cal B}(\overline{B}_{s}^{0}\to\pi^{-}K^{+}) } {\tau_{B_s} {\cal B}(\overline{B}^{0}\to\pi^{+}K^{-}) }=0. 
\end{eqnarray}
In the SU(3) symmetry limit $r_c = 1$.

Using the experimental data from PDG~\cite{Patrignani:2016xqp,Tanabashi:2018oca}:
\begin{eqnarray}
 {\cal B}(\overline{B}_{s}^{0}\to\pi^{-}K^{+}) &=& (5.7\pm0.6)\times 10^{-6},  \;\;\;
A_{CP}(\overline{B}_{s}^{0}\to\pi^{-}K^{+}) = (0.26\pm0.04),  \nonumber\\
 {\cal B}(\overline{B}^{0}\to\pi^{+}K^{-}) &=&(19.6\pm0.5)\times 10^{-6}, \;\;\; A_{CP}(\overline{B}^{0}\to\pi^{+}K^{-}) = -0.082\pm0.006, 
\end{eqnarray}
one finds: 
\begin{eqnarray}
r_c = 1.084\pm0.219
\end{eqnarray}
where all errors have been added in quadrature. 
The resulting $r_c$ value indicates that the U-spin symmetry is well in the case of this decay pair. The exploration in more decay pairs is helpful for further investigation on this symmetry. 

Similar U-spin relations existing in other decays will be studied in the following sections. We will comment on them when specific decay channels are be discussed.

 \subsection{$B\to VV$ decays}

Decay amplitudes for $B\to VV$ channels can be obtained similarly by replacing the pseudo-scalar multiplet $P$ by the vector multiplet $V$ in Eq.~\eqref{eq:B2PP_IRA_tree} and in Eq.~\eqref{eq:B2PP_TDA_tree}.  

\begin{itemize}

\item  Since we have chosen the same parametrization for pseudoscalar and vector mesons, the expanded amplitudes for the $B\to VV$  channels can be obtained  directly  from the $B\to PP$.

\item There are three sets of  amplitudes for $B\to VV$ decays, corresponding to different polarizations. For convenience, one can choose the helicity amplitudes 
$A_0, A_{+}, A_{-}$ defined as:
\begin{eqnarray}
{\cal A}  = S_1 \epsilon_{V_1}^*\cdot \epsilon_{V_2}^* + S_2 \frac{1}{m_B^2}  \epsilon_{V_1}^*\cdot p_B \epsilon_{V_2}^*\cdot p_B -i  S_3\epsilon_{\mu\nu\rho\sigma} p_{V_1}^\mu p_{V_2}^\nu \epsilon_{V_1}^{*\rho}\epsilon_{V_2}^{*\sigma},
\end{eqnarray}
with $\epsilon_{0123}=1$, and  
\begin{eqnarray}
 A_0 = \frac{m_B^2}{2m_{V_1}m_{V_2}} \left(S_1 + \frac{S_2}{2} \right), \;\;\;
 A_{\pm}= S_1 \mp S_3. 
\end{eqnarray} 
 
Thus there are in total $3\times 9=27$ complex amplitudes for both tree and penguin, where ``9" is the number of the polarization combination of final two vectors. These amplitudes correspond to $2\times 54-1=107$ real parameters in theory. Two phases can not be measured through direct measurements of individual  $B$ and $\bar B$ decays, but one of the two can be obtained through the time-dependent analysis.

\item  In principle, all these 107 parameters could be determined through the angular distribution studies in experiment. Each $B\to V(\to P_1P_2)V(\to P_3P_4)$ channel can provide 10 observables. 
The angular distribution is given as:
\begin{eqnarray*}
\frac{d\Gamma}{d\cos\theta_1d\cos\theta_2d\phi}  &\propto& |A_{0}|^2 \cos^2\theta_1 \cos^2\theta_2 +\frac{1}{4} \sin^2\theta_1 \sin^2\theta_2 \left(|A_+|^2 +|A_-|^2 \right) \nonumber\\
&&+ \frac{1}{2} \sin^2\theta_1 \sin^2\theta_2 {\rm Re}(e^{2i\phi} A_{+}A_{-}^*) \nonumber\\
&& -\cos\theta_1 \sin\theta_1 \cos\theta_2 \sin\theta_2 [{\rm Re}(e^{-i\phi}A_{0}A_{+}^* + {\rm Re}(e^{i\phi} A_0 A_{-}^*)]. 
\end{eqnarray*}
Here $\theta_1$ ($\theta_2$) is defined by the flight direction of $P_1(P_3)$ in the rest frame of $V_1(V_2)$ and the flight direction  of $V_1(V_2)$ in the $B$ meson rest frame. $\phi$ is the relative angle between the two decay planes. 

\item Unfortunately,  due to the large amount of input parameters, it is 
a formidable task  to  perform a global fit, and  in particular   only limited data is available~\cite{Patrignani:2016xqp}. A realistic analysis at this stage will pick up only a limited amount of amplitudes.    In this direction, the weak annihilations and hard scattering amplitudes were extracted by fitting relevant data in Ref.~\cite{Chang:2016qyc}, while   the authors in Ref.~\cite{Wang:2017hxe} have performed a factorization-assisted TDA analysis. This allows one to remove some suppressed amplitudes  at the leading order approximation.  In Ref.~\cite{Wang:2017rmh}, the authors have adopted the dynamical analysis in the SCET and performed a flavor SU(3) fit of $B\to VV$ decays. On the other hand, 
recent dynamical improvements exist in Refs.~\cite{Zou:2015iwa,Yan:2018fif} using the perturbative QCD approach and Refs.~\cite{Chang:2017brr} in QCDF. 
 

\end{itemize}

\subsection{$B\to VP$ Decays}

\begin{table}
\renewcommand\arraystretch{1.4}
\newcommand{\tabincell}[2]{\begin{tabular}{@{}#1@{}}#2\end{tabular}}
\caption{$B\rightarrow VP$ decays induced by the $b \rightarrow d$ transition.}
\label{tab:B_VP_bd}%
\begin{tabular}{ccc}
\hline 
\hline 
channel  & IRA  & TDA\tabularnewline
\hline 
$B^{-}\to\rho^{0}\pi^{-}$  & \tabincell{c}{$(A_6^{{T1}}+3 A_{15}^{{T1}}-A_6^{{T2}}-3 A_{15}^{{T2}}-C_3^{{T1}}$\\$-C_6^{{T1}}+5 C_{15}^{{T1}}+C_3^{{T2}}+C_6^{{T2}}+3 C_{15}^{{T2}})/{\sqrt{2}}$} & $(-A_{1}+A_{2}+\text{C}_{1}+P^{u2}-P^{u1}+T_{2})/{\sqrt{2}}$\tabularnewline
\hline 
$B^{-}\to\rho^{-}\pi^{0}$  & \tabincell{c}{$(-A_6^{{T1}}-3 A_{15}^{{T1}}+A_6^{{T2}}+3 A_{15}^{{T2}}+C_3^{{T1}}$\\$+C_6^{{T1}}+3 C_{15}^{{T1}}-C_3^{{T2}}-C_6^{{T2}}+5 C_{15}^{{T2}})/{\sqrt{2}}$} & $(A_{1}-A_{2}+\text{C}_{2}-P^{u2}+P^{u1}+T_{1})/{\sqrt{2}}$\tabularnewline
\hline 
$B^{-}\to\rho^{-}\eta_{q}$  &  \tabincell{c}{$(1/{\sqrt{2}})(A_6^{{T1}}+3 A_{15}^{{T1}}+A_6^{{T2}}+3 A_{15}^{{T2}}+2 B_6^{{T2}}+6 B_{15}^{{T2}}$\\$+C_3^{{T1}}+C_6^{{T1}}+3 C_{15}^{{T1}}+C_3^{{T2}}-C_6^{{T2}}+C_{15}^{{T2}}+2 D_3^{{T2}})$ }&  \tabincell{c}{$(2A_{S}^{u2}+A_{1}+A_{2}+\text{C}_{2}+P^{u2}$\\$+P^{u1}+2S^{u2}+T_{1})/{\sqrt{2}}$}\tabularnewline
\hline 
$B^{-}\to\rho^{-}\eta_{s}$  & $B_6^{{T2}}+3 B_{15}^{{T2}}+C_6^{{T2}}-C_{15}^{{T2}}+D_3^{{T2}}$ & $A_{S}^{u2}+S^{u2}$\tabularnewline
\hline 
$B^{-}\to K^{*0}K^{-}$  & $A_6^{{T2}}+3 A_{15}^{{T2}}+C_3^{{T1}}-C_6^{{T1}}-C_{15}^{{T1}}$ & $A_{1}+P^{u1}$\tabularnewline
\hline 
$B^{-}\to K^{*-}K^{0}$  & $A_6^{{T1}}+3 A_{15}^{{T1}}+C_3^{{T2}}-C_6^{{T2}}-C_{15}^{{T2}}$ & $A_{2}+P^{u2}$\tabularnewline
\hline 
$B^{-}\to\omega\pi^{-}$  & \tabincell{c}{$(1/{\sqrt{2}})(A_6^{{T1}}+3 A_{15}^{{T1}}+A_6^{{T2}}+3 A_{15}^{{T2}}+2 B_6^{{T1}}+6 B_{15}^{{T1}}$\\$+C_3^{{T1}}-C_6^{{T1}}+C_{15}^{{T1}}+C_3^{{T2}}+C_6^{{T2}}+3 C_{15}^{{T2}}+2 D_3^{{T1}})$} &  \tabincell{c}{$(2A_{S}^{u1}+A_{1}+A_{2}+\text{C}_{1}+P^{u2}$\\$+P^{u1}+2S^{u1}+T_{2})/{\sqrt{2}}$}\tabularnewline
\hline 
$B^{-}\to\phi\pi^{-}$  & $B_6^{{T1}}+3 B_{15}^{{T1}}+C_6^{{T1}}-C_{15}^{{T1}}+D_3^{{T1}}$ & $A_{S}^{u1}+S^{u1}$\tabularnewline
\hline 
$\overline{B}^{0}\to\rho^{+}\pi^{-}$  & $A_3^T-2 A_{15}^{{T1}}-A_6^{{T2}}+3 A_{15}^{{T2}}+C_3^{{T2}}+C_6^{{T2}}+3 C_{15}^{{T2}}$ & $P_{A}^{u}+E_{1}+P^{u2}+T_{2}$\tabularnewline
\hline 
$\overline{B}^{0}\to\rho^{0}\pi^{0}$  &  \tabincell{c}{$(1/{2})(2 A_3^T-A_6^{{T1}}+A_{15}^{{T1}}-A_6^{{T2}}+A_{15}^{{T2}}+C_3^{{T1}}$\\$+C_6^{{T1}}-5 C_{15}^{{T1}}+C_3^{{T2}}+C_6^{{T2}}-5 C_{15}^{{T2}})$ }& $(1/{2})(2P_{A}^{u}-C_{1}-C_{2}+E_{1}+E_{2}+P^{u2}+P^{u1})$\tabularnewline
\hline 
$\overline{B}^{0}\to\rho^{0}\eta_{q}$ & \tabincell{c}{$(1/{2})(-A_6^{{T1}}+5 A_{15}^{{T1}}-A_6^{{T2}}+5 A_{15}^{{T2}}-2 B_6^{{T2}}+10 B_{15}^{{T2}}$\\$-C_3^{{T1}}-C_6^{{T1}}+5 C_{15}^{{T1}}-C_3^{{T2}}+C_6^{{T2}}-C_{15}^{{T2}}-2 D_3^{{T2}})$}&  \tabincell{c}{$(1/{2})(C_{1}-C_{2}+2E_{S}^{u2}+E_{1}$\\$+E_{2}-P^{u2}-P^{u1}-2S^{u2})$}\tabularnewline
\hline 
$\overline{B}^{0}\to\rho^{0}\eta_{s}$  & $-(B_6^{{T2}}-5 B_{15}^{{T2}}+C_6^{{T2}}-C_{15}^{{T2}}+D_3^{{T2}})/{\sqrt{2}}$ & $(E_{S}^{u2}-S^{u2})/{\sqrt{2}}$\tabularnewline
\hline 
$\overline{B}^{0}\to\rho^{-}\pi^{+}$  & $A_3^T-A_6^{{T1}}+3 A_{15}^{{T1}}-2 A_{15}^{{T2}}+C_3^{{T1}}+C_6^{{T1}}+3 C_{15}^{{T1}}$ & $P_{A}^{u}+E_{2}+P^{u1}+T_{1}$\tabularnewline
\hline 
$\overline{B}^{0}\to K^{*+}K^{-}$  & $A_3^T+A_6^{{T1}}-A_{15}^{{T1}}-A_6^{{T2}}+3 A_{15}^{{T2}}$ & $P_{A}^{u}+E_{1}$\tabularnewline
\hline 
$\overline{B}^{0}\to K^{*0}\overline{K}^{0}$  & $ A_3^T+A_6^{{T1}}-A_{15}^{{T1}}-2 A_{15}^{{T2}}+C_3^{{T1}}-C_6^{{T1}}-C_{15}^{{T1}}$ & $P_{A}^{u}+P^{u1}$\tabularnewline
\hline 
$\overline{B}^{0}\to\overline{K}^{*0}K^{0}$  & $A_3^T-2 A_{15}^{{T1}}+A_6^{{T2}}-A_{15}^{{T2}}+C_3^{{T2}}-C_6^{{T2}}-C_{15}^{{T2}}$ & $P_{A}^{u}+P^{u2}$\tabularnewline
\hline 
$\overline{B}^{0}\to K^{*-}K^{+}$  & $A_3^T-A_6^{{T1}}+3 A_{15}^{{T1}}+A_6^{{T2}}-A_{15}^{{T2}}$ & $P_{A}^{u}+E_{2}$\tabularnewline
\hline 
$\overline{B}^{0}\to\omega\pi^{0}$  & \tabincell{c}{$(1/{2})(-A_6^{{T1}}+5 A_{15}^{{T1}}-A_6^{{T2}}+5 A_{15}^{{T2}}-2 B_6^{{T1}}+10 B_{15}^{{T1}}-C_3^{{T1}}$\\$+C_6^{{T1}}-C_{15}^{{T1}}-C_3^{{T2}}-C_6^{{T2}}+5 C_{15}^{{T2}}-2 D_3^{{T1}})$ }&  \tabincell{c}{$(1/{2})(-C_{1}+C_{2}+2E_{S}^{u1}+E_{1}$\\$+E_{2}-P^{u2}-P^{u1}-2S^{u1})$}\tabularnewline
\hline 
$\overline{B}^{0}\to\omega\eta_{q}$  &\tabincell{c}{$(1/{2})(2 A_3^T-A_6^{{T1}}+A_{15}^{{T1}}-A_6^{{T2}}+A_{15}^{{T2}}+4 B_3^T-2 B_6^{{T1}}$\\$+2 B_{15}^{{T1}}-2 B_6^{{T2}}+2 B_{15}^{{T2}}+C_3^{{T1}}-C_6^{{T1}}+C_{15}^{{T1}}+C_3^{{T2}}$\\$-C_6^{{T2}}+C_{15}^{{T2}}+2 D_3^{{T1}}+2 D_3^{{T2}})$}&\tabincell{c}{ $(1/{2})(2P_{A}^{u}+C_{1}+C_{2}+2E_{S}^{u1}+2E_{S}^{u2}+E_{1}$\\$+E_{2}+P^{u2}+P^{u1}+2S^{u2}+2S^{u1}+4S_{S}^{u})$}\tabularnewline
\hline 
$\overline{B}^{0}\to\omega\eta_{s}$  & $(2 B_3^T+2 B_6^{{T1}}-2 B_{15}^{{T1}}-B_6^{{T2}}+B_{15}^{{T2}}+C_6^{{T2}}-C_{15}^{{T2}}+D_3^{{T2}})/{\sqrt{2}}$ & $(E_{S}^{u2}+S^{u2}+2S_{S}^{u})/{\sqrt{2}}$\tabularnewline
\hline 
$\overline{B}^{0}\to\phi\pi^{0}$  & $-(B_6^{{T1}}-5 B_{15}^{{T1}}+C_6^{{T1}}-C_{15}^{{T1}}+D_3^{{T1}})/{\sqrt{2}}$ & $(E_{S}^{u1}-S^{u1})/{\sqrt{2}}$\tabularnewline
\hline 
$\overline{B}^{0}\to\phi\eta_{q}$  & $(2 B_3^T-B_6^{{T1}}+B_{15}^{{T1}}+2 B_6^{{T2}}-2 B_{15}^{{T2}}+C_6^{{T1}}-C_{15}^{{T1}}+D_3^{{T1}})/{\sqrt{2}}$ & $(E_{S}^{u1}+S^{u1}+2S_{S}^{u})/{\sqrt{2}}$\tabularnewline
\hline 
$\overline{B}^{0}\to\phi\eta_{s}$  & \tabincell{c}{$A_3^T+A_6^{{T1}}-A_{15}^{{T1}}+A_6^{{T2}}-A_{15}^{{T2}}+B_3^T+B_6^{{T1}}$\\$-B_{15}^{{T1}}+B_6^{{T2}}-B_{15}^{{T2}}$ }& $P_{A}^{u}+S_{S}^{u}$\tabularnewline
\hline 
$\overline{B}_{s}^{0}\to\rho^{0}K^{0}$  & $(A_6^{{T2}}+A_{15}^{{T2}}-C_3^{{T1}}-C_6^{{T1}}+5 C_{15}^{{T1}})/{\sqrt{2}}$ & $(\text{C}_{1}-P^{u1})/{\sqrt{2}}$\tabularnewline
\hline 
$\overline{B}_{s}^{0}\to\rho^{-}K^{+}$  & $-A_6^{{T2}}-A_{15}^{{T2}}+C_3^{{T1}}+C_6^{{T1}}+3 C_{15}^{{T1}}$ & $P^{u1}+T_{1}$\tabularnewline
\hline 
$\overline{B}_{s}^{0}\to K^{*+}\pi^{-}$  & $-A_6^{{T1}}-A_{15}^{{T1}}+C_3^{{T2}}+C_6^{{T2}}+3 C_{15}^{{T2}}$ & $P^{u2}+T_{2}$\tabularnewline
\hline 
$\overline{B}_{s}^{0}\to K^{*0}\pi^{0}$  & $(A_6^{{T1}}+A_{15}^{{T1}}-C_3^{{T2}}-C_6^{{T2}}+5 C_{15}^{{T2}})/{\sqrt{2}}$ & $(C_{2}-P^{u2})/{\sqrt{2}}$\tabularnewline
\hline 
$\overline{B}_{s}^{0}\to K^{*0}\eta_{q}$  & $-(A_6^{{T1}}+A_{15}^{{T1}}+2 B_6^{{T2}}+2 B_{15}^{{T2}}-C_3^{{T2}}+C_6^{{T2}}-C_{15}^{{T2}}-2 D_3^{{T2}})/{\sqrt{2}}$ & $(C_{2}+P^{u2}+2S^{u2})/{\sqrt{2}}$\tabularnewline
\hline 
$\overline{B}_{s}^{0}\to K^{*0}\eta_{s}$  &\tabincell{c}{ $-A_6^{{T2}}-A_{15}^{{T2}}-B_6^{{T2}}-B_{15}^{{T2}}+C_3^{{T1}}-C_6^{{T1}}$\\$-C_{15}^{{T1}}+C_6^{{T2}}-C_{15}^{{T2}}+D_3^{{T2}}$} & $P^{u1}+S^{u2}$\tabularnewline
\hline 
$\overline{B}_{s}^{0}\to\omega K^{0}$  & $-(A_6^{{T2}}+A_{15}^{{T2}}+2 B_6^{{T1}}+2 B_{15}^{{T1}}-C_3^{{T1}}+C_6^{{T1}}-C_{15}^{{T1}}-2 D_3^{{T1}})/{\sqrt{2}}$ & $(C_{1}+P^{u1}+2S^{u1})/{\sqrt{2}}$\tabularnewline
\hline 
$\overline{B}_{s}^{0}\to\phi K^{0}$  & \tabincell{c}{$-A_6^{{T1}}-A_{15}^{{T1}}-B_6^{{T1}}-B_{15}^{{T1}}+C_6^{{T1}}-C_{15}^{{T1}}$\\$+C_3^{{T2}}-C_6^{{T2}}-C_{15}^{{T2}}+D_3^{{T1}}$} & $P^{u2}+S^{u1}$\tabularnewline
\hline 
\hline 
\end{tabular}
\end{table}

\begin{table}
\renewcommand\arraystretch{1.4}
\newcommand{\tabincell}[2]{\begin{tabular}{@{}#1@{}}#2\end{tabular}}
\caption{$B\rightarrow VP$ decays induced by the $b \rightarrow s$ transition.}
\label{tab:B_VP_bs}
\begin{tabular}{ccc}
\hline 
\hline 
channel  & IRA & TDA\tabularnewline
\hline 
$B^{-}\to\rho^{0}K^{-}$  & $(A_{6}^{T_1}+3A_{15}^{T_1}-2C_{6}^{T_1}+4C_{15}^{T_1}+C_{3}^{T_2}+C_{6}^{T_2}+3C_{15}^{T_2})/{\sqrt{2}}$ & $(A_{2}+C_{1}+P^{u2}+T_{2})/{\sqrt{2}}$\tabularnewline
\hline 
$B^{-}\to\rho^{-}\overline{K}^{0}$  & $A_{6}^{T_1}+3A_{15}^{T_1}+C_{3}^{T_2}-C_{6}^{T_2}-C_{15}^{T_2}$ & $A_{2}+P^{u2}$\tabularnewline
\hline 
$B^{-}\to\overline{K}^{*0}\pi^{-}$  & $A_{6}^{T_2}+3A_{15}^{T_2}+C_{3}^{T_1}-C_{6}^{T_1}-C_{15}^{T_1}$ & $A_{1}+P^{u1}$\tabularnewline
\hline 
$B^{-}\to K^{*-}\pi^{0}$  & $(A_{6}^{T_2}+3A_{15}^{T_2}+C_{3}^{T_1}+C_{6}^{T_1}+3C_{15}^{T_1}-2C_{6}^{T_2}+4C_{15}^{T_2})/{\sqrt{2}}$ & $(A_{1}+C_{2}+P^{u1}+T_{1})/{\sqrt{2}}$\tabularnewline
\hline 
$B^{-}\to K^{*-}\eta_{q}$  & \tabincell{c}{$(A_{6}^{T_2}+3A_{15}^{T_2}+2B_{6}^{T_2}+6B_{15}^{T_2}+C_{3}^{T_1}$\\$+C_{6}^{T_1}+3C_{15}^{T_1}+2C_{15}^{T_2}+2D_{3}^{T_2})/{\sqrt{2}}$} & $(2A_{S}^{u2}+A_{1}+C_{2}+P^{u1}+2S^{u2}+T_{1})/{\sqrt{2}}$\tabularnewline
\hline 
$B^{-}\to K^{*-}\eta_{s}$  & $A_{6}^{T_1}+3A_{15}^{T_1}+B_{6}^{T_2}+3B_{15}^{T_2}+C_{3}^{T_2}-2C_{15}^{T_2}+D_{3}^{T_2}$ & $A_{S}^{u2}+A_{2}+P^{u2}+S^{u2}$\tabularnewline
\hline 
$B^{-}\to\omega K^{-}$  & \tabincell{c}{$(A_{6}^{T_1}+3A_{15}^{T_1}+2B_{6}^{T_1}+6B_{15}^{T_1}+2C_{15}^{T_1}$\\$+C_{3}^{T_2}+C_{6}^{T_2}+3C_{15}^{T_2}+2D_{3}^{T_1})/{\sqrt{2}}$} & $(2A_{S}^{u1}+A_{2}+C_{1}+P^{u2}+2S^{u1}+T_{2})/{\sqrt{2}}$\tabularnewline
\hline 
$B^{-}\to\phi K^{-}$  & $A_{6}^{T_2}+3A_{15}^{T_2}+B_{6}^{T_1}+3B_{15}^{T_1}+C_{3}^{T_1}-2C_{15}^{T_1}+D_{3}^{T_1}$ & $A_{S}^{u1}+A_{1}+P^{u1}+S^{u1}$\tabularnewline
\hline 
$\overline{B}^{0}\to\rho^{+}K^{-}$  & $-A_{6}^{T_1}-A_{15}^{T_1}+C_{3}^{T_2}+C_{6}^{T_2}+3C_{15}^{T_2}$ & $P^{u2}+T_{2}$\tabularnewline
\hline 
$\overline{B}^{0}\to\rho^{0}\overline{K}^{0}$  & $(A_{6}^{T_1}+A_{15}^{T_1}-2C_{6}^{T_1}+4C_{15}^{T_1}-C_{3}^{T_2}+C_{6}^{T_2}+C_{15}^{T_2})/{\sqrt{2}}$ & $(C_{1}-P^{u2})/{\sqrt{2}}$\tabularnewline
\hline 
$\overline{B}^{0}\to\overline{K}^{*0}\pi^{0}$  & $(A_{6}^{T_2}+A_{15}^{T_2}-C_{3}^{T_1}+C_{6}^{T_1}+C_{15}^{T_1}-2C_{6}^{T_2}+4C_{15}^{T_2})/{\sqrt{2}}$ & $(C_{2}-P^{u1})/{\sqrt{2}}$\tabularnewline
\hline 
$\overline{B}^{0}\to\overline{K}^{*0}\eta_{q}$  & \tabincell{c}{$-(A_{6}^{T_2}+A_{15}^{T_2}+2B_{6}^{T_2}+2B_{15}^{T_2}-C_{3}^{T_1}$\\$+C_{6}^{T_1}+C_{15}^{T_1}-2C_{15}^{T_2}-2D_{3}^{T_2})/{\sqrt{2}}$} & $(C_{2}+P^{u1}+2S^{u2})/{\sqrt{2}}$\tabularnewline
\hline 
$\overline{B}^{0}\to\overline{K}^{*0}\eta_{s}$  & $-A_{6}^{T_1}-A_{15}^{T_1}-B_{6}^{T_2}-B_{15}^{T_2}+C_{3}^{T_2}-2C_{15}^{T_2}+D_{3}^{T_2}$ & $P^{u2}+S^{u2}$\tabularnewline
\hline 
$\overline{B}^{0}\to K^{*-}\pi^{+}$  & $-A_{6}^{T_2}-A_{15}^{T_2}+C_{3}^{T_1}+C_{6}^{T_1}+3C_{15}^{T_1}$ & $P^{u1}+T_{1}$\tabularnewline
\hline 
$\overline{B}^{0}\to\omega\overline{K}^{0}$  & \tabincell{c}{$-(A_{6}^{T_1}+A_{15}^{T_1}+2B_{6}^{T_1}+2B_{15}^{T_1}-2C_{15}^{T_1}$\\$-C_{3}^{T_2}+C_{6}^{T_2}+C_{15}^{T_2}-2D_{3}^{T_1})/{\sqrt{2}}$} & $(C_{1}+P^{u2}+2S^{u1})/{\sqrt{2}}$\tabularnewline
\hline 
$\overline{B}^{0}\to\phi\overline{K}^{0}$  & $-A_{6}^{T_2}-A_{15}^{T_2}-B_{6}^{T_1}-B_{15}^{T_1}+C_{3}^{T_1}-2C_{15}^{T_1}+D_{3}^{T_1}$ & $P^{u1}+S^{u1}$\tabularnewline
\hline 
$\overline{B}_{s}^{0}\to\rho^{+}\pi^{-}$  & $A_{3}^{T}+A_{6}^{T_1}-A_{15}^{T_1}-A_{6}^{T_2}+3A_{15}^{T_2}$ & $P_{A}^{u}+E_{1}$\tabularnewline
\hline 
$\overline{B}_{s}^{0}\to\rho^{0}\pi^{0}$  & $A_{3}^{T}+A_{15}^{T_1}+A_{15}^{T_2}$ & $(1/{2})(2P_{A}^{u}+E_{1}+E_{2})$\tabularnewline
\hline 
$\overline{B}_{s}^{0}\to\rho^{0}\eta_{q}$  & $2(A_{15}^{T_1}+A_{15}^{T_2}-B_{6}^{T_2}+2B_{15}^{T_2})-A_{6}^{T_1}-A_{6}^{T_2}$ & $(1/{2})(2E_{S}^{u2}+E_{1}+E_{2})$\tabularnewline
\hline 
$\overline{B}_{s}^{0}\to\rho^{0}\eta_{s}$  & $-\sqrt{2}(B_{6}^{T_2}-2B_{15}^{T_2}+C_{6}^{T_1}-2C_{15}^{T_1})$ & $(C_{1}+E_{S}^{u2})/{\sqrt{2}}$\tabularnewline
\hline 
$\overline{B}_{s}^{0}\to\rho^{-}\pi^{+}$  & $A_{3}^{T}-A_{6}^{T_1}+3A_{15}^{T_1}+A_{6}^{T_2}-A_{15}^{T_2}$ & $P_{A}^{u}+E_{2}$\tabularnewline
\hline 
$\overline{B}_{s}^{0}\to K^{*+}K^{-}$  & $A_{3}^{T}-2A_{15}^{T_1}-A_{6}^{T_2}+3A_{15}^{T_2}+C_{3}^{T_2}+C_{6}^{T_2}+3C_{15}^{T_2}$ & $P_{A}^{u}+E_{1}+P^{u2}+T_{2}$\tabularnewline
\hline 
$\overline{B}_{s}^{0}\to K^{*0}\overline{K}^{0}$  & $A_{3}^{T}-2A_{15}^{T_1}+A_{6}^{T_2}-A_{15}^{T_2}+C_{3}^{T_2}-C_{6}^{T_2}-C_{15}^{T_2}$ & $P_{A}^{u}+P^{u2}$\tabularnewline
\hline 
$\overline{B}_{s}^{0}\to\overline{K}^{*0}K^{0}$  & $A_{3}^{T}+A_{6}^{T_1}-A_{15}^{T_1}-2A_{15}^{T_2}+C_{3}^{T_1}-C_{6}^{T_1}-C_{15}^{T_1}$ & $P_{A}^{u}+P^{u1}$\tabularnewline
\hline 
$\overline{B}_{s}^{0}\to K^{*-}K^{+}$  & $A_{3}^{T}-A_{6}^{T_1}+3A_{15}^{T_1}-2A_{15}^{T_2}+C_{3}^{T_1}+C_{6}^{T_1}+3C_{15}^{T_1}$ & $P_{A}^{u}+E_{2}+P^{u1}+T_{1}$\tabularnewline
\hline 
$\overline{B}_{s}^{0}\to\omega\pi^{0}$  & $2(A_{15}^{T_1}+A_{15}^{T_2}-B_{6}^{T_1}+2B_{15}^{T_1})-A_{6}^{T_1}-A_{6}^{T_2}$ & $(1/{2})(2E_{S}^{u1}+E_{1}+E_{2})$\tabularnewline
\hline 
$\overline{B}_{s}^{0}\to\omega\eta_{q}$  & $A_{3}^{T}+A_{15}^{T_1}+A_{15}^{T_2}+2(B_{3}^{T}+B_{15}^{T_1}+B_{15}^{T_2})$ & $P_{A}^{u}+E_{S}^{u1}+E_{S}^{u2}+(E_{1})/{2}+(E_{2})/{2}+2S_{S}^{u}$\tabularnewline
\hline 
$\overline{B}_{s}^{0}\to\omega\eta_{s}$  & $\sqrt{2}(B_{3}^{T}-2B_{15}^{T_1}+B_{15}^{T_2}+C_{15}^{T_1}+D_{3}^{T_1})$ & $(C_{1}+E_{S}^{u2}+2(S^{u1}+S_{S}^{u}))/{\sqrt{2}}$\tabularnewline
\hline 
$\overline{B}_{s}^{0}\to\phi\pi^{0}$  & $-\sqrt{2}(B_{6}^{T_1}-2B_{15}^{T_1}+C_{6}^{T_2}-2C_{15}^{T_2})$ & $(C_{2}+E_{S}^{u1})/{\sqrt{2}}$\tabularnewline
\hline 
$\overline{B}_{s}^{0}\to\phi\eta_{q}$  & $\sqrt{2}(B_{3}^{T}+B_{15}^{T_1}-2B_{15}^{T_2}+C_{15}^{T_2}+D_{3}^{T_2})$ & $(C_{2}+E_{S}^{u1}+2(S^{u2}+S_{S}^{u}))/{\sqrt{2}}$\tabularnewline
\hline 
$\overline{B}_{s}^{0}\to\phi\eta_{s}$  & \tabincell{c}{$A_{3}^{T}-2A_{15}^{T_1}-2A_{15}^{T_2}+B_{3}^{T}-2B_{15}^{T_1}-2B_{15}^{T_2}$\\$+C_{3}^{T_1}-2C_{15}^{T_1}+C_{3}^{T_2}-2C_{15}^{T_2}+D_{3}^{T_1}+D_{3}^{T_2}$} & $P_{A}^{u}+P^{u2}+P^{u1}+S^{u2}+S^{u1}+S_{S}^{u}$\tabularnewline
\hline 
\hline 
\end{tabular}
\end{table}


We now study the   $B \to VP$ decays, whose   amplitudes can be obtained by replacing one of the $P$ in  Eq.~\eqref{eq:B2PP_IRA_tree} and in Eq.~\eqref{eq:B2PP_TDA_tree} by $V$ to obtain the IRA and TDA amplitudes. There are two ways to replace one of the $P$, therefore the amplitudes will be doubled compared with $B \to PP$.
We have IRA and TDA for $B\to VP$  decays  as follows: 
\begin{eqnarray}
{\cal A}_{u}^{IRA} & =&A_{3}^{T}B_{i}(H_{\bar{3}})^{i}P_{k}^{j}V_{j}^{k}+C_{3}^{T1}B_{i}(H_{\bar{3}})^{k}P_{j}^{i}V_{k}^{j}+C_{3}^{T2}B_{i}(H_{\bar{3}})^{k}V_{j}^{i}P_{k}^{j}+B_{3}^{T}B_{i}(H_{\bar{3}})^{i}P_{k}^{k}V_{j}^{j}\nonumber\\
 & &+D_{3}^{T1}B_{i}(H_{\bar{3}})^{j}P_{j}^{i}V_{k}^{k}+D_{3}^{T2}B_{i}(H_{\bar{3}})^{j}V_{j}^{i}P_{k}^{k}+A_{6}^{T1}B_{i}(H_{6})_{k}^{[ij]}P_{j}^{l}V_{l}^{k}+A_{6}^{T2}B_{i}(H_{6})_{k}^{[ij]}V_{j}^{l}P_{l}^{k}\nonumber\\
 & &+C_{6}^{T1}B_{i}(H_{6})_{k}^{[jl]}P_{j}^{i}V_{l}^{k}+C_{6}^{T2}B_{i}(H_{6})_{k}^{[jl]}V_{j}^{i}P_{l}^{k}+B_{6}^{T1}B_{i}(H_{6})_{k}^{[ij]}P_{j}^{k}V_{l}^{l}+B_{6}^{T2}B_{i}(H_{6})_{k}^{[ij]}V_{j}^{k}P_{l}^{l}\nonumber\\ 
 & &+A_{15}^{T1}B_{i}(H_{\overline {15}})_{k}^{\{ij\}}P_{j}^{l}V_{l}^{k}+A_{15}^{T2}B_{i}(H_{\overline {15}})_{k}^{\{ij\}}V_{j}^{l}P_{l}^{k}+C_{15}^{T1}B_{i}(H_{\overline {15}})_{l}^{\{jk\}}P_{j}^{i}V_{k}^{l}+C_{15}^{T2}B_{i}(H_{\overline {15}})_{l}^{\{jk\}}V_{j}^{i}P_{k}^{l}\nonumber\\
 & &+B_{15}^{T1}B_{i}(H_{\overline {15}})_{k}^{\{ij\}}P_{j}^{k}V_{l}^{l}+B_{15}^{T2}B_{i}(H_{\overline {15}})_{k}^{\{ij\}}V_{j}^{k}P_{l}^{l},\\
 {\cal A}_{u}^{TDA} & =&T_{1}B_{i}H_{k}^{jl}P_{j}^{i}V_{l}^{k}+T_{2}B_{i}H_{k}^{jl}V_{j}^{i}P_{l}^{k}+C_{1}B_{i}H_{k}^{lj}P_{j}^{i}V_{l}^{k}+C_{2}B_{i}H_{k}^{lj}V_{j}^{i}P_{l}^{k}\nonumber\\
 & &+A_{1}B_{i}H_{j}^{il}P_{k}^{j}V_{l}^{k}+A_{2}B_{i}H_{j}^{il}V_{k}^{j}P_{l}^{k}+E_{1}B_{i}H_{j}^{li}P_{k}^{j}V_{l}^{k}+E_{2}B_{i}H_{j}^{li}V_{k}^{j}P_{l}^{k}\nonumber\\
 & &+S^{u1}B_{i}H_{l}^{lj}P_{j}^{i}V_{k}^{k}+S^{u2}B_{i}H_{l}^{lj}V_{j}^{i}P_{k}^{k}+P^{u1}B_{i}H_{l}^{lk}P_{j}^{i}V_{k}^{j}+P^{u2}B_{i}H_{l}^{lk}V_{j}^{i}P_{k}^{j}\nonumber\\
 & &+P^{u}_{A}B_{i}H_{l}^{li}P_{k}^{j}V_{j}^{k}+S^{u}_{S}B_{i}H_{l}^{li}P_{j}^{j}V_{k}^{k}+E_{S}^{u1}B_{i}H_{l}^{ji}P_{j}^{l}V_{k}^{k}+E_{S}^{u2}B_{i}H_{l}^{ji}V_{j}^{l}P_{k}^{k}\nonumber\\
 & &+A_{S}^{u1}B_{i}H_{l}^{ij}P_{j}^{l}V_{k}^{k}+A_{S}^{u2}B_{i}H_{l}^{ij}V_{j}^{l}P_{k}^{k}. \label{eq:IRA_B_VP}
\end{eqnarray}

The expanded  amplitudes are given  in Tab.~\ref{tab:B_VP_bd} and Tab.~\ref{tab:B_VP_bs}.
Relations between the two sets of amplitudes are derived as:
\begin{eqnarray}
A_{3}^{T} & =&-\frac{1}{8}(A_{1}+A_{2}-3E_{1}-3E_{2})+P^{u}_{A},\;\;\;
B_{3}^{T} =S^{u}_{S}+\frac{1}{8}(3E_{S}^{u1}+3E_{S}^{u2}-A_{S}^{u1}-A_{S}^{u2})\nonumber\\
C_{3}^{T1} & =&\frac{1}{8}(3T_{1}-C_{1}+3A_{1}-E_{1})+P^{u1},\;\;
C_{3}^{T2} =\frac{1}{8}(3T_{2}-C_{2}+3A_{2}-E_{2})+P^{u2}\nonumber\\ 
D_{3}^{T1} & =&\frac{1}{8}(3C_{1}-T_{1}-E_{S}^{u1}+3A_{S}^{u1})+S^{u1},\;\;\;
D_{3}^{T2}  =\frac{1}{8}(3C_{2}-T_{2}-E_{S}^{u2}+3A_{S}^{u2})+S^{u2}\nonumber\\
A_{6}^{T1} & =&\frac{1}{4}(A_{2}-E_{2}),\;\;\;
A_{6}^{T2} =\frac{1}{4}(A_{1}-E_{1}),\;\;\; 
C_{6}^{T1}  =\frac{1}{4}(T_{1}-C_{1}),\;\;\;
C_{6}^{T2}  =\frac{1}{4}(T_{2}-C_{2})\nonumber\\
A_{15}^{T1} & =&\frac{1}{8}(A_{2}+E_{2}),\;\;\; 
A_{15}^{T2}  =\frac{1}{8}(A_{1}+E_{1}),\;\;\;
C_{15}^{T1} =\frac{1}{8}(T_{1}+C_{1}),\;\;\;
C_{15}^{T2} =\frac{1}{8}(T_{2}+C_{2}),\nonumber\\
B_{6}^{T1} & =&\frac{1}{4}(A_{S}^{u1}-E_{S}^{u1}),\;\;\;
B_{6}^{T2} =\frac{1}{4}(A_{S}^{u2}-E_{S}^{u2}),\;\;\;
B_{15}^{T1}  =\frac{1}{8}(E_{S}^{u1}+A_{S}^{u1}), \;\;\;
B_{15}^{T2}  =\frac{1}{8}(E_{S}^{u2}+A_{S}^{u2}). \label{eq:TDA_B_VP}
\end{eqnarray}
The inverse relations are solved as:
\begin{eqnarray}
A_{1} & =&4A_{15}^{T2}+2A_{6}^{T2}, \;\;\; 
A_{2} =2(2A_{15}^{T1}+A_{6}^{T1}),\;\;\;
T_{1}  =2(2C_{15}^{T1}+C_{6}^{T1}),\;\;\;
T_{2}  =2(2C_{15}^{T2}+C_{6}^{T2}),\nonumber\\
C_{1} & =&2(2C_{15}^{T1}-C_{6}^{T1}),\;\;\;
C_{2} =2(2C_{15}^{T2}-C_{6}^{T2}),\;\;\;
E_{1}  =2(2A_{15}^{T2}-A_{6}^{T2}),\;\;\;
E_{2}  =2(2A_{15}^{T1}-A_{6}^{T1}),\nonumber\\
A_{S}^{u1} & =&2(2B_{15}^{T1}+B_{6}^{T1}),\;\;\;
A_{S}^{u2} =2(2B_{15}^{T2}+B_{6}^{T2}),\;\;\;
E_{S}^{u1} =2(2B_{15}^{T1}-B_{6}^{T1}),\;\;\;
E_{S}^{u2} =2(2B_{15}^{T2}-B_{6}^{T2}),\nonumber\\
P^{u}_{A} & =&-A_{15}^{T1}+A_{6}^{T1}-A_{15}^{T2}+A_{6}^{T2}+A_{3}^{T},\;\;\;
S^{u}_{S}  =-B_{15}^{T1}+B_{6}^{T1}-B_{15}^{T2}+B_{6}^{T2}+B_{3}^{T},\nonumber\\
P^{u1} & =&-A_{15}^{T2}-A_{6}^{T2}-C_{15}^{T1}+C_{3}^{T1}-C_{6}^{T1},\;\;\;
P^{u2}  =-A_{15}^{T1}-A_{6}^{T1}-C_{15}^{T2}+C_{3}^{T2}-C_{6}^{T2},\nonumber\\
S^{u1} & =&-B_{15}^{T1}-B_{6}^{T1}-C_{15}^{T1}+C_{6}^{T1}+D_{3}^{T1},\;\;\;
S^{u2}  =-B_{15}^{T2}-B_{6}^{T2}-C_{15}^{T2}+C_{6}^{T2}+D_{3}^{T2}.
\end{eqnarray}

Unlike the $B\to PP$ and $B\to VV$ case, we are  not able to find any redundant amplitude.  Thus in total, we have 18 complex amplitudes for  ``tree" and ``penguin", respectively. It corresponds to $2\times 36-1=71$ real parameters in theory. A fit with all parameters is not available again, and most of the current analyses have made approximations by neglecting some suppressed amplitudes~\cite{Wang:2008rk,Cheng:2014rfa,Chang:2015wba,Zhou:2016jkv}.

Decay amplitudes for $B\to VP$ decays in Eq.~\eqref{eq:IRA_B_VP} and Eq.~\eqref{eq:TDA_B_VP} are organized in a symmetric way.  Taking the $B^-\to \rho^-\bar K^0$ and $B^-\to\overline K^{*0} \pi^-$ as an example, the IRA amplitudes are related with the replacement $A_6^{T1}\leftrightarrow A_6^{T2}$,  $A_{15}^{T1}\leftrightarrow A_{15}^{T2}$,  $C_3^{T1}\leftrightarrow C_{3}^{T2}$, $C_6^{T1}\leftrightarrow C_{6}^{T2}$, and $C_{15}^{T1}\leftrightarrow C_{15}^{T2}$, while for the TDA amplitudes, the correspondence is:  $A_1\leftrightarrow A_2$ and $P^{u1}\leftrightarrow P^{u2}$.  Other relations can be found in a similar way.

The $B\to VP$ channels related by the U-spin   include: $B^{-}\to\overline{K}^{*0}\pi^{-}$ and $B^{-}\to {K}^{*0} K^{-}$; $B^{-}\to \rho^{-} \overline{K}^{0}$ and $B^{-}\to K^{*-} {K}^{0} $; ${\overline B}^{0}_{s}\to K^{*+} K^{-}$ and ${\overline B}^{0} \to \rho^{+} \pi^{-} $;  ${\overline B}^{0}_{s}\to K^{*-} K^{+}$ and ${\overline B}^{0} \to \rho^{-} \pi^{+} $; 
${\overline B}^{0}_{s}\to \rho^{+} \pi^{-}$ and ${\overline B}^{0} \to K^{*+} K^{-} $;  ${\overline B}^{0}_{s}\to {\overline K}^{*0} K^{0}$ and ${\overline B}^{0} \to {K}^{*0} {\overline K}^{0}$; 
${\overline B}^{0}_{s}\to {K}^{*0} {\overline K}^{0}$ and ${\overline B}^{0} \to {\overline K}^{*0} K^{0}$; 
${\overline B}^{0}_{s}\to \rho^{-} \pi^{+}$ and ${\overline B}^{0} \to K^{*-} K^{+} $;  ${\overline B}^{0} \to {K}^{*-} \pi^{+}$ and ${\overline B}^{0}_{s} \to \rho^{-} K^{+} $;  ${\overline B}^{0} \to \rho^{+} K^{-}$ and ${\overline B}^{0}_{s} \to {K}^{*+} \pi^{-} $.   However on the experimental side, there are not enough  measurements  to examine these relations, in particular the CPV in $B_s$ sector  has received less consideration. We expect the situation will be improved when a large amount of data is available at LHCb, and Belle-II.    


\section{$D\to PP,\; VV,\;PV$ decays}

\begin{table}
\renewcommand\arraystretch{1.3}
\caption{Decay amplitudes for two-body $D\to PP$ decays. Decay amplitudes
for two-body $D\to PP$ decays. The CKM factor should be multiplied:
$V_{cs}V_{ud}^{*}$ for Cabibblo-Allowed decays; $V_{cs}V_{us}^{*}$
for singly Cabibbo-suppressed modes and $V_{cd}V_{us}^{*}$ for doubly
Cabibbo-suppressed modes. }
\label{tab:D_PP} %
\begin{tabular}{cccccc}
\hline 
\hline 
{$V_{cs}V_{ud}^{*}$}  & {IRA}  & {TDA}  &  &  & \tabularnewline
\hline 
$D^{0}\to\pi^{+}K^{-}$  & $-A_{6}^{T}+A_{15}^{T}+C_{6}^{T}+C_{15}^{T}$ & $\text{E}+\text{T}$ &  &  & \tabularnewline
\hline 
$D^{0}\to\pi^{0}\overline{K}^{0}$  & $(A_{6}^{T}-A_{15}^{T}-C_{6}^{T}+C_{15}^{T})/{\sqrt{2}}$ & $(\text{C}-\text{E})/{\sqrt{2}}$ &  &  & \tabularnewline
\hline 
$D^{0}\to\overline{K}^{0}\eta_{q}$  & $(-A_{6}^{T}+A_{15}^{T}-2B_{6}^{T}+2B_{15}^{T}-C_{6}^{T}+C_{15}^{T})/{\sqrt{2}}$ & $(\text{C}+2\text{E}_{S}^{u}+\text{E})/{\sqrt{2}}$ &  &  & \tabularnewline
\hline 
$D^{0}\to\overline{K}^{0}\eta_{s}$  & $-A_{6}^{T}+A_{15}^{T}-B_{6}^{T}+B_{15}^{T}$ & $\text{E}_{S}^{u}+\text{E}$ &  &  & \tabularnewline
\hline 
$D^{+}\to\pi^{+}\overline{K}^{0}$  & $2C_{15}^{T}$ & $\text{C}+\text{T}$ &  &  & \tabularnewline
\hline 
$D_{s}^{+}\to\pi^{+}\eta_{q}$  & $\sqrt{2}\left(A_{6}^{T}+A_{15}^{T}+B_{6}^{T}+B_{15}^{T}\right)$ & $\sqrt{2}\left(A_{S}^{u}+\text{A}\right)$ &  &  & \tabularnewline
\hline 
$D_{s}^{+}\to\pi^{+}\eta_{s}$  & $B_{6}^{T}+B_{15}^{T}+C_{6}^{T}+C_{15}^{T}$ & $A_{S}^{u}+\text{T}$ &  &  & \tabularnewline
\hline 
$D_{s}^{+}\to K^{+}\overline{K}^{0}$  & $A_{6}^{T}+A_{15}^{T}-C_{6}^{T}+C_{15}^{T}$ & $\text{A}+\text{C}$ &  &  & \tabularnewline
\hline 
\hline 
{$V_{cs}V_{us}^{*}$}  & {IRA}  & {TDA}  &  &  & \tabularnewline
\hline 
$D^{0}\to\pi^{+}\pi^{-}$  & $A_{6}^{T}-A_{15}^{T}-C_{6}^{T}-C_{15}^{T}$ & $-\text{E}-\text{T}$ &  &  & \tabularnewline
\hline 
$D^{0}\to\pi^{0}\pi^{0}$  & $A_{6}^{T}-A_{15}^{T}-C_{6}^{T}+C_{15}^{T}$ & $\text{C}-\text{E}$ &  &  & \tabularnewline
\hline 
$D^{0}\to\pi^{0}\eta_{q}$  & $-A_{6}^{T}+A_{15}^{T}-B_{6}^{T}+B_{15}^{T}$ & $\text{E}_{S}^{u}+\text{E}$ &  &  & \tabularnewline
\hline 
$D^{0}\to\pi^{0}\eta_{s}$  & $(-B_{6}^{T}+B_{15}^{T}-C_{6}^{T}+C_{15}^{T})/{\sqrt{2}}$ & $(\text{C}+\text{E}_{S}^{u})/{\sqrt{2}}$ &  &  & \tabularnewline
\hline 
$D^{0}\to K^{+}K^{-}$  & $-A_{6}^{T}+A_{15}^{T}+C_{6}^{T}+C_{15}^{T}$ & $\text{E}+\text{T}$ &  &  & \tabularnewline
\hline 
$D^{0}\to\eta_{q}\eta_{q}$  & $A_{6}^{T}-A_{15}^{T}+2B_{6}^{T}-2B_{15}^{T}+C_{6}^{T}-C_{15}^{T}$ & $-\text{C}-2\text{E}_{S}^{u}-\text{E}$ &  &  & \tabularnewline
\hline 
$D^{0}\to\eta_{q}\eta_{s}$  & $(-B_{6}^{T}+B_{15}^{T}-C_{6}^{T}+C_{15}^{T})/{\sqrt{2}}$ & $(\text{C}+\text{E}_{S}^{u})/{\sqrt{2}}$ &  &  & \tabularnewline
\hline 
$D^{0}\to\eta_{s}\eta_{s}$  & $-A_{6}^{T}+A_{15}^{T}-B_{6}^{T}+B_{15}^{T}$ & $\text{E}_{S}^{u}+\text{E}$ &  &  & \tabularnewline
\hline 
$D^{+}\to\pi^{+}\pi^{0}$  & $\sqrt{2}C_{15}^{T}$ & $(\text{C}+\text{T})/{\sqrt{2}}$ &  &  & \tabularnewline
\hline 
$D^{+}\to\pi^{+}\eta_{q}$  & $-\sqrt{2}\left(A_{6}^{T}+A_{15}^{T}+B_{6}^{T}+B_{15}^{T}+C_{15}^{T}\right)$ & $-(2A_{S}^{u}+2\text{A}+\text{C}+\text{T})/{\sqrt{2}}$ &  &  & \tabularnewline
\hline 
$D^{+}\to\pi^{+}\eta_{s}$  & $-B_{6}^{T}-B_{15}^{T}-C_{6}^{T}+C_{15}^{T}$ & $\text{C}-A_{S}^{u}$ &  &  & \tabularnewline
\hline 
$D^{+}\to K^{+}\overline{K}^{0}$  & $-A_{6}^{T}-A_{15}^{T}+C_{6}^{T}+C_{15}^{T}$ & $\text{T}-\text{A}$ &  &  & \tabularnewline
\hline 
$D_{s}^{+}\to\pi^{+}K^{0}$  & $A_{6}^{T}+A_{15}^{T}-C_{6}^{T}-C_{15}^{T}$ & $\text{A}-\text{T}$ &  &  & \tabularnewline
\hline 
$D_{s}^{+}\to\pi^{0}K^{+}$  & $(A_{6}^{T}+A_{15}^{T}-C_{6}^{T}+C_{15}^{T})/{\sqrt{2}}$ & $(\text{A}+\text{C})/{\sqrt{2}}$ &  &  & \tabularnewline
\hline 
$D_{s}^{+}\to K^{+}\eta_{q}$  & $(A_{6}^{T}+A_{15}^{T}+2B_{6}^{T}+2B_{15}^{T}+C_{6}^{T}-C_{15}^{T})/{\sqrt{2}}$ & $(2A_{S}^{u}+\text{A}-\text{C})/{\sqrt{2}}$ &  &  & \tabularnewline
\hline 
$D_{s}^{+}\to K^{+}\eta_{s}$  & $A_{6}^{T}+A_{15}^{T}+B_{6}^{T}+B_{15}^{T}+2C_{15}^{T}$ & $A_{S}^{u}+\text{A}+\text{C}+\text{T}$ &  &  & \tabularnewline
\hline 
\hline 
{$V_{cd}V_{us}^{*}$}  & {IRA}  & {TDA}  &  &  & \tabularnewline
\hline 
$D^{0}\to\pi^{0}K^{0}$  & $(A_{6}^{T}-A_{15}^{T}-C_{6}^{T}+C_{15}^{T})/{\sqrt{2}}$ & $(\text{C}-\text{E})/{\sqrt{2}}$ &  &  & \tabularnewline
\hline 
$D^{0}\to\pi^{-}K^{+}$  & $-A_{6}^{T}+A_{15}^{T}+C_{6}^{T}+C_{15}^{T}$ & $\text{E}+\text{T}$ &  &  & \tabularnewline
\hline 
$D^{0}\to K^{0}\eta_{q}$  & $(-A_{6}^{T}+A_{15}^{T}-2B_{6}^{T}+2B_{15}^{T}-C_{6}^{T}+C_{15}^{T})/{\sqrt{2}}$ & $(\text{C}+2\text{E}_{S}^{u}+\text{E})/{\sqrt{2}}$ &  &  & \tabularnewline
\hline 
$D^{0}\to K^{0}\eta_{s}$  & $-A_{6}^{T}+A_{15}^{T}-B_{6}^{T}+B_{15}^{T}$ & $\text{E}_{S}^{u}+\text{E}$ &  &  & \tabularnewline
\hline 
$D^{+}\to\pi^{+}K^{0}$  & $A_{6}^{T}+A_{15}^{T}-C_{6}^{T}+C_{15}^{T}$ & $\text{A}+\text{C}$ &  &  & \tabularnewline
\hline 
$D^{+}\to\pi^{0}K^{+}$  & $(A_{6}^{T}+A_{15}^{T}-C_{6}^{T}-C_{15}^{T})/{\sqrt{2}}$ & $(\text{A}-\text{T})/{\sqrt{2}}$ &  &  & \tabularnewline
\hline 
$D^{+}\to K^{+}\eta_{q}$  & $(A_{6}^{T}+A_{15}^{T}+2B_{6}^{T}+2B_{15}^{T}+C_{6}^{T}+C_{15}^{T})/{\sqrt{2}}$ & $(2A_{S}^{u}+\text{A}+\text{T})/{\sqrt{2}}$ &  &  & \tabularnewline
\hline 
$D^{+}\to K^{+}\eta_{s}$  & $A_{6}^{T}+A_{15}^{T}+B_{6}^{T}+B_{15}^{T}$ & $A_{S}^{u}+\text{A}$ &  &  & \tabularnewline
\hline 
$D_{s}^{+}\to K^{+}K^{0}$  & $2C_{15}^{T}$ & $\text{C}+\text{T}$ &  &  & \tabularnewline
\hline 
\hline 
\end{tabular}
\end{table}  
\begin{table}
\renewcommand\arraystretch{1.4}
\caption{Decay amplitudes for two-body Cabibbo-Allowed $D\to VP$ decays.}
\label{tab:D_VP_Cabibblo_Allowed}%
\begin{tabular}{cccccc}
\hline 
\hline 
{channel}  & {IRA}  & {TDA}  &  &  & \tabularnewline
\hline 
$D^{0}\to\rho^{+}K^{-}$  & $-A_{6}^{\text{T1}}+A_{15}^{\text{T1}}+C_{6}^{\text{T1}}+C_{15}^{\text{T1}}$ & $T_{1}+E_{2}$ &  &  & \tabularnewline
\hline 
$D^{0}\to\rho^{0}\overline{K}^{0}$  & $(A_{6}^{\text{T1}}-A_{15}^{\text{T1}}-C_{6}^{\text{T2}}+C_{15}^{\text{T2}})/{\sqrt{2}}$ & $(C_{2}-E_{2})/{\sqrt{2}}$ &  &  & \tabularnewline
\hline 
$D^{0}\to\overline{K}^{*0}\pi^{0}$  & $(A_{6}^{\text{T2}}-A_{15}^{\text{T2}}-C_{6}^{\text{T1}}+C_{15}^{\text{T1}})/{\sqrt{2}}$ & $(C_{1}-E_{1})/{\sqrt{2}}$ &  &  & \tabularnewline
\hline 
$D^{0}\to\overline{K}^{*0}\eta_{q}$  & $(-A_{6}^{\text{T2}}+A_{15}^{\text{T2}}-2B_{6}^{\text{T2}}+2B_{15}^{\text{T2}}-C_{6}^{\text{T1}}+C_{15}^{\text{T1}})/{\sqrt{2}}$ & $(C_{1}+2E_{S}^{\text{u2}}+E_{1})/{\sqrt{2}}$ &  &  & \tabularnewline
\hline 
$D^{0}\to\overline{K}^{*0}\eta_{s}$  & $-A_{6}^{\text{T1}}+A_{15}^{\text{T1}}-B_{6}^{\text{T2}}+B_{15}^{\text{T2}}$ & $E_{S}^{\text{u2}}+E_{2}$ &  &  & \tabularnewline
\hline 
$D^{0}\to K^{*-}\pi^{+}$  & $-A_{6}^{\text{T2}}+A_{15}^{\text{T2}}+C_{6}^{\text{T2}}+C_{15}^{\text{T2}}$ & $T_{2}+E_{1}$ &  &  & \tabularnewline
\hline 
$D^{0}\to\omega\overline{K}^{0}$  & $(-A_{6}^{\text{T1}}+A_{15}^{\text{T1}}-2B_{6}^{\text{T1}}+2B_{15}^{\text{T1}}-C_{6}^{\text{T2}}+C_{15}^{\text{T2}})/{\sqrt{2}}$ & $(C_{2}+2E_{S}^{\text{u1}}+E_{2})/{\sqrt{2}}$ &  &  & \tabularnewline
\hline 
$D^{0}\to\phi\overline{K}^{0}$  & $-A_{6}^{\text{T2}}+A_{15}^{\text{T2}}-B_{6}^{\text{T1}}+B_{15}^{\text{T1}}$ & $E_{S}^{\text{u1}}+E_{1}$ &  &  & \tabularnewline
\hline 
$D^{+}\to\rho^{+}\overline{K}^{0}$  & $C_{6}^{\text{T1}}+C_{15}^{\text{T1}}-C_{6}^{\text{T2}}+C_{15}^{\text{T2}}$ & $C_{2}+T_{1}$ &  &  & \tabularnewline
\hline 
$D^{+}\to\overline{K}^{*0}\pi^{+}$  & $-C_{6}^{\text{T1}}+C_{15}^{\text{T1}}+C_{6}^{\text{T2}}+C_{15}^{\text{T2}}$ & $C_{1}+T_{2}$ &  &  & \tabularnewline
\hline 
$D_{s}^{+}\to\rho^{+}\pi^{0}$  & $(A_{6}^{\text{T1}}+A_{15}^{\text{T1}}-A_{6}^{\text{T2}}-A_{15}^{\text{T2}})/{\sqrt{2}}$ & $(A_{2}-A_{1})/{\sqrt{2}}$ &  &  & \tabularnewline
\hline 
$D_{s}^{+}\to\rho^{+}\eta_{q}$  & $(A_{6}^{\text{T1}}+A_{15}^{\text{T1}}+A_{6}^{\text{T2}}+A_{15}^{\text{T2}}+2\left(B_{6}^{\text{T2}}+B_{15}^{\text{T2}}\right))/{\sqrt{2}}$ & $(2A_{S}^{\text{u2}}+A_{1}+A_{2})/{\sqrt{2}}$ &  &  & \tabularnewline
\hline 
$D_{s}^{+}\to\rho^{+}\eta_{s}$  & $B_{6}^{\text{T2}}+B_{15}^{\text{T2}}+C_{6}^{\text{T1}}+C_{15}^{\text{T1}}$ & $A_{S}^{\text{u2}}+T_{1}$ &  &  & \tabularnewline
\hline 
$D_{s}^{+}\to\rho^{0}\pi^{+}$  & $(-A_{6}^{\text{T1}}-A_{15}^{\text{T1}}+A_{6}^{\text{T2}}+A_{15}^{\text{T2}})/{\sqrt{2}}$ & $(A_{1}-A_{2})/{\sqrt{2}}$ &  &  & \tabularnewline
\hline 
$D_{s}^{+}\to K^{*+}\overline{K}^{0}$  & $A_{6}^{\text{T2}}+A_{15}^{\text{T2}}-C_{6}^{\text{T2}}+C_{15}^{\text{T2}}$ & $A_{1}+C_{2}$ &  &  & \tabularnewline
\hline 
$D_{s}^{+}\to\overline{K}^{*0}K^{+}$  & $A_{6}^{\text{T1}}+A_{15}^{\text{T1}}-C_{6}^{\text{T1}}+C_{15}^{\text{T1}}$ & $A_{2}+C_{1}$ &  &  & \tabularnewline
\hline 
$D_{s}^{+}\to\omega\pi^{+}$  & $(A_{6}^{\text{T1}}+A_{15}^{\text{T1}}+A_{6}^{\text{T2}}+A_{15}^{\text{T2}}+2\left(B_{6}^{\text{T1}}+B_{15}^{\text{T1}}\right))/{\sqrt{2}}$ & $(2A_{S}^{\text{u1}}+A_{1}+A_{2})/{\sqrt{2}}$ &  &  & \tabularnewline
\hline 
$D_{s}^{+}\to\phi\pi^{+}$  & $B_{6}^{\text{T1}}+B_{15}^{\text{T1}}+C_{6}^{\text{T2}}+C_{15}^{\text{T2}}$ & $A_{S}^{\text{u1}}+T_{2}$ &  &  & \tabularnewline
\hline 
\hline 
\end{tabular}
\end{table}
\begin{table}
\renewcommand\arraystretch{1.4}
\newcommand{\tabincell}[2]{\begin{tabular}{@{}#1@{}}#2\end{tabular}}
\caption{Decay amplitudes for two-body Singly Cabibblo-Suppressed $D\to VP$
decays.}
\label{tab:D_VP_Singly_Cabibblo_Suppressed}%
\begin{tabular}{cccccc}
\hline 
\hline 
{channel}  & {IRA}  & {TDA}  &  &  & \tabularnewline
\hline 
$D^{0}\to\rho^{+}\pi^{-}$  & $A_{6}^{\text{T1}}-A_{15}^{\text{T1}}-C_{6}^{\text{T1}}-C_{15}^{\text{T1}}$ & $-T_{1}-E_{2}$ &  &  & \tabularnewline
\hline 
$D^{0}\to\rho^{0}\pi^{0}$  & $(1/{2})\left(A_{6}^{\text{T1}}-A_{15}^{\text{T1}}+A_{6}^{\text{T2}}-A_{15}^{\text{T2}}-C_{6}^{\text{T1}}+C_{15}^{\text{T1}}-C_{6}^{\text{T2}}+C_{15}^{\text{T2}}\right)$ & $(1/{2})\left(C_{1}+C_{2}-E_{1}-E_{2}\right)$ &  &  & \tabularnewline
\hline 
$D^{0}\to\rho^{0}\eta_{q}$  &\tabincell{c}{  $(1/{2})(-A_{6}^{T1}+A_{15}^{T1}-A_{6}^{T2}+A_{15}^{T2}-2B_{6}^{T2}$\\$+2B_{15}^{T2}-C_{6}^{T1}+C_{15}^{T1}+C_{6}^{T2}-C_{15}^{T2})$ }& $(1/{2})(C_{1}-C_{2}+2E_{s}^{u2}+E_{1}+E_{2})$ &  &  & \tabularnewline
\hline 
$D^{0}\to\rho^{0}\eta_{s}$  & $(-B_{6}^{\text{T2}}+B_{15}^{\text{T2}}-C_{6}^{\text{T2}}+C_{15}^{\text{T2}})/{\sqrt{2}}$ & $(C_{2}+E_{S}^{\text{u2}})/{\sqrt{2}}$ &  &  & \tabularnewline
\hline 
$D^{0}\to\rho^{-}\pi^{+}$  & $A_{6}^{\text{T2}}-A_{15}^{\text{T2}}-C_{6}^{\text{T2}}-C_{15}^{\text{T2}}$ & $-T_{2}-E_{1}$ &  &  & \tabularnewline
\hline 
$D^{0}\to K^{*+}K^{-}$  & $-A_{6}^{\text{T1}}+A_{15}^{\text{T1}}+C_{6}^{\text{T1}}+C_{15}^{\text{T1}}$ & $T_{1}+E_{2}$ &  &  & \tabularnewline
\hline 
$D^{0}\to K^{*0}\overline{K}^{0}$  & $-A_{6}^{\text{T1}}+A_{15}^{\text{T1}}+A_{6}^{\text{T2}}-A_{15}^{\text{T2}}$ & $E_{2}-E_{1}$ &  &  & \tabularnewline
\hline 
$D^{0}\to\overline{K}^{*0}K^{0}$  & $A_{6}^{\text{T1}}-A_{15}^{\text{T1}}-A_{6}^{\text{T2}}+A_{15}^{\text{T2}}$ & $E_{1}-E_{2}$ &  &  & \tabularnewline
\hline 
$D^{0}\to K^{*-}K^{+}$  & $-A_{6}^{\text{T2}}+A_{15}^{\text{T2}}+C_{6}^{\text{T2}}+C_{15}^{\text{T2}}$ & $T_{2}+E_{1}$ &  &  & \tabularnewline
\hline 
$D^{0}\to\omega\pi^{0}$  & \tabincell{c}{ $(1/{2})(-A_{6}^{T1}+A_{15}^{T1}-A_{6}^{T2}+A_{15}^{T2}-2B_{6}^{T1}$\\$+2B_{15}^{T1}+C_{6}^{T1}-C_{15}^{T1}-C_{6}^{T2}+C_{15}^{T2})$} & $(1/{2})(-C_{1}+C_{2}+2E_{s}^{u1}+E_{1}+E_{2})$ &  &  & \tabularnewline
\hline 
$D^{0}\to\omega\eta_{q}$  & \tabincell{c}{ $(1/{2})(A_{6}^{T1}-A_{15}^{T1}+A_{6}^{T2}-A_{15}^{T2}+2B_{6}^{T1}-2B_{15}^{T1}$\\$+2B_{6}^{T2}-2B_{15}^{T2}+C_{6}^{T1}-C_{15}^{T1}+C_{6}^{T2}-C_{15}^{T2})$} & $(1/{2})(-C_{1}-C_{2}-2E_{s}^{u1}-2E_{s}^{u2}-E_{1}-E_{2})$ &  &  & \tabularnewline
\hline 
$D^{0}\to\omega\eta_{s}$  & $(-2B_{6}^{\text{T1}}+2B_{15}^{\text{T1}}+B_{6}^{\text{T2}}-B_{15}^{\text{T2}}-C_{6}^{\text{T2}}+C_{15}^{\text{T2}})/{\sqrt{2}}$ & $(C_{2}+2E_{S}^{\text{u1}}-E_{S}^{\text{u2}})/{\sqrt{2}}$ &  &  & \tabularnewline
\hline 
$D^{0}\to\phi\pi^{0}$  & $(-B_{6}^{\text{T1}}+B_{15}^{\text{T1}}-C_{6}^{\text{T1}}+C_{15}^{\text{T1}})/{\sqrt{2}}$ & $(C_{1}+E_{S}^{\text{u1}})/{\sqrt{2}}$ &  &  & \tabularnewline
\hline 
$D^{0}\to\phi\eta_{q}$  & $(B_{6}^{\text{T1}}-B_{15}^{\text{T1}}-2B_{6}^{\text{T2}}+2B_{15}^{\text{T2}}-C_{6}^{\text{T1}}+C_{15}^{\text{T1}})/{\sqrt{2}}$ & $(C_{1}-E_{S}^{\text{u1}}+2E_{S}^{\text{u2}})/{\sqrt{2}}$ &  &  & \tabularnewline
\hline 
$D^{0}\to\phi\eta_{s}$  & $-A_{6}^{\text{T1}}+A_{15}^{\text{T1}}-A_{6}^{\text{T2}}+A_{15}^{\text{T2}}-B_{6}^{\text{T1}}+B_{15}^{\text{T1}}-B_{6}^{\text{T2}}+B_{15}^{\text{T2}}$ & $E_{S}^{\text{u1}}+E_{S}^{\text{u2}}+E_{1}+E_{2}$ &  &  & \tabularnewline
\hline 
$D^{+}\to\rho^{+}\pi^{0}$  & $(-A_{6}^{\text{T1}}-A_{15}^{\text{T1}}+A_{6}^{\text{T2}}+A_{15}^{\text{T2}}+C_{6}^{\text{T1}}+C_{15}^{\text{T1}}-C_{6}^{\text{T2}}+C_{15}^{\text{T2}})/{\sqrt{2}}$ & $(A_{1}-A_{2}+C_{2}+T_{1})/{\sqrt{2}}$ &  &  & \tabularnewline
\hline 
$D^{+}\to\rho^{+}\eta_{q}$  & \tabincell{c}{$-(A_{6}^{\text{T1}}+A_{15}^{\text{T1}}+A_{6}^{\text{T2}}+A_{15}^{\text{T2}}+2B_{6}^{\text{T2}}$\\$+2B_{15}^{\text{T2}}+C_{6}^{\text{T1}}+C_{15}^{\text{T1}}-C_{6}^{\text{T2}}+C_{15}^{\text{T2}})/{\sqrt{2}}$} & $-(2A_{S}^{\text{u2}}+A_{1}+A_{2}+C_{2}+T_{1})/{\sqrt{2}}$ &  &  & \tabularnewline
\hline 
$D^{+}\to\rho^{+}\eta_{s}$  & $-B_{6}^{\text{T2}}-B_{15}^{\text{T2}}-C_{6}^{\text{T2}}+C_{15}^{\text{T2}}$ & $C_{2}-A_{S}^{\text{u2}}$ &  &  & \tabularnewline
\hline 
$D^{+}\to\rho^{0}\pi^{+}$  & $(A_{6}^{\text{T1}}+A_{15}^{\text{T1}}-A_{6}^{\text{T2}}-A_{15}^{\text{T2}}-C_{6}^{\text{T1}}+C_{15}^{\text{T1}}+C_{6}^{\text{T2}}+C_{15}^{\text{T2}})/{\sqrt{2}}$ & $(-A_{1}+A_{2}+C_{1}+T_{2})/{\sqrt{2}}$ &  &  & \tabularnewline
\hline 
$D^{+}\to K^{*+}\overline{K}^{0}$  & $-A_{6}^{\text{T2}}-A_{15}^{\text{T2}}+C_{6}^{\text{T1}}+C_{15}^{\text{T1}}$ & $T_{1}-A_{1}$ &  &  & \tabularnewline
\hline 
$D^{+}\to\overline{K}^{*0}K^{+}$  & $-A_{6}^{\text{T1}}-A_{15}^{\text{T1}}+C_{6}^{\text{T2}}+C_{15}^{\text{T2}}$ & $T_{2}-A_{2}$ &  &  & \tabularnewline
\hline 
$D^{+}\to\omega\pi^{+}$  & \tabincell{c}{$-(A_{6}^{\text{T1}}+A_{15}^{\text{T1}}+A_{6}^{\text{T2}}+A_{15}^{\text{T2}}+2B_{6}^{\text{T1}}$\\$+2B_{15}^{\text{T1}}-C_{6}^{\text{T1}}+C_{15}^{\text{T1}}+C_{6}^{\text{T2}}+C_{15}^{\text{T2}})/{\sqrt{2}}$} & $-(2A_{S}^{\text{u1}}+A_{1}+A_{2}+C_{1}+T_{2})/{\sqrt{2}}$ &  &  & \tabularnewline
\hline 
$D^{+}\to\phi\pi^{+}$  & $-B_{6}^{\text{T1}}-B_{15}^{\text{T1}}-C_{6}^{\text{T1}}+C_{15}^{\text{T1}}$ & $C_{1}-A_{S}^{\text{u1}}$ &  &  & \tabularnewline
\hline 
$D_{s}^{+}\to\rho^{+}K^{0}$  & $A_{6}^{\text{T2}}+A_{15}^{\text{T2}}-C_{6}^{\text{T1}}-C_{15}^{\text{T1}}$ & $A_{1}-T_{1}$ &  &  & \tabularnewline
\hline 
$D_{s}^{+}\to\rho^{0}K^{+}$  & $(A_{6}^{\text{T2}}+A_{15}^{\text{T2}}-C_{6}^{\text{T1}}+C_{15}^{\text{T1}})/{\sqrt{2}}$ & $(A_{1}+C_{1})/{\sqrt{2}}$ &  &  & \tabularnewline
\hline 
$D_{s}^{+}\to K^{*+}\pi^{0}$  & $(A_{6}^{\text{T1}}+A_{15}^{\text{T1}}-C_{6}^{\text{T2}}+C_{15}^{\text{T2}})/{\sqrt{2}}$ & $(A_{2}+C_{2})/{\sqrt{2}}$ &  &  & \tabularnewline
\hline 
$D_{s}^{+}\to K^{*+}\eta_{q}$  & $(A_{6}^{\text{T1}}+A_{15}^{\text{T1}}+2B_{6}^{\text{T2}}+2B_{15}^{\text{T2}}+C_{6}^{\text{T2}}-C_{15}^{\text{T2}})/{\sqrt{2}}$ & $(2A_{S}^{\text{u2}}+A_{2}-C_{2})/{\sqrt{2}}$ &  &  & \tabularnewline
\hline 
$D_{s}^{+}\to K^{*+}\eta_{s}$  & $A_{6}^{\text{T2}}+A_{15}^{\text{T2}}+B_{6}^{\text{T2}}+B_{15}^{\text{T2}}+C_{6}^{\text{T1}}+C_{15}^{\text{T1}}-C_{6}^{\text{T2}}+C_{15}^{\text{T2}}$ & $A_{S}^{\text{u2}}+A_{1}+C_{2}+T_{1}$ &  &  & \tabularnewline
\hline 
$D_{s}^{+}\to K^{*0}\pi^{+}$  & $A_{6}^{\text{T1}}+A_{15}^{\text{T1}}-C_{6}^{\text{T2}}-C_{15}^{\text{T2}}$ & $A_{2}-T_{2}$ &  &  & \tabularnewline
\hline 
$D_{s}^{+}\to\omega K^{+}$  & $(A_{6}^{\text{T2}}+A_{15}^{\text{T2}}+2B_{6}^{\text{T1}}+2B_{15}^{\text{T1}}+C_{6}^{\text{T1}}-C_{15}^{\text{T1}})/{\sqrt{2}}$ & $(2A_{S}^{\text{u1}}+A_{1}-C_{1})/{\sqrt{2}}$ &  &  & \tabularnewline
\hline 
$D_{s}^{+}\to\phi K^{+}$  & $A_{6}^{\text{T1}}+A_{15}^{\text{T1}}+B_{6}^{\text{T1}}+B_{15}^{\text{T1}}-C_{6}^{\text{T1}}+C_{15}^{\text{T1}}+C_{6}^{\text{T2}}+C_{15}^{\text{T2}}$ & $A_{S}^{\text{u1}}+A_{2}+C_{1}+T_{2}$ &  &  & \tabularnewline
\hline 
\hline 
\end{tabular}
\end{table}

\begin{table}
\renewcommand\arraystretch{1.4}
\caption{Decay amplitudes for two-body Doubly Cabibblo-Suppressed $D\to VP$
decays.}
\label{tab:D_VP_Doubly_Cabibblo_Suppressed}%
\begin{tabular}{cccccc}
\hline 
\hline 
{channel}  & {IRA}  & {TDA}  &  &  & \tabularnewline
\hline 
$D^{0}\to\rho^{0}K^{0}$  & $(A_{6}^{\text{T2}}-A_{15}^{\text{T2}}-C_{6}^{\text{T2}}+C_{15}^{\text{T2}})/{\sqrt{2}}$ & $(C_{2}-E_{1})/{\sqrt{2}}$ &  &  & \tabularnewline
\hline 
$D^{0}\to\rho^{-}K^{+}$  & $-A_{6}^{\text{T2}}+A_{15}^{\text{T2}}+C_{6}^{\text{T2}}+C_{15}^{\text{T2}}$ & $T_{2}+E_{1}$ &  &  & \tabularnewline
\hline 
$D^{0}\to K^{*+}\pi^{-}$  & $-A_{6}^{\text{T1}}+A_{15}^{\text{T1}}+C_{6}^{\text{T1}}+C_{15}^{\text{T1}}$ & $T_{1}+E_{2}$ &  &  & \tabularnewline
\hline 
$D^{0}\to K^{*0}\pi^{0}$  & $(A_{6}^{\text{T1}}-A_{15}^{\text{T1}}-C_{6}^{\text{T1}}+C_{15}^{\text{T1}})/{\sqrt{2}}$ & $(C_{1}-E_{2})/{\sqrt{2}}$ &  &  & \tabularnewline
\hline 
$D^{0}\to K^{*0}\eta_{q}$  & $(-A_{6}^{\text{T1}}+A_{15}^{\text{T1}}-2B_{6}^{\text{T2}}+2B_{15}^{\text{T2}}-C_{6}^{\text{T1}}+C_{15}^{\text{T1}})/{\sqrt{2}}$ & $(C_{1}+2E_{S}^{\text{u2}}+E_{2})/{\sqrt{2}}$ &  &  & \tabularnewline
\hline 
$D^{0}\to K^{*0}\eta_{s}$  & $-A_{6}^{\text{T2}}+A_{15}^{\text{T2}}-B_{6}^{\text{T2}}+B_{15}^{\text{T2}}$ & $E_{S}^{\text{u2}}+E_{1}$ &  &  & \tabularnewline
\hline 
$D^{0}\to\omega K^{0}$  & $(-A_{6}^{\text{T2}}+A_{15}^{\text{T2}}-2B_{6}^{\text{T1}}+2B_{15}^{\text{T1}}-C_{6}^{\text{T2}}+C_{15}^{\text{T2}})/{\sqrt{2}}$ & $(C_{2}+2E_{S}^{\text{u1}}+E_{1})/{\sqrt{2}}$ &  &  & \tabularnewline
\hline 
$D^{0}\to\phi K^{0}$  & $-A_{6}^{\text{T1}}+A_{15}^{\text{T1}}-B_{6}^{\text{T1}}+B_{15}^{\text{T1}}$ & $E_{S}^{\text{u1}}+E_{2}$ &  &  & \tabularnewline
\hline 
$D^{+}\to\rho^{+}K^{0}$  & $A_{6}^{\text{T2}}+A_{15}^{\text{T2}}-C_{6}^{\text{T2}}+C_{15}^{\text{T2}}$ & $A_{1}+C_{2}$ &  &  & \tabularnewline
\hline 
$D^{+}\to\rho^{0}K^{+}$  & $(A_{6}^{\text{T2}}+A_{15}^{\text{T2}}-C_{6}^{\text{T2}}-C_{15}^{\text{T2}})/{\sqrt{2}}$ & $(A_{1}-T_{2})/{\sqrt{2}}$ &  &  & \tabularnewline
\hline 
$D^{+}\to K^{*+}\pi^{0}$  & $(A_{6}^{\text{T1}}+A_{15}^{\text{T1}}-C_{6}^{\text{T1}}-C_{15}^{\text{T1}})/{\sqrt{2}}$ & $(A_{2}-T_{1})/{\sqrt{2}}$ &  &  & \tabularnewline
\hline 
$D^{+}\to K^{*+}\eta_{q}$  & $(A_{6}^{\text{T1}}+A_{15}^{\text{T1}}+2B_{6}^{\text{T2}}+2B_{15}^{\text{T2}}+C_{6}^{\text{T1}}+C_{15}^{\text{T1}})/{\sqrt{2}}$ & $(2A_{S}^{\text{u2}}+A_{2}+T_{1})/{\sqrt{2}}$ &  &  & \tabularnewline
\hline 
$D^{+}\to K^{*+}\eta_{s}$  & $A_{6}^{\text{T2}}+A_{15}^{\text{T2}}+B_{6}^{\text{T2}}+B_{15}^{\text{T2}}$ & $A_{S}^{\text{u2}}+A_{1}$ &  &  & \tabularnewline
\hline 
$D^{+}\to K^{*0}\pi^{+}$  & $A_{6}^{\text{T1}}+A_{15}^{\text{T1}}-C_{6}^{\text{T1}}+C_{15}^{\text{T1}}$ & $A_{2}+C_{1}$ &  &  & \tabularnewline
\hline 
$D^{+}\to\omega K^{+}$  & $(A_{6}^{\text{T2}}+A_{15}^{\text{T2}}+2B_{6}^{\text{T1}}+2B_{15}^{\text{T1}}+C_{6}^{\text{T2}}+C_{15}^{\text{T2}})/{\sqrt{2}}$ & $(2A_{S}^{\text{u1}}+A_{1}+T_{2})/{\sqrt{2}}$ &  &  & \tabularnewline
\hline 
$D^{+}\to\phi K^{+}$  & $A_{6}^{\text{T1}}+A_{15}^{\text{T1}}+B_{6}^{\text{T1}}+B_{15}^{\text{T1}}$ & $A_{S}^{\text{u1}}+A_{2}$ &  &  & \tabularnewline
\hline 
$D_{s}^{+}\to K^{*+}K^{0}$  & $C_{6}^{\text{T1}}+C_{15}^{\text{T1}}-C_{6}^{\text{T2}}+C_{15}^{\text{T2}}$ & $C_{2}+T_{1}$ &  &  & \tabularnewline
\hline 
$D_{s}^{+}\to K^{*0}K^{+}$  & $-C_{6}^{\text{T1}}+C_{15}^{\text{T1}}+C_{6}^{\text{T2}}+C_{15}^{\text{T2}}$ & $C_{1}+T_{2}$ &  &  & \tabularnewline
\hline 
\hline 
\end{tabular}
\end{table}


Using the effective Hamiltonian in Eqs.~\eqref{eq:H3615_c_allowed} and \eqref{eq:H3615_c_doubly_suprressed}, one can easily obtain the $SU(3)$ decay amplitudes in a similar fashion as that for $B\to PP,\;VV,\;PV$. For $D$ decays we will only discuss   tree contributions which are written  as ${\cal A} = V_{cs/d}V^*_{ud/s} {\cal A}_{u}^{IRA, TDA}$.  The penguin amplitudes are   suppressed and thus neglected in this work, however it is necessary to stress  penguin amplitudes are mandatory for   the CP violation.  The SU(3) amplitudes for $D \to PP$ are parametrized as
\begin{eqnarray} 
 {\cal A}^{IRA}_u &=&A_6^T D_i (H_{ 6})^{[ij]}_k P_j^lP_l^k 
  +C_6^T D_i (H_{ 6})^{[jl]}_k P^i_j P_l^k +B_6^T D_i (H_{ 6})^{[ij]}_k P_j^kP_l^l \nonumber\\
  && 
  +A_{15}^T D_i (H_{\overline{15}})^{\{ij\}}_k P_j^lP_l^k 
  +C_{15}^T D_i (H_{\overline{15}})^{\{jk\}}_l P^i_j P_k^l +B_{15}^T D_i (H_{\overline{15}})^{\{ij\}}_k P_j^kP_l^l,  \label{eq:D2PP_IRA_tree}\\ 
{\cal A}^{TDA}_u &=&  T   D_i H^{jl}_k  P^{i}_j  P^k_l   +C   D_i H^{lj}_k P^{i}_j  P^k_l + A   D_i H^{il}_j   P^j_k P^{k}_l  + E   D_i  H^{li}_j P^j_k P^{k}_l\nonumber\\
&& +E^{u}_{S} D_i H^{ji}_{l}  P^{l}_j  P^k_k+A^{u}_{S} D_i  H^{ij}_{l}   P^{l}_j    P^k_k. \label{eq:D2PP_TDA_tree}
\end{eqnarray}
The expanded amplitudes are given in Tab.~\ref{tab:D_PP}. 
The amplitudes $A_{6}^T$ can be incorporated in  $B_{6}^{T\prime}$ and $C_{6}^{T\prime}$, and then we have 5 independent amplitudes for $D\to PP$: 
\begin{eqnarray} 
A_{15}^T&=& \frac{A+E}{2},  \;\;\;B_{15}^T= \frac{A^{u}_{S}+E^{u}_{S}}{2}, \;\; 
C_{15}^T= \frac{T+C}{2},  
B_{6}^{\prime T}= \frac{A^{u}_{S}-E^{u}_{S}+A-E}{2}, \;\;\; 
C_{6}^{\prime T}= \frac{T-C-A+E}{2},  
\end{eqnarray}
with the inverse relation:
\begin{eqnarray}
 T+E &=& A_{15}^T + C_{6}^{\prime T}+C_{15}^T, \;\;  C-E= -A_{15}^T - C_{6}^{\prime T}+C_{15}^T, \;\;
 A+E= 2A_{15}^T, \nonumber\\
 A^{u}_{S}-E&=& -A_{15}^T +B_{6}^{\prime T} +B_{15}^T,  \;\;  E^{u}_{S}+E= A_{15}^T -B_{6}^{\prime T} +B_{15}^T. 
\end{eqnarray}
Since   one amplitude is redundant, fits with all 6 complex amplitudes should  not be resolved in principle.  This has been indicated  by the strong correlation of parameters  in the fits in Ref.~\cite{Li:2012cfa}.

Again for $D \to VV$ decays there are three sets of amplitudes similar as the $D \to PP$,  and thus  we   have 15 independent amplitudes in total.

\begin{figure}
\begin{center}
\includegraphics[scale=0.5]{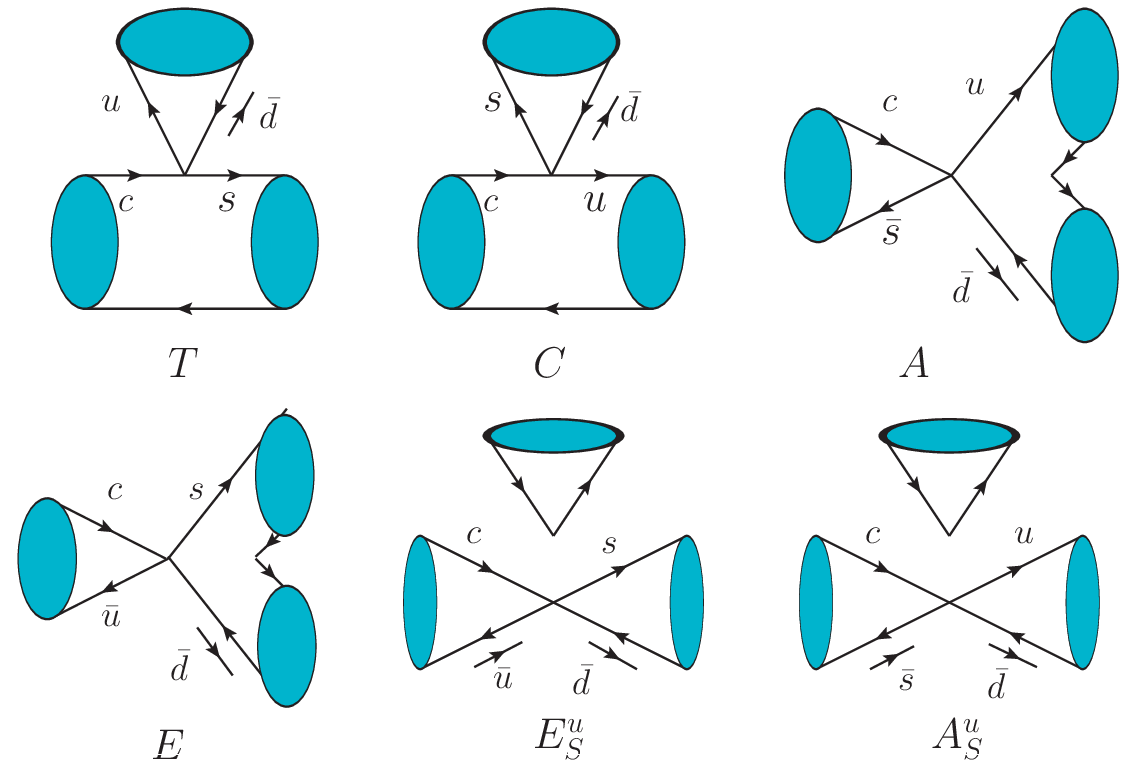}
\end{center}
\caption{Topological diagrams for tree amplitudes in Cabibbo-allowed  $D\to PP$ and $D\to VV$ decays.  The penguin amplitudes are   suppressed by both CKM matrix elements and Wilson coefficients, and thus neglected in this work, however it is necessary to point out penguin amplitudes are mandatory to study the CP violation. The $s$ can be replaced by $d$ and $\bar d$ can be replaced by $\bar s$ to obtain the singly-Cabibbo suppressed and doubly-Cabibbo suppressed decays.     } \label{fig:Feynman_D_PP}
\end{figure}

The IRA and TDA for $D\to VP$ decays are given as: 
\begin{eqnarray}
{\cal A}_{u}^{IRA} & =&A_{6}^{T1}D_{i}(H_{6})_{k}^{[ij]}P_{j}^{l}V_{l}^{k}+A_{6}^{T2}D_{i}(H_{6})_{k}^{[ij]}V_{j}^{l}P_{l}^{k}+C_{6}^{T1}D_{i}(H_{6})_{k}^{[jl]}P_{j}^{i}V_{l}^{k}+C_{6}^{T2}D_{i}(H_{6})_{k}^{[jl]}V_{j}^{i}P_{l}^{k}\nonumber\\
 & &+B_{6}^{T1}D_{i}(H_{6})_{k}^{[ij]}P_{j}^{k}V_{l}^{l}+B_{6}^{T2}D_{i}(H_{6})_{k}^{[ij]}V_{j}^{k}P_{l}^{l}+A_{15}^{T1}D_{i}(H_{\overline {15}})_{k}^{\{ij\}}P_{j}^{l}V_{l}^{k}+A_{15}^{T2}D_{i}(H_{\overline {15}})_{k}^{\{ij\}}V_{j}^{l}P_{l}^{k}\nonumber\\
 & &+C_{15}^{T1}D_{i}(H_{\overline {15}})_{l}^{\{jk\}}P_{j}^{i}V_{k}^{l}+C_{15}^{T2}D_{i}(H_{\overline {15}})_{l}^{\{jk\}}V_{j}^{i}P_{k}^{l}+B_{15}^{T1}D_{i}(H_{\overline {15}})_{k}^{\{ij\}}P_{j}^{k}V_{l}^{l}+B_{15}^{T2}D_{i}(H_{\overline {15}})_{k}^{\{ij\}}V_{j}^{k}P_{l}^{l},\\
 {\cal A}_{u}^{TDA} & =&T_{1}D_{i}H_{k}^{jl}P_{j}^{i}V_{l}^{k}+T_{2}D_{i}H_{k}^{jl}V_{j}^{i}P_{l}^{k}+C_{1}D_{i}H_{k}^{lj}P_{j}^{i}V_{l}^{k}+C_{2}D_{i}H_{k}^{lj}V_{j}^{i}P_{l}^{k}\nonumber\\
 & &+A_{1}D_{i}H_{j}^{il}P_{k}^{j}V_{l}^{k}+A_{2}D_{i}H_{j}^{il}V_{k}^{j}P_{l}^{k}+E_{1}D_{i}H_{j}^{li}P_{k}^{j}V_{l}^{k}+E_{2}D_{i}H_{j}^{li}V_{k}^{j}P_{l}^{k}\nonumber\\
 & &+E^{u1}_{S}D_{i}H_{l}^{ji}P_{j}^{l}V_{k}^{k}+E^{u2}_{S}D_{i}H_{l}^{ji}V_{j}^{l}P_{k}^{k}+A^{u1}_{S}D_{i}H_{l}^{ij}P_{j}^{l}V_{k}^{k}+A^{u2}_{S}D_{i}H_{l}^{ij}V_{j}^{l}P_{k}^{k}. 
\end{eqnarray} 
The above amplitudes are also expanded in a symmetric way. 
The expanded  amplitudes are collected  in Tab.~\ref{tab:D_VP_Cabibblo_Allowed}, Tab.~\ref{tab:D_VP_Singly_Cabibblo_Suppressed} and Tab.~\ref{tab:D_VP_Doubly_Cabibblo_Suppressed} for the different transitions. 
Relations between the two sets of amplitudes are derived as: 
\begin{eqnarray}
A_{6}^{T1} & =&\frac{1}{2}(A_{2}-E_{2}),\;\;\;
A_{6}^{T2} =\frac{1}{2}(A_{1}-E_{1}),\;\;\;
B_{6}^{T1} =\frac{1}{2}(A^{u1}_{S}-E^{u1}_{S}),\;\;\;
B_{6}^{T2} =\frac{1}{2}(A^{u2}_{S}-E^{u2}_{S})\nonumber\\
C_{6}^{T1} & =&\frac{1}{2}(T_{1}-C_{1}),\;\;\;
C_{6}^{T2}  =\frac{1}{2}(T_{2}-C_{2}),\;\;\;
A_{15}^{T1}  =\frac{1}{2}(A_{2}+E_{2}),\;\;\;
A_{15}^{T2} =\frac{1}{2}(A_{1}+E_{1})\nonumber\\
B_{15}^{T1} &=&\frac{1}{2}(E^{u1}_{S}+A^{u1}_{S}),\;\;\;
B_{15}^{T2}  =\frac{1}{2}(E^{u2}_{S}+A^{u2}_{S}),\;\;\;
C_{15}^{T1}  =\frac{1}{2}(T_{1}+C_{1}),\;\;\;
C_{15}^{T2}  =\frac{1}{2}(T_{2}+C_{2}). 
\end{eqnarray}
The inverse relations are solved as:
\begin{eqnarray}
A_{1} & =&A_{15}^{T2}+A_{6}^{T2},\;\;\;
A_{2} =A_{15}^{T1}+A_{6}^{T1},\;\;\;
T_{1} =C_{15}^{T1}+C_{6}^{T1},\;\;\;
T_{2} =C_{15}^{T2}+C_{6}^{T2}\nonumber\\
C_{1} & =&C_{15}^{T1}-C_{6}^{T1},\;\;\;
C_{2}  =C_{15}^{T2}-C_{6}^{T2},\;\;\;
E_{1}  =A_{15}^{T2}-A_{6}^{T2},\;\;\;
E_{2} =A_{15}^{T1}-A_{6}^{T1}\nonumber\\
A^{u1}_{S} & =&B_{15}^{T1}+B_{6}^{T1},\;\;\;
A^{u2}_{S}  =B_{15}^{T2}+B_{6}^{T2},\;\;\;
E^{u1}_{S} =B_{15}^{T1}-B_{6}^{T1},\;\;\;
E^{u2}_{S} =B_{15}^{T2}-B_{6}^{T2}. 
\end{eqnarray}

It is interesting  to explore the useful  relations for decay widths from the amplitudes listed in Tab.~\ref{tab:D_VP_Cabibblo_Allowed}, Tab.~\ref{tab:D_VP_Singly_Cabibblo_Suppressed} and Tab.~\ref{tab:D_VP_Doubly_Cabibblo_Suppressed}. 
For Cabibblo Allowed channels, we find   $\Gamma(D^+_s\to \rho^+ \pi^0 )$ = $\Gamma(D^+_s\to \rho^0 \pi^+ )$.
For singly Cabibblo suppressed channels, one has: 
\begin{eqnarray}
\Gamma(D^0\to \rho^+ \pi^- )= \Gamma(D^0\to K^{*+} K^- ),&\ \ \Gamma(D^0\to \rho^- \pi^+ )= \Gamma(D^0\to K^{*-} K^+ ),\nonumber\\ 
\Gamma(D^+\to K^{*+} \overline K^0 )= \Gamma(D^+_s\to \rho^+ K^0 ),&\ \ \Gamma(D^+\to \overline K^{*0} K^+ )= \Gamma(D^+_s\to K^{*0} \pi^+ ),\nonumber\\ 
\Gamma(D^0\to \overline K^{*0} K^0 )=\Gamma(D^0\to K^{*0} \overline K^0 ).&
\end{eqnarray}
We refer the reader to Refs.~\cite{Li:2012cfa,Cheng:2016ejf,Cheng:2012xb,Grossman:2012ry} for some explorations of the implications on decay rates and CP asymmetries, and Refs.~\cite{Ablikim:2018ydy,Aaij:2016cfh} for the experimental analyses.

It is necessary to notice that since  the QCD scale is comparable to charm quark mass, finite quark mass difference in s and d quarks may lead to  sizable SU(3) symmetry breaking effect.  A notable  example to explore   SU(3) symmetry breaking effects is the $D^0\to K^0\bar K^0$. This channel has vanishing branching fraction if exact SU(3)  symmetry holds and the $\bar 3$ contribution proportional to $V_{cb}V^*_{ub}$ is neglected, which can also been seen from Tab.~\ref{tab:D_PP}. The experimental data for branching ratios of  $D^0\to K^0\bar K^0$ and $D^0\to K^+K^-$ are given as~\cite{Patrignani:2016xqp,Tanabashi:2018oca}:
\begin{eqnarray}
{\cal B}(D^0\to K^+K^-) &=& (3.97\pm0.07)\times 10^{-3}, \nonumber\\
{\cal B}(D^0\to K^0 \bar K^0)&=& (3.40\pm0.12)\times 10^{-4}. 
\end{eqnarray}
The decay amplitude of  $D^0\to K^+K^-$ is $E+T$, where $T$ is  color-allowed  and could be estimated using the factorization framework.  As $|V_{cb}V^*_{ub}/V_{cs}V_{sd}^*| < 1.3\times 10^{-3}$, it is unlikely that the above non-zero branching ratio is caused by  the $\bar 3 $ penguin contribution. The above data shows that the SU(3) symmetry breaking effects in some channels can be as large as $30\%$ at the amplitude level.  Though the above estimate is channel-dependent, it indicates that  symmetry breaking effects must carefully treated in $D$ meson and charmed baryon decays in a systematic way~\cite{Li:2012cfa,Cheng:2012xb,Grossman:2012ry,Hiller:2012xm,Muller:2015rna,Muller:2015lua}.

\section{$B_c\to DP,\;DV$ decays}


\begin{table}
\renewcommand\arraystretch{1.3}
\caption{Decay amplitudes for $B_c\to DP$ decays. }\label{tab:Bc_DP}\begin{tabular}{ccccccccc}\hline\hline
{$b\to d$} &{IRA}  & {TDA}  &{$b\to s$} &{IRA}  & {TDA}   \\\hline
$B_c^-\to \overline D^0  \pi^-  $ & $ A_6^T+3 A_{15}^T+B_3^T $ & $ P^{u}+T $ & $B_c^-\to \overline D^0  K^{-}  $ & $ A_6^T+3 A_{15}^T+B_3^T $ & $ P^{u}+T $   \\\hline
$B_c^-\to D^-  \pi^0  $ & $ (-A_6^T+5 A_{15}^T-B_3^T)/{\sqrt{2}} $ & $ (C-P^{u})/{\sqrt{2}} $ & $B_c^-\to D^-  \overline K^{0}  $ & $ -A_6^T-A_{15}^T+B_3^T $ & $ P^{u} $  \\\hline
$B_c^-\to D^-  \eta_q  $ & $ (2 A_3^T-A_6^T+A_{15}^T+B_3^T)/{\sqrt{2}} $ & $ (C+P^{u}+2 S^{u})/{\sqrt{2}} $&$B_c^-\to D^-_s  \pi^0  $ & $ \sqrt{2} \left(2 A_{15}^T-A_6^T\right) $ & $ C/{\sqrt{2}} $  \\\hline
$B_c^-\to D^-  \eta_s  $ & $ A_3^T+A_6^T-A_{15}^T $ & $ S^{u} $ & $B_c^-\to D^-_s  \eta_q  $ & $ \sqrt{2} \left(A_3^T+A_{15}^T\right) $ & $ (C+2 S^{u})/{\sqrt{2}} $\\\hline
$B_c^-\to D^-_s  K^{0}  $ & $ -A_6^T-A_{15}^T+B_3^T $ & $ P^{u} $ & $B_c^-\to D^-_s  \eta_s  $ & $ A_3^T-2 A_{15}^T+B_3^T $ & $ P^{u}+S^{u} $  \\\hline
\hline 
\end{tabular}
\end{table}

\begin{figure}
\begin{center}
\includegraphics[scale=0.5]{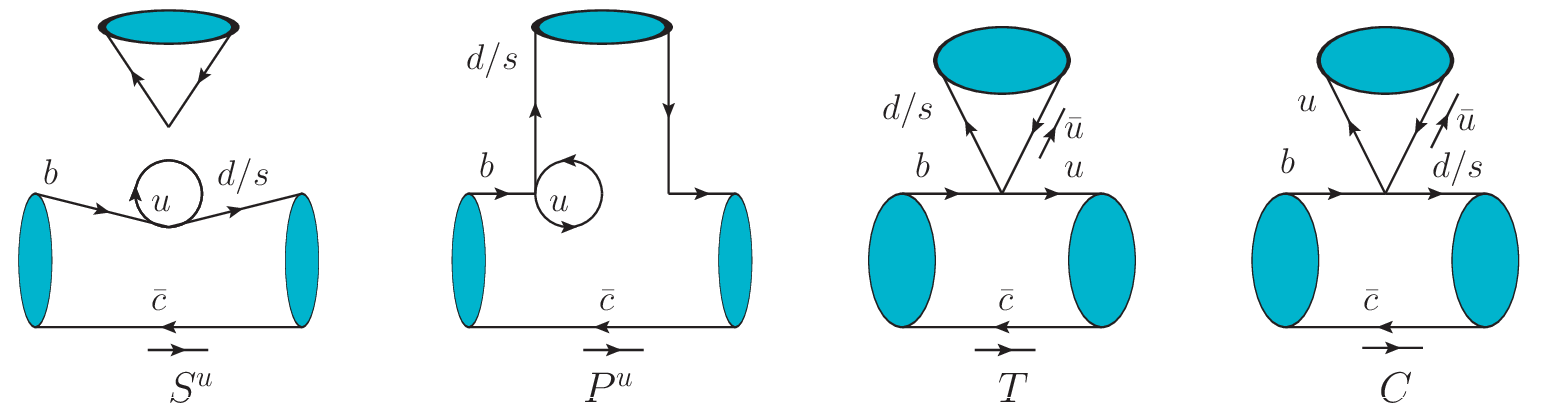}
\end{center}
\caption{Feynman diagrams for tree amplitudes in the $B_c\to DP,\;DV$ decays.  }\label{fig:Bc}
\end{figure}

The  effective Hamiltonian for $b$ quark decays can induce $B_c\to DP, DV$ transitions. The corresponding topological diagrams are given in Fig.~\ref{fig:Bc}.
The IRA and TDA  for $B_{c}\to DP$ decays are given as: 
\begin{eqnarray} 
{\cal A}_{u}^{IRA}&=&A_{3}^T B_{c}D_{i}H_{\bar{3}}^{i}P_{j}^{j}+B_{3}^T B_{c}D_{i}H_{\bar{3}}^{j}P_{j}^{i}+A_{6}^TB_{c}D_{i}(H_{6})_{j}^{[ik]}P_{k}^{j}+A_{15}^T B_{c}D_{i}(H_{\overline {15}})_{j}^{\{ik\}}P_{k}^{j},\\
{\cal A}_{u}^{TDA}&=&S^{u}B_{c}D_{i}H_{l}^{li}P_{j}^{j}+P^{u}B_{c}D_{i}H_{l}^{lj}P_{j}^{i}+T B_{c}D_{i}H_{l}^{ik}P_{k}^{l}+CB_{c}D_{i}H_{l}^{ki}P_{k}^{l}. 
\end{eqnarray}
The expanded  amplitudes can be found in Tab.~\ref{tab:Bc_DP}. 
Relations between the two sets of amplitudes are given as: 
\begin{eqnarray}
A_3^T & =&S^{u}-\frac{1}{8}T+\frac{3}{8}C,\;\;\;
B_3^T =P^{u}+\frac{3}{8}T-\frac{1}{8}C,\;\;\;
A_6^T =\frac{1}{4}T-\frac{1}{4}C,\;\;\;
A_{15}^T =\frac{1}{8}T+\frac{1}{8}C. 
\end{eqnarray} 
``Penguin" amplitudes are obtained similarly:  
\begin{eqnarray}
 A_{3,6,15}^T\to A_{3,6,15}^P, \;\;\;  B_{3}^T\to B_{3}^P, \;\;\; S^{u}\to S ,\;\;\; P^{u}\to P,\;\;\; T\to P_{T},\;\;\; C\to P_{C}.
\end{eqnarray}
Including the ``penguins", one has 8 complex amplitudes in total. 

Again decay amplitudes for $B_c\to DV$ can be obtained by replacing the pseudoscalars by their vector counterparts.  
The U-spin related channels include:  $B_{c}^{-}\to\overline{D}^{0} K^{-}$ and $B_{c}^{-}\to \overline{D}^{0} {\pi}^{-}$; 
$B_{c}^{-}\to {D}^{-} {\overline K}_{0}$ and $B_{c}^{-}\to {D}^{-}_{s} {K}_{0}$;  $B_{c}^{-}\to {\overline D}^{0} {K}^{*-}$ and $B_{c}^{-}\to {\overline D}^{0} {\rho}^{-}$;   $B_{c}^{-}\to {D}^{-} {\overline K}^{*0}$ and $B_{c}^{-}\to {D}^{-}_{s} {\overline K}^{*0}$.

In Ref.~\cite{Aaij:2017kea}, the LHCb collaboration  has measured the product:
\begin{eqnarray}
\frac{f(B_c)}{f(B^+)} \times {\cal B}(B_c^+\to D^0K^+) = (9.3^{+2.8}_{-2.5}\pm0.6)\times 10^{-7}, 
\end{eqnarray}
where the $f(B_c)$ and $f(B^+)$ are the production rates of $B_c^+$ and $B^+$, respectively. With the measured ratio~\cite{1411.2943}:
\begin{eqnarray}
\frac{f(B_c)}{f(B^+)}\sim 0.004-0.012, 
\end{eqnarray}
one can obtain an estimated branching fraction:
\begin{eqnarray}
{\cal B}(B_c^+\to D^0K^+) \sim 7.8\times 10^{-5}-2.3\times 10^{-4}. 
\end{eqnarray}
On theoretical side, model-dependent analyses give  $1.3\times 10^{-7}$~\cite{1102.5399}, and $6.6\times 10^{-5}$~\cite{0905.0945}, while a phenomenological study implies the ${\cal B}(B_c^+\to D^0K^+)\sim [4.4-9]\times 10^{-5}$~\cite{Chen:2017jrr}.  Since this transition is induced by  $b\to s$, the large branching fraction may imply a large penguin amplitude $P$.  Such a scenario can be tested by measuring  the corresponding  $(B_c^+\to D^+  K^0)$,  which has the same penguin amplitude.  Model-dependent calculations of other $B_c$ decays can be found in Refs.~\cite{Rui:2011qc,Zou:2012sy,Wang:2008xt,Wang:2009mi}. 
Some recent SU(3)  analyses of two-body $B_c$ decays can be found in Ref.~\cite{1708.07504,Zhu:2018epc}.  Compared to these studies, we have included all penguin amplitudes.

For the $B_c^-$ meson, the charm quark can also decays, with the final state $BP$ or $BV$~\cite{Aaij:2013cda}. Since the heavy bottom quark plays as a spectator, the decay modes are simpler. For example, for Cabibbo-allowed decay modes, there are only two channels: $B_c^-\to \pi^- \overline B_s^0$ and $B_c^-\to \rho^- \overline B_s^0$.  Thus we expect that the SU(3) symmetry will not provide much information in these decays.

It is necessary to point out that the charmless two-body $B_c$ decays are purely annihilation,  and the typical branching fractions are below the order $10^{-6}$~\cite{Liu:2009qa,Xiao:2013lia,Chang:2017ivy}. Since there are not too many channels,  it is less  useful to apply the flavor SU(3) symmetry to these modes.

\section{ Antitriplet Bottom Baryon Decay into a Baryon and a Meson}

In this and next sections, we discuss weak decays of baryons with a heavy $b$ and $c$ quark. Charmed or   bottom baryons with two light quarks  can form an anti-triplet or a sextet. Most  members of the sextet  can decay via strong interactions or electromagnetic interactions. The only exceptions are $\Omega_{b}$ and $\Omega_{c}$.  In the following we will concentrate on the anti-triplet baryons, whose weak decays are induced by the effective Hamiltonian $H^b_{eff}$ and $H^c_{eff}$.

\subsection{$T_b: (\Lambda_b,\Xi_{b}^0, \Xi_b^-)$ Decay into a decuplet baryon $T_{10}$ and a light meson}

\begin{figure}
\begin{center}
\includegraphics[scale=0.4]{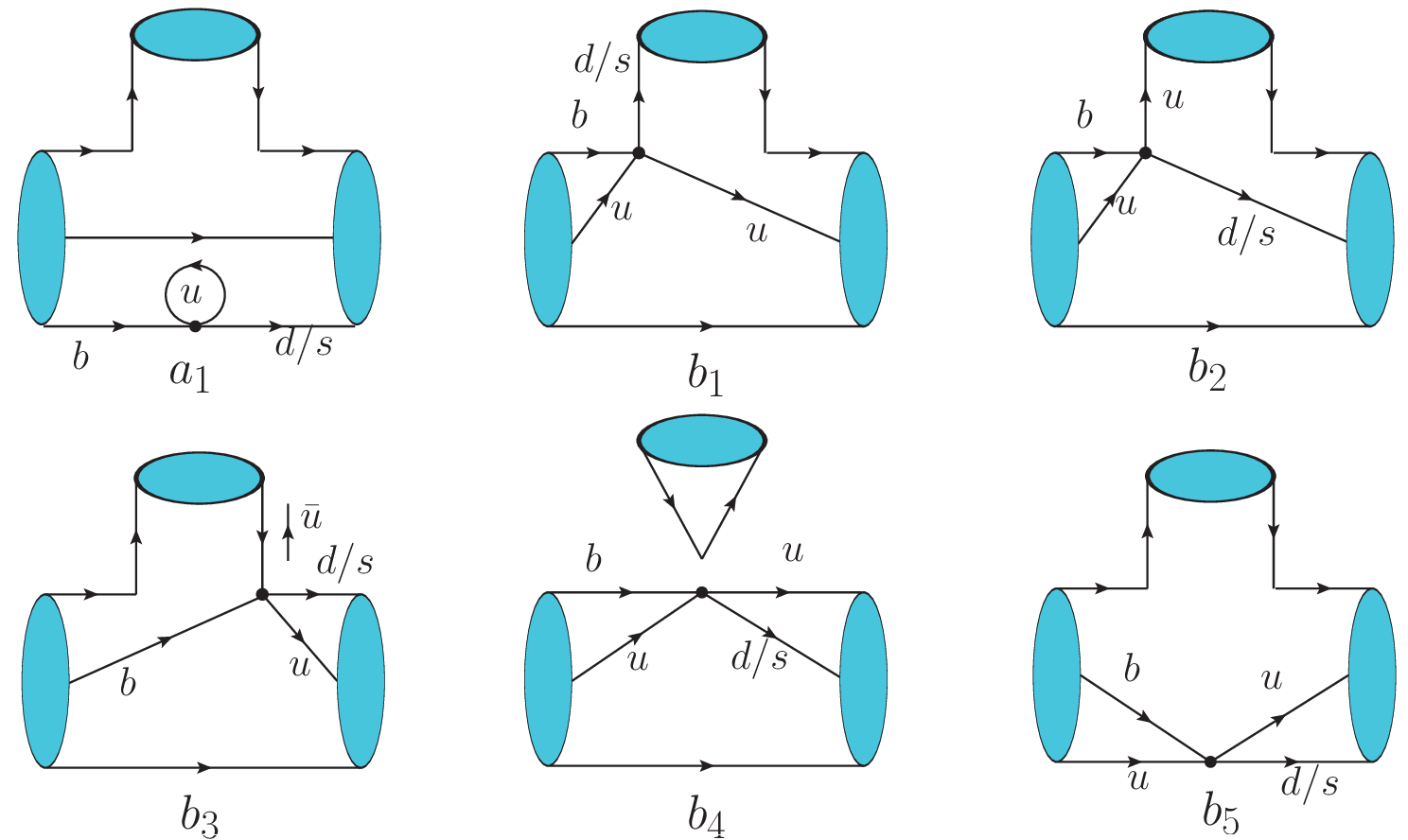}
\end{center}
\caption{Topology diagrams for the bottom baryon decays into a decuplet baryon and a light meson. We have explicitly specified the quark flavors for the four-quark interaction vertex, while the unspecified quarks can be $u,d,s$.   }\label{fig:Feynman_Lambdab_T10}
\end{figure}

\begin{table}
\renewcommand\arraystretch{1.2}
\caption{Decay amplitudes for $T_b\to T_{10} P$ decays. Only those amplitudes proportional to $V_{ub}V_{uq}^*$ are shown, while ``penguin"  amplitudes proportional to $V_{tb}V_{tq}^*$ are similar. }\label{tab:Lambdab_B10P}\begin{tabular}{cccccc}\hline\hline
{$b\to d$} &{IRA}  & {TDA}   \\\hline
$\Lambda_b^0\to \Delta^{+}  \pi^-  $ & $ (A_3^T-A_6^T-5 A_{15}^T+6 B_{15}^T-6 D_{15}^T)/{\sqrt{3}} $ & $ (a_1-b_1+b_3-b_5)/{\sqrt{3}} $ \\\hline
$\Lambda_b^0\to \Delta^{0}  \pi^0  $ & $ \sqrt{2/{3}} \left(-A_3^T+A_6^T+A_{15}^T-2 B_{15}^T+2 D_{15}^T\right) $ & $ (-2 a_1+b_1-b_2-b_3+b_5)/{\sqrt{6}} $ \\\hline
$\Lambda_b^0\to \Delta^{0}  \eta_q  $ & $ -4 \sqrt{2/{3}} \left(A_{15}^T+B_{15}^T+2 C_{15}^T+D_{15}^T\right) $ & $ -(b_1+b_2+b_3+2 b_4+b_5)/{\sqrt{6}} $ \\\hline
$\Lambda_b^0\to \Delta^{0}  \eta_s  $ & $ -(8 C_{15}^T)/{\sqrt{3}} $ & $ -b_{4}/{\sqrt{3}} $ \\\hline
$\Lambda_b^0\to \Delta^{-}  \pi^+  $ & $ -A_3^T+A_6^T-3 A_{15}^T+2 B_{15}^T-2 D_{15}^T $ & $ -a_1-b_2 $ \\\hline
$\Lambda_b^0\to \Sigma^{\prime0}  K^{0}  $ & $ -(-A_3^T+A_6^T+5 A_{15}^T+2 B_{15}^T+6 D_{15}^T)/{\sqrt{6}} $ & $ (a_1-b_1-b_5)/{\sqrt{6}} $ \\\hline
$\Lambda_b^0\to \Sigma^{\prime-}  K^{+}  $ & $ (-A_3^T+A_6^T-3 A_{15}^T+2 B_{15}^T-2 D_{15}^T)/{\sqrt{3}} $ & $ -(a_1+b_2)/{\sqrt{3}} $ \\\hline
$\Xi_b^0\to \Delta^{+}  K^{-}  $ & $ (A_3^T+A_6^T-A_{15}^T+6 B_{15}^T-6 D_{15}^T)/{\sqrt{3}} $ & $ (a_1+b_3-b_5)/{\sqrt{3}} $ \\\hline
$\Xi_b^0\to \Delta^{0}  \overline K^{0}  $ & $ (A_3^T+A_6^T-A_{15}^T-2 B_{15}^T-6 D_{15}^T)/{\sqrt{3}} $ & $ (a_1-b_5)/{\sqrt{3}} $ \\\hline
$\Xi_b^0\to \Sigma^{\prime+}  \pi^-  $ & $ -(2 \left(A_6^T+2 A_{15}^T\right))/{\sqrt{3}} $ & $ -b_{1}/{\sqrt{3}} $ \\\hline
$\Xi_b^0\to \Sigma^{\prime0}  \pi^0  $ & $ (-A_3^T+3 A_6^T+A_{15}^T-6 B_{15}^T-2 D_{15}^T)/{2 \sqrt{3}} $ & $ -(a_1-b_1+b_2+b_3)/{2 \sqrt{3}} $ \\\hline
$\Xi_b^0\to \Sigma^{\prime0}  \eta_q  $ & $ -(A_3^T+A_6^T+7 A_{15}^T+6 B_{15}^T+16 C_{15}^T+2 D_{15}^T)/{2 \sqrt{3}} $ & $ -(a_1+b_1+b_2+b_3+2 b_4)/{2 \sqrt{3}} $ \\\hline
$\Xi_b^0\to \Sigma^{\prime0}  \eta_s  $ & $ (A_3^T+A_6^T-A_{15}^T-2 B_{15}^T-8 C_{15}^T-6 D_{15}^T)/{\sqrt{6}} $ & $ (a_1-b_4-b_5)/{\sqrt{6}} $ \\\hline
$\Xi_b^0\to \Sigma^{\prime-}  \pi^+  $ & $ (-A_3^T+A_6^T-3 A_{15}^T+2 B_{15}^T-2 D_{15}^T)/{\sqrt{3}} $ & $ -(a_1+b_2)/{\sqrt{3}} $ \\\hline
$\Xi_b^0\to \Xi^{\prime0}  K^{0}  $ & $ -(2 \left(A_6^T+2 A_{15}^T\right))/{\sqrt{3}} $ & $ -b_{1}/{\sqrt{3}} $ \\\hline
$\Xi_b^0\to \Xi^{\prime-}  K^{+}  $ & $ (-A_3^T+A_6^T-3 A_{15}^T+2 B_{15}^T-2 D_{15}^T)/{\sqrt{3}} $ & $ -(a_1+b_2)/{\sqrt{3}} $ \\\hline
$\Xi_b^-\to \Delta^{0}  K^{-}  $ & $ (A_3^T+A_6^T-A_{15}^T+6 B_{15}^T+2 D_{15}^T)/{\sqrt{3}} $ & $ (a_1+b_3)/{\sqrt{3}} $ \\\hline
$\Xi_b^-\to \Delta^{-}  \overline K^{0}  $ & $ A_3^T+A_6^T-A_{15}^T-2 B_{15}^T+2 D_{15}^T $ & $ a_1 $ \\\hline
$\Xi_b^-\to \Sigma^{\prime0}  \pi^-  $ & $ (-A_3^T-A_6^T+A_{15}^T-6 B_{15}^T-2 D_{15}^T)/{\sqrt{6}} $ & $ -(a_1+b_3)/{\sqrt{6}} $ \\\hline
$\Xi_b^-\to \Sigma^{\prime-}  \pi^0  $ & $ (A_3^T+A_6^T-A_{15}^T-2 B_{15}^T+2 D_{15}^T)/{\sqrt{6}} $ & $ a_{1}/{\sqrt{6}} $ \\\hline
$\Xi_b^-\to \Sigma^{\prime-}  \eta_q  $ & $ (-A_3^T-A_6^T+A_{15}^T+2 B_{15}^T-2 D_{15}^T)/{\sqrt{6}} $ & $ -a_{1}/{\sqrt{6}} $ \\\hline
$\Xi_b^-\to \Sigma^{\prime-}  \eta_s  $ & $ (A_3^T+A_6^T-A_{15}^T-2 B_{15}^T+2 D_{15}^T)/{\sqrt{3}} $ & $ a_{1}/{\sqrt{3}} $ \\\hline
$\Xi_b^-\to \Xi^{\prime-}  K^{0}  $ & $ (-A_3^T-A_6^T+A_{15}^T+2 B_{15}^T-2 D_{15}^T)/{\sqrt{3}} $ & $ -a_{1}/{\sqrt{3}} $ \\\hline
\hline
{$b\to s$}  & {IRA}  & {TDA}  &  &  & \tabularnewline
\hline 
$\Lambda_{b}^{0}\to\Delta^{+}K^{-}$  & $-(2\left(A_{6}^{T}+2A_{15}^{T}\right))/{\sqrt{3}}$  & $-b_{1}/{\sqrt{3}}$  &  &  & \tabularnewline
\hline 
$\Lambda_{b}^{0}\to\Delta^{0}\overline{K}^{0}$  & $-(2\left(A_{6}^{T}+2A_{15}^{T}\right))/{\sqrt{3}}$  & $-b_{1}/{\sqrt{3}}$  &  &  & \tabularnewline
\hline 
$\Lambda_{b}^{0}\to\Sigma^{\prime+}\pi^{-}$  & $(A_{3}^{T}+A_{6}^{T}-A_{15}^{T}+6B_{15}^{T}-6D_{15}^{T})/{\sqrt{3}}$  & $(a_{1}+b_{3}-b_{5})/{\sqrt{3}}$  &  &  & \tabularnewline
\hline 
$\Lambda_{b}^{0}\to\Sigma^{\prime0}\pi^{0}$  & $-(A_{3}^{T}+A_{15}^{T}+2B_{15}^{T}-2D_{15}^{T})/{\sqrt{3}}$  & $-(2a_{1}+b_{2}+b_{3}-b_{5})/{2\sqrt{3}}$  &  &  & \tabularnewline
\hline 
$\Lambda_{b}^{0}\to\Sigma^{\prime0}\eta_{q}$  & $(A_{6}^{T}-2A_{15}^{T}-4\left(B_{15}^{T}+2C_{15}^{T}+D_{15}^{T}\right))/{\sqrt{3}}$  & $-(b_{2}+b_{3}+2b_{4}+b_{5})/{2\sqrt{3}}$  &  &  & \tabularnewline
\hline 
$\Lambda_{b}^{0}\to\Sigma^{\prime0}\eta_{s}$  & $-\sqrt{2/{3}}\left(A_{6}^{T}+2A_{15}^{T}+4C_{15}^{T}\right)$  & $-(b_{1}+b_{4})/{\sqrt{6}}$  &  &  & \tabularnewline
\hline 
$\Lambda_{b}^{0}\to\Sigma^{\prime-}\pi^{+}$  & $(-A_{3}^{T}+A_{6}^{T}-3A_{15}^{T}+2B_{15}^{T}-2D_{15}^{T})/{\sqrt{3}}$  & $-(a_{1}+b_{2})/{\sqrt{3}}$  &  &  & \tabularnewline
\hline 
$\Lambda_{b}^{0}\to\Xi^{\prime0}K^{0}$  & $(A_{3}^{T}+A_{6}^{T}-A_{15}^{T}-2B_{15}^{T}-6D_{15}^{T})/{\sqrt{3}}$  & $(a_{1}-b_{5})/{\sqrt{3}}$  &  &  & \tabularnewline
\hline 
$\Lambda_{b}^{0}\to\Xi^{\prime-}K^{+}$  & $(-A_{3}^{T}+A_{6}^{T}-3A_{15}^{T}+2B_{15}^{T}-2D_{15}^{T})/{\sqrt{3}}$  & $-(a_{1}+b_{2})/{\sqrt{3}}$  &  &  & \tabularnewline
\hline 
$\Xi_{b}^{0}\to\Sigma^{\prime+}K^{-}$  & $(A_{3}^{T}-A_{6}^{T}-5A_{15}^{T}+6B_{15}^{T}-6D_{15}^{T})/{\sqrt{3}}$  & $(a_{1}-b_{1}+b_{3}-b_{5})/{\sqrt{3}}$  &  &  & \tabularnewline
\hline 
$\Xi_{b}^{0}\to\Sigma^{\prime0}\overline{K}^{0}$  & $-(-A_{3}^{T}+A_{6}^{T}+5A_{15}^{T}+2B_{15}^{T}+6D_{15}^{T})/{\sqrt{6}}$  & $(a_{1}-b_{1}-b_{5})/{\sqrt{6}}$  &  &  & \tabularnewline
\hline 
$\Xi_{b}^{0}\to\Xi^{\prime0}\pi^{0}$  & $-(A_{3}^{T}-A_{6}^{T}+3A_{15}^{T}+6B_{15}^{T}+2D_{15}^{T})/{\sqrt{6}}$  & $-(a_{1}+b_{2}+b_{3})/{\sqrt{6}}$  &  &  & \tabularnewline
\hline 
$\Xi_{b}^{0}\to\Xi^{\prime0}\eta_{q}$  & $-(A_{3}^{T}-A_{6}^{T}+3A_{15}^{T}+6B_{15}^{T}+16C_{15}^{T}+2D_{15}^{T})/{\sqrt{6}}$  & $-(a_{1}+b_{2}+b_{3}+2b_{4})/{\sqrt{6}}$  &  &  & \tabularnewline
\hline 
$\Xi_{b}^{0}\to\Xi^{\prime0}\eta_{s}$  & $-(-A_{3}^{T}+A_{6}^{T}+5A_{15}^{T}+2B_{15}^{T}+8C_{15}^{T}+6D_{15}^{T})/{\sqrt{3}}$  & $(a_{1}-b_{1}-b_{4}-b_{5})/{\sqrt{3}}$  &  &  & \tabularnewline
\hline 
$\Xi_{b}^{0}\to\Xi^{\prime-}\pi^{+}$  & $(-A_{3}^{T}+A_{6}^{T}-3A_{15}^{T}+2B_{15}^{T}-2D_{15}^{T})/{\sqrt{3}}$  & $-(a_{1}+b_{2})/{\sqrt{3}}$  &  &  & \tabularnewline
\hline 
$\Xi_{b}^{0}\to\Omega^{-}K^{+}$  & $-A_{3}^{T}+A_{6}^{T}-3A_{15}^{T}+2B_{15}^{T}-2D_{15}^{T}$  & $-a_{1}-b_{2}$  &  &  & \tabularnewline
\hline 
$\Xi_{b}^{-}\to\Sigma^{\prime0}K^{-}$  & $(A_{3}^{T}+A_{6}^{T}-A_{15}^{T}+6B_{15}^{T}+2D_{15}^{T})/{\sqrt{6}}$  & $(a_{1}+b_{3})/{\sqrt{6}}$  &  &  & \tabularnewline
\hline 
$\Xi_{b}^{-}\to\Sigma^{\prime-}\overline{K}^{0}$  & $(A_{3}^{T}+A_{6}^{T}-A_{15}^{T}-2B_{15}^{T}+2D_{15}^{T})/{\sqrt{3}}$  & $a_{1}/{\sqrt{3}}$  &  &  & \tabularnewline
\hline 
$\Xi_{b}^{-}\to\Xi^{\prime0}\pi^{-}$  & $(-A_{3}^{T}-A_{6}^{T}+A_{15}^{T}-6B_{15}^{T}-2D_{15}^{T})/{\sqrt{3}}$  & $-(a_{1}+b_{3})/{\sqrt{3}}$  &  &  & \tabularnewline
\hline 
$\Xi_{b}^{-}\to\Xi^{\prime-}\pi^{0}$  & $(A_{3}^{T}+A_{6}^{T}-A_{15}^{T}-2B_{15}^{T}+2D_{15}^{T})/{\sqrt{6}}$  & $a_{1}/{\sqrt{6}}$  &  &  & \tabularnewline
\hline 
$\Xi_{b}^{-}\to\Xi^{\prime-}\eta_{q}$  & $(-A_{3}^{T}-A_{6}^{T}+A_{15}^{T}+2B_{15}^{T}-2D_{15}^{T})/{\sqrt{6}}$  & $-a_{1}/{\sqrt{6}}$  &  &  & \tabularnewline
\hline 
$\Xi_{b}^{-}\to\Xi^{\prime-}\eta_{s}$  & $(A_{3}^{T}+A_{6}^{T}-A_{15}^{T}-2B_{15}^{T}+2D_{15}^{T})/{\sqrt{3}}$  & $a_{1}/{\sqrt{3}}$  &  &  & \tabularnewline
\hline 
$\Xi_{b}^{-}\to\Omega^{-}K^{0}$  & $-A_{3}^{T}-A_{6}^{T}+A_{15}^{T}+2B_{15}^{T}-2D_{15}^{T}$  & $-a_{1}$  &  &  & \tabularnewline
\hline 
\hline 
\end{tabular}
\end{table}

The IRA amplitudes  for the $T_b$ decays  into a decuplet baryon and a light meson  can be parametrized  as:
\begin{eqnarray}
{\cal A}_{u}^{IRA} & = & A_3^T T_{b\bar 3}^{[ij]}  H_{\bar{3}}^{k} (\overline T_{10})_{ikl}P_{j}^{l} + A_6^T T_{b\bar 3}^{[ij]}  (H_{6})_{j}^{[kl]} (\overline T_{10})_{ikm}P_{l}^{m}  +A_{15}^T T_{b\bar 3}^{[ij]}  (H_{15})_{j}^{\{kl\}} (\overline T_{10})_{ikm}P_{l}^{m}\nonumber \\
&&+ B_{15}^T T_{b\bar 3}^{[ij]}  (H_{15})_{m}^{\{kl\}} (\overline T_{10})_{ikl}P_{j}^{m} + C_{15}^T T_{b\bar 3}^{[ij]}  (H_{15})_{j}^{\{kl\}} (\overline T_{10})_{ikl}P_{m}^{m} + D_{15}^T T_{b\bar 3}^{[ij]}  (H_{15})_{j}^{\{kl\}} (\overline T_{10})_{klm}P_{i}^{m}. 
\end{eqnarray}
The TDA amplitudes are shown in Fig.~\ref{fig:Feynman_Lambdab_T10} with the parametrization: 
\begin{eqnarray}
{\cal A}_{u}^{TDA}&=&  a_1  T_{b\bar 3}^{[ij]}  H^{mk}_{m} (\overline T_{10})_{ikl}P_{j}^{l}  + b_1 T_{b\bar 3}^{[ij]}  H_{j}^{kl} (\overline T_{10})_{ikm}P_{l}^{m}  +b_2 T_{b\bar 3}^{[ij]}  H_{j}^{lk} (\overline T_{10})_{ikm}P_{l}^{m}\nonumber \\
&&+ b_3 T_{b\bar 3}^{[ij]}  H_{m}^{kl} (\overline T_{10})_{ikl}P_{j}^{m} + b_4 T_{b\bar 3}^{[ij]}  H_{j}^{kl} (\overline T_{10})_{ikl}P_{m}^{m} + b_5 T_{b\bar 3}^{[ij]}  H_{j}^{kl} (\overline T_{10})_{klm}P_{i}^{m}. 
\end{eqnarray}
We find 
relations between the two sets of amplitudes  as:
\begin{eqnarray}
a_1= A_3^T+A_6^T-A_{15}^T-2B_{15}^T+2D_{15}^T, \;\;\; b_1= 4A_{15}^T+2A_{6}^T,\nonumber\\
b_2=4A_{15}^T -2A_{6}^T, \;\;\;  b_3= 8B_{15}^T,\;\;b_4= 8C_{15}^T,\;\;b_5= 8D_{15}^T. 
\end{eqnarray}
The expanded amplitudes for individual  decay modes can be found  in Tab.~\ref{tab:Lambdab_B10P}.

\begin{table}
\renewcommand\arraystretch{1.2}
 \caption{U-spin relations for $T_b \to T_{10} V$. If the final state contains a light pseudoscalar meson, the U-spin relations can be obtained similarly except that $\eta_q$ and $\eta_s$ mix.  }
\label{tab:U-spinLbB10V}%
\begin{tabular}{cccccc}
\hline 
\hline 
$b\to d$ & $b\to s$ & $r$ & $b\to d$ & $b\to s$ & $r$\tabularnewline
\hline 
${\cal A}(\Lambda_{b}^{0}\to\Delta^{+}\rho^{-})$  & ${\cal A}(\Xi_{b}^{0}\to\Sigma^{\prime+}K^{*-})$ & $1$ & ${\cal A}(\Xi_{b}^{-}\to\Sigma^{\prime-}\omega)$  & ${\cal A}(\Xi_{b}^{-}\to\Sigma^{\prime-}\overline{K}^{*0})$ & $-\frac{1}{\sqrt{2}}$\tabularnewline
\hline 
${\cal A}(\Lambda_{b}^{0}\to\Delta^{-}\rho^{+})$  & ${\cal A}(\Lambda_{b}^{0}\to\Sigma^{\prime-}\rho^{+})$ & $\sqrt{3}$ & ${\cal A}(\Xi_{b}^{-}\to\Sigma^{\prime-}\omega)$  & ${\cal A}(\Xi_{b}^{-}\to\Xi^{\prime-}\rho^{0})$ & $-1$\tabularnewline
\hline 
${\cal A}(\Lambda_{b}^{0}\to\Delta^{-}\rho^{+})$  & ${\cal A}(\Lambda_{b}^{0}\to\Xi^{\prime-}K^{*+})$ & $\sqrt{3}$ & ${\cal A}(\Xi_{b}^{-}\to\Sigma^{\prime-}\omega)$  & ${\cal A}(\Xi_{b}^{-}\to\Xi^{\prime-}\omega)$ & $1$\tabularnewline
\hline 
${\cal A}(\Lambda_{b}^{0}\to\Delta^{-}\rho^{+})$  & ${\cal A}(\Xi_{b}^{0}\to\Xi^{\prime-}\rho^{+})$ & $\sqrt{3}$ & ${\cal A}(\Xi_{b}^{-}\to\Sigma^{\prime-}\omega)$  & ${\cal A}(\Xi_{b}^{-}\to\Xi^{\prime-}\phi)$ & $-\frac{1}{\sqrt{2}}$\tabularnewline
\hline 
${\cal A}(\Lambda_{b}^{0}\to\Delta^{-}\rho^{+})$  & ${\cal A}(\Xi_{b}^{0}\to\Omega^{-}K^{*+})$ & $1$ & ${\cal A}(\Xi_{b}^{-}\to\Sigma^{\prime-}\omega)$  & ${\cal A}(\Xi_{b}^{-}\to\Omega^{-}K^{*0})$ & $\frac{1}{\sqrt{6}}$\tabularnewline
\hline 
${\cal A}(\Lambda_{b}^{0}\to\Sigma^{\prime0}K^{*0})$  & ${\cal A}(\Xi_{b}^{0}\to\Sigma^{\prime0}\overline{K}^{*0})$ & $1$ & ${\cal A}(\Xi_{b}^{-}\to\Sigma^{\prime-}\phi)$  & ${\cal A}(\Xi_{b}^{-}\to\Sigma^{\prime-}\overline{K}^{*0})$ & $1$\tabularnewline
\hline 
${\cal A}(\Lambda_{b}^{0}\to\Sigma^{\prime-}K^{*+})$  & ${\cal A}(\Lambda_{b}^{0}\to\Sigma^{\prime-}\rho^{+})$ & $1$ & ${\cal A}(\Xi_{b}^{-}\to\Sigma^{\prime-}\phi)$  & ${\cal A}(\Xi_{b}^{-}\to\Xi^{\prime-}\rho^{0})$ & $\sqrt{2}$\tabularnewline
\hline 
${\cal A}(\Lambda_{b}^{0}\to\Sigma^{\prime-}K^{*+})$  & ${\cal A}(\Lambda_{b}^{0}\to\Xi^{\prime-}K^{*+})$ & $1$ & ${\cal A}(\Xi_{b}^{-}\to\Sigma^{\prime-}\phi)$  & ${\cal A}(\Xi_{b}^{-}\to\Xi^{\prime-}\omega)$ & $-\sqrt{2}$\tabularnewline
\hline 
${\cal A}(\Lambda_{b}^{0}\to\Sigma^{\prime-}K^{*+})$  & ${\cal A}(\Xi_{b}^{0}\to\Xi^{\prime-}\rho^{+})$ & $1$ & ${\cal A}(\Xi_{b}^{-}\to\Sigma^{\prime-}\phi)$  & ${\cal A}(\Xi_{b}^{-}\to\Xi^{\prime-}\phi)$ & $1$\tabularnewline
\hline 
${\cal A}(\Lambda_{b}^{0}\to\Sigma^{\prime-}K^{*+})$  & ${\cal A}(\Xi_{b}^{0}\to\Omega^{-}K^{*+})$ & $\frac{1}{\sqrt{3}}$ & ${\cal A}(\Xi_{b}^{-}\to\Sigma^{\prime-}\phi)$  & ${\cal A}(\Xi_{b}^{-}\to\Omega^{-}K^{*0})$ & $-\frac{1}{\sqrt{3}}$\tabularnewline
\hline 
${\cal A}(\Xi_{b}^{0}\to\Delta^{+}K^{*-})$  & ${\cal A}(\Lambda_{b}^{0}\to\Sigma^{\prime+}\rho^{-})$ & $1$ & ${\cal A}(\Xi_{b}^{-}\to\Xi^{\prime-}K^{*0})$  & ${\cal A}(\Xi_{b}^{-}\to\Sigma^{\prime-}\overline{K}^{*0})$ & $-1$\tabularnewline
\hline 
${\cal A}(\Xi_{b}^{0}\to\Delta^{0}\overline{K}^{*0})$  & ${\cal A}(\Lambda_{b}^{0}\to\Xi^{\prime0}K^{*0})$ & $1$ & ${\cal A}(\Xi_{b}^{-}\to\Xi^{\prime-}K^{*0})$  & ${\cal A}(\Xi_{b}^{-}\to\Xi^{\prime-}\rho^{0})$ & $-\sqrt{2}$\tabularnewline
\hline 
${\cal A}(\Xi_{b}^{0}\to\Sigma^{\prime+}\rho^{-})$  & ${\cal A}(\Lambda_{b}^{0}\to\Delta^{+}K^{*-})$ & $1$ & ${\cal A}(\Xi_{b}^{-}\to\Xi^{\prime-}K^{*0})$  & ${\cal A}(\Xi_{b}^{-}\to\Xi^{\prime-}\omega)$ & $\sqrt{2}$\tabularnewline
\hline 
${\cal A}(\Xi_{b}^{0}\to\Sigma^{\prime+}\rho^{-})$  & ${\cal A}(\Lambda_{b}^{0}\to\Delta^{0}\overline{K}^{*0})$ & $1$ & ${\cal A}(\Xi_{b}^{-}\to\Xi^{\prime-}K^{*0})$  & $-1{\cal A}(\Xi_{b}^{-}\to\Xi^{\prime-}\phi)$ & $-1$\tabularnewline
\hline 
${\cal A}(\Xi_{b}^{0}\to\Sigma^{\prime-}\rho^{+})$  & ${\cal A}(\Lambda_{b}^{0}\to\Sigma^{\prime-}\rho^{+})$ & $1$ & ${\cal A}(\Xi_{b}^{-}\to\Xi^{\prime-}K^{*0})$  & ${\cal A}(\Xi_{b}^{-}\to\Omega^{-}K^{*0})$ & $\frac{1}{\sqrt{3}}$\tabularnewline
\hline 
${\cal A}(\Xi_{b}^{0}\to\Sigma^{\prime-}\rho^{+})$  & ${\cal A}(\Lambda_{b}^{0}\to\Xi^{\prime-}K^{*+})$ & $1$ & ${\cal A}(\Xi_{b}^{-}\to\Sigma^{\prime0}\rho^{-})$  & ${\cal A}(\Xi_{b}^{-}\to\Sigma^{\prime0}K^{*-})$ & $-1$\tabularnewline
\hline 
${\cal A}(\Xi_{b}^{0}\to\Sigma^{\prime-}\rho^{+})$  & ${\cal A}(\Xi_{b}^{0}\to\Xi^{\prime-}\rho^{+})$ & $1$ & ${\cal A}(\Xi_{b}^{-}\to\Sigma^{\prime0}\rho^{-})$  & ${\cal A}(\Xi_{b}^{-}\to\Xi^{\prime0}\rho^{-})$ & $\frac{1}{\sqrt{2}}$\tabularnewline
\hline 
${\cal A}(\Xi_{b}^{0}\to\Sigma^{\prime-}\rho^{+})$  & ${\cal A}(\Xi_{b}^{0}\to\Omega^{-}K^{*+})$ & $\frac{1}{\sqrt{3}}$ & ${\cal A}(\Xi_{b}^{-}\to\Sigma^{\prime-}\rho^{0})$  & ${\cal A}(\Xi_{b}^{-}\to\Sigma^{\prime-}\overline{K}^{*0})$ & $\frac{1}{\sqrt{2}}$\tabularnewline
\hline 
${\cal A}(\Xi_{b}^{0}\to\Xi^{\prime0}K^{*0})$  & ${\cal A}(\Lambda_{b}^{0}\to\Delta^{+}K^{*-})$ & $1$ & ${\cal A}(\Xi_{b}^{-}\to\Sigma^{\prime-}\rho^{0})$  & ${\cal A}(\Xi_{b}^{-}\to\Xi^{\prime-}\rho^{0})$ & $1$\tabularnewline
\hline 
${\cal A}(\Xi_{b}^{0}\to\Xi^{\prime0}K^{*0})$  & ${\cal A}(\Lambda_{b}^{0}\to\Delta^{0}\overline{K}^{*0})$ & $1$ & ${\cal A}(\Xi_{b}^{-}\to\Sigma^{\prime-}\rho^{0})$  & ${\cal A}(\Xi_{b}^{-}\to\Xi^{\prime-}\omega)$ & $-1$\tabularnewline
\hline 
${\cal A}(\Xi_{b}^{0}\to\Xi^{\prime-}K^{*+})$  & ${\cal A}(\Lambda_{b}^{0}\to\Sigma^{\prime-}\rho^{+})$ & $1$ & ${\cal A}(\Xi_{b}^{-}\to\Sigma^{\prime-}\rho^{0})$  & ${\cal A}(\Xi_{b}^{-}\to\Xi^{\prime-}\phi)$ & $\frac{1}{\sqrt{2}}$\tabularnewline
\hline 
${\cal A}(\Xi_{b}^{0}\to\Xi^{\prime-}K^{*+})$  & ${\cal A}(\Lambda_{b}^{0}\to\Xi^{\prime-}K^{*+})$ & $1$ & ${\cal A}(\Xi_{b}^{-}\to\Sigma^{\prime-}\rho^{0})$  & ${\cal A}(\Xi_{b}^{-}\to\Omega^{-}K^{*0})$ & $-\frac{1}{\sqrt{6}}$\tabularnewline
\hline 
${\cal A}(\Xi_{b}^{0}\to\Xi^{\prime-}K^{*+})$  & ${\cal A}(\Xi_{b}^{0}\to\Xi^{\prime-}\rho^{+})$ & $1$ & ${\cal A}(\Xi_{b}^{-}\to\Delta^{-}\overline{K}^{*0})$  & ${\cal A}(\Xi_{b}^{-}\to\Sigma^{\prime-}\overline{K}^{*0})$ & $\sqrt{3}$\tabularnewline
\hline 
${\cal A}(\Xi_{b}^{0}\to\Xi^{\prime-}K^{*+})$  & ${\cal A}(\Xi_{b}^{0}\to\Omega^{-}K^{*+})$ & $\frac{1}{\sqrt{3}}$ & ${\cal A}(\Xi_{b}^{-}\to\Delta^{-}\overline{K}^{*0})$  & ${\cal A}(\Xi_{b}^{-}\to\Xi^{\prime-}\rho^{0})$ & $\sqrt{6}$\tabularnewline
\hline 
${\cal A}(\Xi_{b}^{-}\to\Delta^{0}K^{*-})$  & ${\cal A}(\Xi_{b}^{-}\to\Sigma^{\prime0}K^{*-})$ & $\sqrt{2}$ & ${\cal A}(\Xi_{b}^{-}\to\Delta^{-}\overline{K}^{*0})$  & ${\cal A}(\Xi_{b}^{-}\to\Xi^{\prime-}\omega)$ & $-\sqrt{6}$\tabularnewline
\hline 
${\cal A}(\Xi_{b}^{-}\to\Delta^{0}K^{*-})$  & ${\cal A}(\Xi_{b}^{-}\to\Xi^{\prime0}\rho^{-})$ & $-1$ & ${\cal A}(\Xi_{b}^{-}\to\Delta^{-}\overline{K}^{*0})$  & ${\cal A}(\Xi_{b}^{-}\to\Xi^{\prime-}\phi)$ & $\sqrt{3}$\tabularnewline
\hline 
 &  &  & ${\cal A}(\Xi_{b}^{-}\to\Delta^{-}\overline{K}^{*0})$  & ${\cal A}(\Xi_{b}^{-}\to\Omega^{-}K^{*0})$ & $-1$\tabularnewline
\hline 
\hline 
\end{tabular}
\end{table}

A few remarks are given in order. 
\begin{itemize}
\item 
As the two light quarks in the initial state are antisymmetric in the flavor space while they are symmetric in  the final state. An  overlap of wave functions of the initial and final baryons is zero~\cite{Mannel:2011xg}, which lead to vanishing decay amplitudes unless hard scattering interactions occur~\cite{Wang:2011uv}. In other words, there is no  ``factorizable" contribution in the transition.  In addition, all diagrams in Fig.~\ref{fig:Feynman_Lambdab_T10} are suppressed by powers of   $1/N_c$ compared to the $T_b\to T_{8}P$. 
This will indicate that branching fractions for these decays  are likely smaller than the relevant $B$ decays and $T_b\to T_{8}P$ decays, where $T_8$ represents the   octet baryon.

\item For the $T_b\to T_{10}P$,  one can construct the amplitudes with the spinors, and a general form is:
\begin{eqnarray}
{\cal A} =p_{T_{b},\mu}\bar u^\mu(p_{T_{10}}) (A+B\gamma_5) u(p_{T_{b}}), \label{eq:spinor_B10P}
\end{eqnarray}
where $A$ and $B$ are two nonperturbative coefficients containing the CKM factors, and have the same flavor structure with ${\cal A}_{u,t}$.  Thus
in total, one has $6\times 2 \times 2=24$ complex amplitudes in theory.

\item 
Since the initial baryon and final baryons \red{can be} polarized, it is convenient to express the decays with helicity amplitudes: 
\begin{eqnarray}
{\cal A}(S_{in} \to S_{f1}\ S_{f2}), 
\end{eqnarray}
where $S_{in}$ and $S_{f1},\ S_{f2}$ are polarizations of initial and final states.
The  two sets of helicity amplitudes   for $T_b\to T_{10}P$   can be derived using the parametrization~in Eq.~\eqref{eq:spinor_B10P}: 
\begin{eqnarray}
{\cal A }\left(\frac{1}{2} \to \frac{1}{2} \  0\right) &=&\sqrt{\frac{2}{3}}\frac{m_{T_b}}{m_{T_{10}}}p_{cm}N_{T_{10}}N_{T_b}\left(A-B\frac{p_{cm}}{E_{T_{10}}+m_{T_{10}}}\right), \\
{\cal A }\left(-\frac{1}{2} \to -\frac{1}{2} \  0\right) &=&\sqrt{\frac{2}{3}}\frac{m_{T_b}}{m_{T_{10}}}p_{cm}N_{T_{10}}N_{T_b}\left(A+B\frac{p_{cm}}{E_{T_{10}}+m_{T_{10}}}\right).
\end{eqnarray} 
Here $E_{T_{10}}$ and $p_{cm}$ are the energy and 3-momentum magnitude of $T_{10}$ in the rest frame of $T_b$. $N_{T_{10}}$ and $N_{T_b}$ are normalization factors of $T_{10}$ and $T_b$ spinors:
\begin{eqnarray}
p_{cm} & =&\frac{1}{2m_{T_b}}\sqrt{(m_{T_b}^{2}-(m_{T_{10}}+m_{P})^{2})(m_{T_b}^{2}-(m_{T_{10}}-m_{P})^{2})},\;\;\;
E_{T_{10}}  =\frac{m_{T_{10}}^{2}+m_{T_b}^{2}-m_{P}^{2}}{2m_{T_b}},\nonumber\\
N_{T_{10}} & =&\sqrt{\frac{(m_{T_{10}}+m_{T_b})^{2}-m_{P}^{2}}{2m_{T_b}}},\;\;\;
N_{T_b}  =\sqrt{2m_{T_b}}.\label{ampparameter}
\end{eqnarray}

\item For  $T_b\to T_{10}V$, one can construct the amplitudes with the spinors and polarization vector~\footnote{One may expect a term which looks like $\epsilon_{\mu\nu\alpha\beta} \epsilon^{*\mu}\bar u^\nu(p_{T_{10}}) (G'\sigma^{\alpha\beta}+H'\sigma^{\alpha\beta}\gamma_5) u(p_{b})$. Actually such term cam be absorbed into the term $ \epsilon^{*}_\mu \bar u^\mu(p_{T_{10}}) (E'+F'\gamma_5) u(p_{b})$ by using the fact that the spinor-vector $u^\mu(p_{T_{10}})$, as a irreducible representation of $1/2 \otimes 1$, must satisfy $\gamma_\mu u^\mu(p_{T_{10}})=0$. }:
\begin{eqnarray}
{\cal A} &=&  \epsilon^{*}\cdot p_{T_{b}}  p_{T_{b},\mu}\bar u^\mu(p_{T_{10}}) (A'+B'\gamma_5) u(p_{T_{b}}) + \epsilon^{*\nu}  p_{T_{b},\mu}\bar u^\mu(p_{T_{10}}) (C'\gamma_\nu +D'\gamma_\nu \gamma_5) u(p_{T_{b}}) \nonumber\\
&&+   \epsilon^{*}_\mu \bar u^\mu(p_{T_{10}}) (E'+F'\gamma_5) u(p_{T_{b}}). \label{eq:spinor_B10V}
\end{eqnarray}

There are six different polarization configurations. The helicity amplitudes are given as:
\begin{eqnarray}
{\cal A}(\frac{1}{2}\to\frac{1}{2}\ 0) & = & \sqrt{\frac{2}{3}}\frac{m_{T_b}}{m_{T_{10}}}p_{cm}N_{T_{10}}N_{T_b}[-\frac{m_{T_b}p_{cm}}{m_{V}}\left(A^{\prime}-B^{\prime}\frac{p_{cm}}{E_{T_{10}}+m_{T_{10}}}\right)\nonumber\\
 & + & C^{\prime}\frac{p_{cm}}{m_{V}}\left(\frac{m_{T_b}-E_{T_{10}}}{E_{T_{10}}+m_{T_{10}}}-1\right)+D^{\prime}\left(\frac{m_{T_b}-E_{T_{10}}}{m_{V}}-\frac{p_{cm}^{2}}{m_{V}(E_{T_{10}}+m_{T_{10}})}\right)\nonumber\\
 & + & \left(-\frac{p_{cm}}{m_{V}m_{T_b}}+\frac{E_{T_{10}}(m_{T_b}-E_{T_{10}})}{m_{V}m_{T_b}p_{cm}}\right)\left(E^{\prime}-F^{\prime}\frac{p_{cm}}{E_{T_{10}}+m_{T_{10}}}\right)],\nonumber\\
{\cal A}(-\frac{1}{2}\to-\frac{1}{2}\ 0) & = & \sqrt{\frac{2}{3}}\frac{m_{T_b}}{m_{T_{10}}}p_{cm}N_{T_{10}}N_{T_b}[-\frac{m_{T_b}p_{cm}}{m_{V}}\left(A^{\prime}+B^{\prime}\frac{p_{cm}}{E_{T_{10}}+m_{T_{10}}}\right)\nonumber\\
 & + & C^{\prime}\frac{p_{cm}}{m_{V}}\left(\frac{m_{T_b}-E_{T_{10}}}{E_{T_{10}}+m_{T_{10}}}-1\right)-D^{\prime}\left(\frac{m_{T_b}-E_{T_{10}}}{m_{V}}-\frac{p_{cm}^{2}}{m_{V}(E_{T_{10}}+m_{T_{10}})}\right)\nonumber\\
 & + & \left(-\frac{p_{cm}}{m_{V}m_{T_b}}+\frac{E_{T_{10}}(m_{T_b}-E_{T_{10}})}{m_{V}m_{T_b}p_{cm}}\right)\left(E^{\prime}+F^{\prime}\frac{p_{cm}}{E_{T_{10}}+m_{T_{10}}}\right)],\nonumber\\
{\cal A}(\frac{1}{2}\to-\frac{1}{2}\ 1) & = & \frac{1}{\sqrt{3}}N_{T_b}N_{T_{10}}[\frac{2p_{cm}m_{T_b}}{m_{T_{10}}}\left(D^{\prime}-C^{\prime}\frac{p}{E_{T_{10}}+m_{T_{10}}}\right)+\left(E^{\prime}-F^{\prime}\frac{p}{E_{T_{10}}+m_{T_{10}}}\right)],\nonumber\\
{\cal A}(-\frac{1}{2}\to \frac{1}{2}\ -1) & = & \frac{1}{\sqrt{3}}N_{T_b}N_{T_{10}}[-\frac{2p_{cm}m_{T_b}}{m_{T_{10}}}\left(D^{\prime}+C^{\prime}\frac{p}{E_{T_{10}}+m_{T_{10}}}\right)+\left(E^{\prime}+F^{\prime}\frac{p}{E_{T_{10}}+m_{T_{10}}}\right)],\nonumber\\
{\cal A}(\frac{1}{2}\to \frac{3}{2}\ -1) & = &N_{T_b}N_{T_{10}}\left(E^{\prime}-F^{\prime}\frac{p_{cm}}{E_{T_{10}}+m_{T_{10}}}\right),\nonumber\\
{\cal A}(-\frac{1}{2}\to -\frac{3}{2}\ 1) & = &N_{T_b}N_{T_{10}}\left(E^{\prime}+F^{\prime}\frac{p_{cm}}{E_{T_{10}}+m_{T_{10}}}\right).
\end{eqnarray}
The definitions of $E_{T_{10}}$, $p_{cm}$, $N_{T_{10}}$ and $N_{T_{b}}$ are the same as Eq.\eqref{ampparameter} except replacing $m_P$ by $m_V$. 
Again all these amplitudes can be determined  from the angular distributions of the four-body decays $T_b\to T_{10}(\to T_{8}P_1)V(\to P_2P_3)$. 

\item 
Branching fractions for $T_b$  decays into a proton with three charged pion/kaons are found at the order $10^{-5}$ in Ref.~\cite{Aaij:2017pgy}.  A plausible scenario  is  that the  $T_b \to T_{10}V$ contribute    significantly to the $T_b$ decaying into a proton and three charged light mesons.  If this is true,  we expect that with more data in future, a detailed analysis will determine the decay widths of $T_b \to T_{10}V$. Then the flavor SU(3) symmetry can be examined, and meanwhile it will also shed light on the $CP$ and $T$ violation in baryonic transitions by using the triplet product asymmetries~\cite{Bediaga:2018lhg,Aaij:2018lsx}. 

\item Through the results in Tab.~\ref{tab:Lambdab_B10P},  we can find  the relations both for decays into $T_{8} P$ and $T_{8} V$. Here only the channels with one vector octet in final states can be listed \eqref{eq:T10Vbdwidthrela}, \eqref{eq:T10Vbswidthrela}. For channels with one pseudoscalar in final states the relations are almost the same,  obtained by replacing the vector multiplets $V$ by the pseudo-scalar multiplets $P$.  However, $\eta_q$ and $\eta_s$ are unphysical states so that the decay width relations involving them should be removed.
 
For $b \to d$ transitions, one has: 
\begin{eqnarray}
\Gamma(\Lambda_{b}^{0}\to\Delta^{-}\rho^{+})=3\Gamma(\Lambda_{b}^{0}\to\Sigma^{\prime-}K^{*+}),&\ \Gamma(\Xi_{b}^{-}\to\Delta^{-}\overline{K}^{*0})=6\Gamma(\Xi_{b}^{-}\to\Sigma^{\prime-}\omega),&\nonumber\\
\Gamma(\Lambda_{b}^{0}\to\Delta^{-}\rho^{+})=3\Gamma(\Xi_{b}^{0}\to\Sigma^{\prime-}\rho^{+}),&\ \Gamma(\Xi_{b}^{-}\to\Delta^{-}\overline{K}^{*0})=3\Gamma(\Xi_{b}^{-}\to\Sigma^{\prime-}\phi),&\nonumber\\
\Gamma(\Lambda_{b}^{0}\to\Delta^{-}\rho^{+})=3\Gamma(\Xi_{b}^{0}\to\Xi^{\prime-}K^{*+}),&\ \Gamma(\Xi_{b}^{-}\to\Delta^{-}\overline{K}^{*0})=3\Gamma(\Xi_{b}^{-}\to\Xi^{\prime-}K^{*0}),&\nonumber\\
\Gamma(\Lambda_{b}^{0}\to\Sigma^{\prime-}K^{*+})=\Gamma(\Xi_{b}^{0}\to\Sigma^{\prime-}\rho^{+}),&\ \Gamma(\Xi_{b}^{-}\to\Sigma^{\prime-}\rho^{0})=\Gamma(\Xi_{b}^{-}\to\Sigma^{\prime-}\omega),&\nonumber\\
\Gamma(\Lambda_{b}^{0}\to\Sigma^{\prime-}K^{*+})=\Gamma(\Xi_{b}^{0}\to\Xi^{\prime-}K^{*+}),&\ \Gamma(\Xi_{b}^{-}\to\Sigma^{\prime-}\rho^{0})=\frac{1}{2}\Gamma(\Xi_{b}^{-}\to\Sigma^{\prime-}\phi),&\nonumber\\
\Gamma(\Xi_{b}^{0}\to\Sigma^{\prime+}\rho^{-})=\Gamma(\Xi_{b}^{0}\to\Xi^{\prime0}K^{*0}),&\ \Gamma(\Xi_{b}^{-}\to\Sigma^{\prime-}\rho^{0})=\frac{1}{2}\Gamma(\Xi_{b}^{-}\to\Xi^{\prime-}K^{*0}),&\nonumber\\
\Gamma(\Xi_{b}^{0}\to\Sigma^{\prime-}\rho^{+})=\Gamma(\Xi_{b}^{0}\to\Xi^{\prime-}K^{*+}),&\ \Gamma(\Xi_{b}^{-}\to\Sigma^{\prime-}\omega)=\frac{1}{2}\Gamma(\Xi_{b}^{-}\to\Sigma^{\prime-}\phi),&\nonumber\\
\Gamma(\Xi_{b}^{-}\to\Delta^{0}K^{*-})=2\Gamma(\Xi_{b}^{-}\to\Sigma^{\prime0}\rho^{-}),&\ \Gamma(\Xi_{b}^{-}\to\Sigma^{\prime-}\omega)=\frac{1}{2}\Gamma(\Xi_{b}^{-}\to\Xi^{\prime-}K^{*0}),&\nonumber\\
\Gamma(\Xi_{b}^{-}\to\Delta^{-}\overline{K}^{*0})=6\Gamma(\Xi_{b}^{-}\to\Sigma^{\prime-}\rho^{0}),&\ \Gamma(\Xi_{b}^{-}\to\Sigma^{\prime-}\phi)=\Gamma(\Xi_{b}^{-}\to\Xi^{\prime-}K^{*0}).\label{eq:T10Vbdwidthrela}
\end{eqnarray}
For $b \to s$ transition, we have the relations for decay widths: 
\begin{eqnarray}
\Gamma(\Lambda_{b}^{0}\to\Delta^{+}K^{*-})=\Gamma(\Lambda_{b}^{0}\to\Delta^{0}\overline{K}^{*0}),&\ \Gamma(\Xi_{b}^{-}\to\Sigma^{\prime-}\overline{K}^{*0})=2\Gamma(\Xi_{b}^{-}\to\Xi^{\prime-}\omega),\nonumber\\
\Gamma(\Lambda_{b}^{0}\to\Sigma^{\prime-}\rho^{+})=\Gamma(\Lambda_{b}^{0}\to\Xi^{\prime-}K^{*+}),&\ \Gamma(\Xi_{b}^{-}\to\Sigma^{\prime-}\overline{K}^{*0})=\Gamma(\Xi_{b}^{-}\to\Xi^{\prime-}\phi),\nonumber\\
\Gamma(\Lambda_{b}^{0}\to\Sigma^{\prime-}\rho^{+})=\Gamma(\Xi_{b}^{0}\to\Xi^{\prime-}\rho^{+}),&\ \Gamma(\Xi_{b}^{-}\to\Sigma^{\prime-}\overline{K}^{*0})=\frac{1}{3}\Gamma(\Xi_{b}^{-}\to\Omega^{-}K^{*0}),\nonumber\\
\Gamma(\Lambda_{b}^{0}\to\Sigma^{\prime-}\rho^{+})=\frac{1}{3}\Gamma(\Xi_{b}^{0}\to\Omega^{-}K^{*+}),&\ \Gamma(\Xi_{b}^{-}\to\Xi^{\prime-}\rho^{0})=\Gamma(\Xi_{b}^{-}\to\Xi^{\prime-}\omega),\nonumber\\
\Gamma(\Lambda_{b}^{0}\to\Xi^{\prime-}K^{*+})=\Gamma(\Xi_{b}^{0}\to\Xi^{\prime-}\rho^{+}),&\ \Gamma(\Xi_{b}^{-}\to\Xi^{\prime-}\rho^{0})=\frac{1}{2}\Gamma(\Xi_{b}^{-}\to\Xi^{\prime-}\phi),\nonumber\\
\Gamma(\Lambda_{b}^{0}\to\Xi^{\prime-}K^{*+})=\frac{1}{3}\Gamma(\Xi_{b}^{0}\to\Omega^{-}K^{*+}),&\ \Gamma(\Xi_{b}^{-}\to\Xi^{\prime-}\rho^{0})=\frac{1}{6}\Gamma(\Xi_{b}^{-}\to\Omega^{-}K^{*0}),\nonumber\\
\Gamma(\Xi_{b}^{0}\to\Xi^{\prime-}\rho^{+})=\frac{1}{3}\Gamma(\Xi_{b}^{0}\to\Omega^{-}K^{*+}),&\ \Gamma(\Xi_{b}^{-}\to\Xi^{\prime-}\omega)=\frac{1}{2}\Gamma(\Xi_{b}^{-}\to\Xi^{\prime-}\phi),\nonumber\\
\Gamma(\Xi_{b}^{-}\to\Sigma^{\prime0}K^{*-})=\frac{1}{2}\Gamma(\Xi_{b}^{-}\to\Xi^{\prime0}\rho^{-}),&\ \Gamma(\Xi_{b}^{-}\to\Xi^{\prime-}\omega)=\frac{1}{6}\Gamma(\Xi_{b}^{-}\to\Omega^{-}K^{*0}),\nonumber\\
\Gamma(\Xi_{b}^{-}\to\Sigma^{\prime-}\overline{K}^{*0})=2\Gamma(\Xi_{b}^{-}\to\Xi^{\prime-}\rho^{0}),&\ \Gamma(\Xi_{b}^{-}\to\Xi^{\prime-}\phi)=\frac{1}{3}\Gamma(\Xi_{b}^{-}\to\Omega^{-}K^{*0}).\label{eq:T10Vbswidthrela}
\end{eqnarray}

\item As discussed in the previous section, charmless $b\to d$ and $b\to s$ transitions can be connected by U-spin. 
In Table~\ref{tab:U-spinLbB10V}, we collect the $T_b\to T_{10}V$ decay pairs related by $U$-spin, while results for the final state with a light pseudoscalar meson can be obtained similarly. CP asymmetries for these pairs satisfy relation in Eq.(\ref{CPR}). Inspired from $B$ decay data~\cite{Patrignani:2016xqp,Tanabashi:2018oca}, we expect 
CP asymmetries for these decays are at the order 10\%. Experimental measurements of these  relations are important to test flavor $SU(3)$ symmetry and  the CKM description of  CP violation in SM.

\end{itemize}

\subsection{$T_b (\Lambda_b,\Xi_{b}^0, \Xi_b^-)$ Decay into an octet baryon and a meson}

\begin{table}
\renewcommand\arraystretch{1.4}
\newcommand{\tabincell}[2]{\begin{tabular}{@{}#1@{}}#2\end{tabular}}
\caption{Decay amplitudes for  $T_b\to T_8 P$ decays governed  by the $b\to d$ transition. Results for vector meson final state are similar. }\label{tab:Lambdab_B8P_bd}\begin{tabular}{cccccc}\hline\hline
channel  & IRA  & TDA \tabularnewline
\hline 
$\Lambda_{b}^{0}\to\Lambda^{0}K^{0}$  & \tabincell{c}{$(-2B_{3}^{T}+D_{3}^{T}-B_{6}^{T}+2C_{6}^{T}+2E_{6}^{T}+3D_{6}^{T}$\\$-B_{15}^{T}+2C_{15}^{T}+2E_{15}^{T}+3D_{15}^{T})/{\sqrt{6}}$} & \tabincell{c}{$(2\bar{a}_{4}+\bar{a}_{6}-\bar{a}_{10}-2\bar{a}_{12}-\bar{a}_{13}+\bar{a}_{14}$\\$-4\bar{b}_{2}-2\bar{b}_{4}-2\bar{b}_{5}-\bar{b}_{6}+\bar{b}_{7})/{\sqrt{6}}$}\tabularnewline
\hline 
$\Lambda_{b}^{0}\to\Sigma^{0}K^{0}$  & $(-D_{3}^{T}+B_{6}^{T}+D_{6}^{T}+B_{15}^{T}+5D_{15}^{T})/{\sqrt{2}}$ & $(\bar{a}_{6}+\bar{a}_{10}+\bar{a}_{13}+\bar{a}_{14}+\bar{b}_{6}+\bar{b}_{7})/{\sqrt{2}}$\tabularnewline
\hline 
$\Lambda_{b}^{0}\to\Sigma^{-}K^{+}$  & $D_{3}^{T}-B_{6}^{T}-D_{6}^{T}-B_{15}^{T}+3D_{15}^{T}$ & $\bar{a}_{8}+\bar{a}_{11}-\bar{b}_{6}-\bar{b}_{7}$\tabularnewline
\hline 
$\Lambda_{b}^{0}\to{p}\pi^{-}$  & $B_{3}^{T}-C_{6}^{T}+E_{6}^{T}-C_{15}^{T}+3E_{15}^{T}$ &  \tabincell{c}{$2\bar{a}_{1}-\bar{a}_{4}-\bar{a}_{6}+\bar{a}_{12}-\bar{a}_{14}+\bar{a}_{15}$\\$+\bar{a}_{18}-\bar{a}_{19}+2\bar{b}_{2}+\bar{b}_{4}+\bar{b}_{5}-\bar{b}_{7}$}\tabularnewline
\hline 
$\Lambda_{b}^{0}\to{n}\pi^{0}$  & $(-B_{3}^{T}+C_{6}^{T}-E_{6}^{T}+C_{15}^{T}+5E_{15}^{T})/{\sqrt{2}}$ & \tabincell{c}{$(2\bar{a}_{2}+\bar{a}_{4}-\bar{a}_{8}-\bar{a}_{10}-\bar{a}_{11}-\bar{a}_{12}-\bar{a}_{13}+\bar{a}_{16}$\\$+\bar{a}_{17}+\bar{a}_{19}-2\bar{b}_{2}-\bar{b}_{4}-\bar{b}_{5}+\bar{b}_{7})/{\sqrt{2}}$}\tabularnewline
\hline 
$\Lambda_{b}^{0}\to{n}\eta_{q}$  & $(2A_{3}^{T}+B_{3}^{T}-2A_{6}^{T}-C_{6}^{T}-E_{6}^{T}-2A_{15}^{T}-C_{15}^{T}+E_{15}^{T})/{\sqrt{2}}$ &  \tabincell{c}{$(1/{\sqrt{2}})(2\bar{a}_{2}-2\bar{a}_{3}-\bar{a}_{4}-\bar{a}_{8}+2\bar{a}_{9}+\bar{a}_{10}-\bar{a}_{11}$\\$+\bar{a}_{12}+\bar{a}_{13}+\bar{a}_{16}+\bar{a}_{17}+\bar{a}_{19}+4\bar{b}_{1}+2\bar{b}_{2}$\\$+2\bar{b}_{3}+\bar{b}_{4}+\bar{b}_{5}+2\bar{b}_{6}+\bar{b}_{7})$}\tabularnewline
\hline 
$\Lambda_{b}^{0}\to{n}\eta_{s}$  & \tabincell{c}{$A_{3}^{T}+D_{3}^{T}-A_{6}^{T}-B_{6}^{T}+E_{6}^{T}$\\$+D_{6}^{T}-A_{15}^{T}-B_{15}^{T}-E_{15}^{T}-D_{15}^{T}$} & $-\bar{a}_{3}+\bar{a}_{9}+2\bar{b}_{1}+\bar{b}_{3}$\tabularnewline
\hline 
$\Xi_{b}^{0}\to\Lambda^{0}\pi^{0}$  & \tabincell{c}{$(B_{3}^{T}+D_{3}^{T}+B_{6}^{T}+C_{6}^{T}+E_{6}^{T}+3D_{6}^{T}$\\$-5B_{15}^{T}-5C_{15}^{T}-5E_{15}^{T}+3D_{15}^{T})/{2\sqrt{3}}$} &  \tabincell{c}{$(1/{2\sqrt{3}})(-2\bar{a}_{2}-\bar{a}_{4}+\bar{a}_{6}-\bar{a}_{7}+\bar{a}_{8}+2\bar{a}_{10}+2\bar{a}_{11}-\bar{a}_{16}$\\$-\bar{a}_{17}+\bar{a}_{18}-2\bar{a}_{19}+2\bar{b}_{2}+\bar{b}_{4}+\bar{b}_{5}-\bar{b}_{6}-2\bar{b}_{7})$}\tabularnewline
\hline 
$\Xi_{b}^{0}\to\Lambda^{0}\eta_{q}$  & \tabincell{c}{$-(2A_{3}^{T}+B_{3}^{T}+2C_{3}^{T}+D_{3}^{T}-6A_{6}^{T}-B_{6}^{T}-C_{6}^{T}-E_{6}^{T}$\\$+3D_{6}^{T}+6A_{15}^{T}+B_{15}^{T}+C_{15}^{T}+E_{15}^{T}+3D_{15}^{T})/{2\sqrt{3}}$} & \tabincell{c}{ $-(1/{2\sqrt{3}})(2\bar{a}_{2}-2\bar{a}_{3}-\bar{a}_{4}+2\bar{a}_{5}+\bar{a}_{6}+\bar{a}_{7}-\bar{a}_{8}$\\$+4\bar{a}_{9}+2\bar{a}_{10}-2\bar{a}_{11}+\bar{a}_{16}+\bar{a}_{17}-\bar{a}_{18}+2\bar{a}_{19}$\\$+4\bar{b}_{1}+2\bar{b}_{2}+2\bar{b}_{3}+\bar{b}_{4}-\bar{b}_{5}+\bar{b}_{6}+2\bar{b}_{7})$}\tabularnewline
\hline 
$\Xi_{b}^{0}\to\Lambda^{0}\eta_{s}$  & \tabincell{c}{$-(A_{3}^{T}-2C_{3}^{T}-3A_{6}^{T}-2B_{6}^{T}-2C_{6}^{T}+E_{6}^{T}$\\$+3A_{15}^{T}+2B_{15}^{T}+2C_{15}^{T}-E_{15}^{T})/{\sqrt{6}}$} & \tabincell{c}{$(\bar{a}_{3}-\bar{a}_{5}-2\bar{a}_{9}-2\bar{a}_{12}-\bar{a}_{13}+\bar{a}_{14}$\\$-2\bar{b}_{1}-\bar{b}_{3}-2\bar{b}_{5}-\bar{b}_{6}+\bar{b}_{7})/{\sqrt{6}}$}\tabularnewline
\hline 
$\Xi_{b}^{0}\to\Sigma^{+}\pi^{-}$  & $-B_{3}^{T}-C_{3}^{T}+B_{6}^{T}-E_{6}^{T}-3B_{15}^{T}+2C_{15}^{T}-3E_{15}^{T}$ & $-2\bar{a}_{1}+\bar{a}_{4}+\bar{a}_{6}-\bar{a}_{15}-2\bar{b}_{2}-\bar{b}_{4}$\tabularnewline
\hline 
$\Xi_{b}^{0}\to\Sigma^{0}\pi^{0}$  &\tabincell{c}{  $(1/2)(-B_{3}^{T}-2C_{3}^{T}-D_{3}^{T}+B_{6}^{T}+C_{6}^{T}$\\$-E_{6}^{T}+D_{6}^{T}-B_{15}^{T}-C_{15}^{T}+5E_{15}^{T}+5D_{15}^{T})$} & \tabincell{c}{$(1/2)(2\bar{a}_{2}+\bar{a}_{4}+\bar{a}_{6}-\bar{a}_{7}-\bar{a}_{8}+\bar{a}_{16}$\\$+\bar{a}_{17}+\bar{a}_{18}-2\bar{b}_{2}-\bar{b}_{4}+\bar{b}_{5}+\bar{b}_{6})$}\tabularnewline
\hline 
$\Xi_{b}^{0}\to\Sigma^{0}\eta_{q}$  &\tabincell{c}{  $(1/2)(2A_{3}^{T}+B_{3}^{T}+D_{3}^{T}+2A_{6}^{T}+B_{6}^{T}+C_{6}^{T}-E_{6}^{T}$\\$-D_{6}^{T}-10A_{15}^{T}-5B_{15}^{T}-5C_{15}^{T}+E_{15}^{T}-5D_{15}^{T})$ }&\tabincell{c}{  $(1/2)(2\bar{a}_{2}-2\bar{a}_{3}-\bar{a}_{4}-2\bar{a}_{5}-\bar{a}_{6}-\bar{a}_{7}-\bar{a}_{8}+\bar{a}_{16}$\\$+\bar{a}_{17}+\bar{a}_{18}+4\bar{b}_{1}+2\bar{b}_{2}+2\bar{b}_{3}+\bar{b}_{4}+\bar{b}_{5}+\bar{b}_{6})$}\tabularnewline
\hline 
$\Xi_{b}^{0}\to\Sigma^{0}\eta_{s}$  & $(A_{3}^{T}+A_{6}^{T}+E_{6}^{T}-5A_{15}^{T}-E_{15}^{T})/{\sqrt{2}}$ & $(-\bar{a}_{3}-\bar{a}_{5}+\bar{a}_{13}+\bar{a}_{14}+2\bar{b}_{1}+\bar{b}_{3}+\bar{b}_{6}+\bar{b}_{7})/{\sqrt{2}}$\tabularnewline
\hline 
$\Xi_{b}^{0}\to\Sigma^{-}\pi^{+}$  & $-C_{3}^{T}-D_{3}^{T}+C_{6}^{T}+D_{6}^{T}+2B_{15}^{T}-3C_{15}^{T}-3D_{15}^{T}$ & $-\bar{a}_{7}-\bar{a}_{8}+\bar{b}_{5}+\bar{b}_{6}$\tabularnewline
\hline 
$\Xi_{b}^{0}\to{p}K^{-}$  & $-C_{3}^{T}+B_{6}^{T}-C_{6}^{T}-3B_{15}^{T}+C_{15}^{T}$ & $\bar{a}_{12}-\bar{a}_{14}+\bar{a}_{18}-\bar{a}_{19}+\bar{b}_{5}-\bar{b}_{7}$\tabularnewline
\hline 
$\Xi_{b}^{0}\to{n}\overline{K}^{0}$  & $-C_{3}^{T}-D_{3}^{T}-C_{6}^{T}-D_{6}^{T}+2B_{15}^{T}+C_{15}^{T}+D_{15}^{T}$ & $\bar{a}_{12}+\bar{a}_{13}+\bar{b}_{5}+\bar{b}_{6}$\tabularnewline
\hline 
$\Xi_{b}^{0}\to\Xi^{-}K^{+}$  & $-C_{3}^{T}-B_{6}^{T}+C_{6}^{T}+B_{15}^{T}-3C_{15}^{T}$ & $-\bar{a}_{7}+\bar{a}_{11}+\bar{b}_{5}-\bar{b}_{7}$\tabularnewline
\hline 
$\Xi_{b}^{0}\to\Xi^{0}K^{0}$  & $-B_{3}^{T}-C_{3}^{T}-B_{6}^{T}+E_{6}^{T}+B_{15}^{T}+2C_{15}^{T}+E_{15}^{T}$ & $\bar{a}_{4}-\bar{a}_{10}-2\bar{b}_{2}-\bar{b}_{4}$\tabularnewline
\hline 
$\Xi_{b}^{-}\to\Lambda^{0}\pi^{-}$  & \tabincell{c}{$(B_{3}^{T}+D_{3}^{T}+B_{6}^{T}+C_{6}^{T}+E_{6}^{T}+3D_{6}^{T}$\\$+3B_{15}^{T}+3C_{15}^{T}+3E_{15}^{T}+3D_{15}^{T})/{\sqrt{6}}$} & \tabincell{c}{$(2\bar{a}_{1}+\bar{a}_{15}-\bar{a}_{17}+\bar{a}_{18}-2\bar{a}_{19}$\\$+2\bar{b}_{2}+\bar{b}_{4}+\bar{b}_{5}-\bar{b}_{6}-2\bar{b}_{7})/{\sqrt{6}}$}\tabularnewline
\hline 
$\Xi_{b}^{-}\to\Sigma^{0}\pi^{-}$  & \tabincell{c}{$(B_{3}^{T}-D_{3}^{T}-B_{6}^{T}+C_{6}^{T}+E_{6}^{T}+D_{6}^{T}$\\$-3B_{15}^{T}+3C_{15}^{T}+3E_{15}^{T}+5D_{15}^{T})/{\sqrt{2}}$} & $(2\bar{a}_{1}+\bar{a}_{15}+\bar{a}_{17}+\bar{a}_{18}+2\bar{b}_{2}+\bar{b}_{4}+\bar{b}_{5}+\bar{b}_{6})/{\sqrt{2}}$\tabularnewline
\hline 
$\Xi_{b}^{-}\to\Sigma^{-}\pi^{0}$  &  \tabincell{c}{$-(B_{3}^{T}-D_{3}^{T}-B_{6}^{T}+C_{6}^{T}+E_{6}^{T}+D_{6}^{T}$\\$-3B_{15}^{T}+3C_{15}^{T}-5E_{15}^{T}-3D_{15}^{T})/{\sqrt{2}}$} & $(2\bar{a}_{2}+\bar{a}_{16}-2\bar{b}_{2}-\bar{b}_{4}-\bar{b}_{5}-\bar{b}_{6})/{\sqrt{2}}$\tabularnewline
\hline 
$\Xi_{b}^{-}\to\Sigma^{-}\eta_{q}$  & \tabincell{c}{$(2A_{3}^{T}+B_{3}^{T}+D_{3}^{T}+2A_{6}^{T}+B_{6}^{T}+C_{6}^{T}-E_{6}^{T}-D_{6}^{T}$\\$+6A_{15}^{T}+3B_{15}^{T}+3C_{15}^{T}+E_{15}^{T}+3D_{15}^{T})/{\sqrt{2}}$} & $(2\bar{a}_{2}+\bar{a}_{16}+4\bar{b}_{1}+2\bar{b}_{2}+2\bar{b}_{3}+\bar{b}_{4}+\bar{b}_{5}+\bar{b}_{6})/{\sqrt{2}}$\tabularnewline
\hline 
$\Xi_{b}^{-}\to\Sigma^{-}\eta_{s}$  & $A_{3}^{T}+A_{6}^{T}+E_{6}^{T}+3A_{15}^{T}-E_{15}^{T}$ & $2\bar{b}_{1}+\bar{b}_{3}+\bar{b}_{6}+\bar{b}_{7}$\tabularnewline
\hline 
$\Xi_{b}^{-}\to{n}K^{-}$  & $D_{3}^{T}+B_{6}^{T}+D_{6}^{T}+3B_{15}^{T}-D_{15}^{T}$ & $-\bar{a}_{17}-\bar{a}_{19}-\bar{b}_{6}-\bar{b}_{7}$\tabularnewline
\hline 
$\Xi_{b}^{-}\to\Xi^{-}K^{0}$  & $B_{3}^{T}+C_{6}^{T}-E_{6}^{T}+3C_{15}^{T}-E_{15}^{T}$ & $2\bar{b}_{2}+\bar{b}_{4}+\bar{b}_{5}-\bar{b}_{7}$\tabularnewline
\hline 
\hline 
\end{tabular}
\end{table}

\begin{table}
\renewcommand\arraystretch{1.4}
\newcommand{\tabincell}[2]{\begin{tabular}{@{}#1@{}}#2\end{tabular}}
\caption{Decay amplitudes for two-body $T_b\to T_8 P$ decays induced by the $b\to s$ transition.}\label{tab:Lambdab_B8P_bs}\begin{tabular}{cccccc}\hline\hline
channel  & IRA  & TDA \tabularnewline
\hline 
$\Lambda_{b}^{0}\to\Lambda^{0}\pi^{0}$  & $-(-2(E_{6}^{T}+B_{15}^{T}+C_{15}^{T}-2E_{15}^{T})+B_{6}^{T}+C_{6}^{T})/{\sqrt{3}}$ &  \tabincell{c}{$(-4\bar{a}_{2}+\bar{a}_{7}+2\bar{a}_{8}+\bar{a}_{11}+\bar{a}_{12}+2\bar{a}_{13}+\bar{a}_{14}$\\$-2\bar{a}_{16}-2\bar{a}_{17}-\bar{a}_{18}-\bar{a}_{19})/{2\sqrt{3}}$}\tabularnewline
\hline 
$\Lambda_{b}^{0}\to\Lambda^{0}\eta_{q}$  & $(-2A_{3}^{T}+C_{3}^{T}+6A_{15}^{T}+B_{15}^{T}+C_{15}^{T}-2E_{15}^{T})/{\sqrt{3}}$ & \tabincell{c}{$-1/{2\sqrt{3}}(4\bar{a}_{2}-4\bar{a}_{3}-2\bar{a}_{5}-\bar{a}_{7}-2\bar{a}_{8}+2\bar{a}_{9}-\bar{a}_{11}$\\$+\bar{a}_{12}+2\bar{a}_{13}+\bar{a}_{14}+2\bar{a}_{16}+2\bar{a}_{17}+\bar{a}_{18}$\\$+\bar{a}_{19}+8\bar{b}_{1}+4\bar{b}_{3}+2\bar{b}_{5}+4\bar{b}_{6}+2\bar{b}_{7})$}\tabularnewline
\hline 
$\Lambda_{b}^{0}\to\Lambda^{0}\eta_{s}$  & \tabincell{c}{$-\sqrt{\frac{2}{3}}(A_{3}^{T}+B_{3}^{T}+C_{3}^{T}+D_{3}^{T}$\\$-3A_{15}^{T}-2B_{15}^{T}-2C_{15}^{T}-2E_{15}^{T}-3D_{15}^{T})$} &  \tabincell{c}{$(2\bar{a}_{3}+2\bar{a}_{4}+\bar{a}_{5}+\bar{a}_{6}-\bar{a}_{9}-\bar{a}_{10}$\\$-4\bar{b}_{1}-4\bar{b}_{2}-2\bar{b}_{3}-2\bar{b}_{4})/{\sqrt{6}}$}\tabularnewline
\hline 
$\Lambda_{b}^{0}\to\Sigma^{+}\pi^{-}$  & $C_{3}^{T}-B_{6}^{T}+C_{6}^{T}+3B_{15}^{T}-C_{15}^{T}$ & $-\bar{a}_{12}+\bar{a}_{14}-\bar{a}_{18}+\bar{a}_{19}-\bar{b}_{5}+\bar{b}_{7}$\tabularnewline
\hline 
$\Lambda_{b}^{0}\to\Sigma^{0}\pi^{0}$  & $C_{3}^{T}+B_{15}^{T}+C_{15}^{T}$ & $1/2(\bar{a}_{7}-\bar{a}_{11}-\bar{a}_{12}+\bar{a}_{14}-\bar{a}_{18}+\bar{a}_{19}-2\bar{b}_{5}+2\bar{b}_{7})$\tabularnewline
\hline 
$\Lambda_{b}^{0}\to\Sigma^{0}\eta_{q}$  & $-2A_{6}^{T}-B_{6}^{T}-C_{6}^{T}+4A_{15}^{T}+2B_{15}^{T}+2C_{15}^{T}$ & $1/2(2\bar{a}_{5}+\bar{a}_{7}+2\bar{a}_{9}-\bar{a}_{11}+\bar{a}_{12}-\bar{a}_{14}-\bar{a}_{18}+\bar{a}_{19})$\tabularnewline
\hline 
$\Lambda_{b}^{0}\to\Sigma^{0}\eta_{s}$  & $\sqrt{2}(-A_{6}^{T}+D_{6}^{T}+2(A_{15}^{T}+D_{15}^{T}))$ & $(\bar{a}_{5}+\bar{a}_{6}+\bar{a}_{9}+\bar{a}_{10})/{\sqrt{2}}$\tabularnewline
\hline 
$\Lambda_{b}^{0}\to\Sigma^{-}\pi^{+}$  & $C_{3}^{T}+B_{6}^{T}-C_{6}^{T}-B_{15}^{T}+3C_{15}^{T}$ & $\bar{a}_{7}-\bar{a}_{11}-\bar{b}_{5}+\bar{b}_{7}$\tabularnewline
\hline 
$\Lambda_{b}^{0}\to{p}K^{-}$  & $B_{3}^{T}+C_{3}^{T}-B_{6}^{T}+E_{6}^{T}+3B_{15}^{T}-2C_{15}^{T}+3E_{15}^{T}$ & $2\bar{a}_{1}-\bar{a}_{4}-\bar{a}_{6}+\bar{a}_{15}+2\bar{b}_{2}+\bar{b}_{4}$\tabularnewline
\hline 
$\Lambda_{b}^{0}\to{n}\overline{K}^{0}$  & $B_{3}^{T}+C_{3}^{T}+B_{6}^{T}-E_{6}^{T}-B_{15}^{T}-2C_{15}^{T}-E_{15}^{T}$ & $-\bar{a}_{4}+\bar{a}_{10}+2\bar{b}_{2}+\bar{b}_{4}$\tabularnewline
\hline 
$\Lambda_{b}^{0}\to\Xi^{-}K^{+}$  & $C_{3}^{T}+D_{3}^{T}-C_{6}^{T}-D_{6}^{T}-2B_{15}^{T}+3C_{15}^{T}+3D_{15}^{T}$ & $\bar{a}_{7}+\bar{a}_{8}-\bar{b}_{5}-\bar{b}_{6}$\tabularnewline
\hline 
$\Lambda_{b}^{0}\to\Xi^{0}K^{0}$  & $C_{3}^{T}+D_{3}^{T}+C_{6}^{T}+D_{6}^{T}-2B_{15}^{T}-C_{15}^{T}-D_{15}^{T}$ & $-\bar{a}_{12}-\bar{a}_{13}-\bar{b}_{5}-\bar{b}_{6}$\tabularnewline
\hline 
$\Xi_{b}^{0}\to\Lambda^{0}\overline{K}^{0}$  & \tabincell{c}{$(-B_{3}^{T}+2D_{3}^{T}-2B_{6}^{T}+C_{6}^{T}+E_{6}^{T}$\\$-2B_{15}^{T}+C_{15}^{T}+E_{15}^{T}-6D_{15}^{T})/{\sqrt{6}}$} &  \tabincell{c}{$(\bar{a}_{4}-\bar{a}_{6}-2\bar{a}_{10}-\bar{a}_{12}-2\bar{a}_{13}-\bar{a}_{14}$\\$-2\bar{b}_{2}-\bar{b}_{4}-\bar{b}_{5}-2\bar{b}_{6}-\bar{b}_{7})/{\sqrt{6}}$}\tabularnewline
\hline 
$\Xi_{b}^{0}\to\Sigma^{+}K^{-}$  & $-B_{3}^{T}+C_{6}^{T}-E_{6}^{T}+C_{15}^{T}-3E_{15}^{T}$ &\tabincell{c}{ $-2\bar{a}_{1}+\bar{a}_{4}+\bar{a}_{6}-\bar{a}_{12}+\bar{a}_{14}-\bar{a}_{15}-\bar{a}_{18}$\\$+\bar{a}_{19}-2\bar{b}_{2}-\bar{b}_{4}-\bar{b}_{5}+\bar{b}_{7}$}\tabularnewline
\hline 
$\Xi_{b}^{0}\to\Sigma^{0}\overline{K}^{0}$  & $(B_{3}^{T}-C_{6}^{T}-E_{6}^{T}-2D_{6}^{T}-C_{15}^{T}-E_{15}^{T}-4D_{15}^{T})/{\sqrt{2}}$ & $-(\bar{a}_{4}+\bar{a}_{6}-\bar{a}_{12}+\bar{a}_{14}-2\bar{b}_{2}-\bar{b}_{4}-\bar{b}_{5}+\bar{b}_{7})/{\sqrt{2}}$\tabularnewline
\hline 
$\Xi_{b}^{0}\to\Xi^{-}\pi^{+}$  & $-D_{3}^{T}+B_{6}^{T}+D_{6}^{T}+B_{15}^{T}-3D_{15}^{T}$ & $-\bar{a}_{8}-\bar{a}_{11}+\bar{b}_{6}+\bar{b}_{7}$\tabularnewline
\hline 
$\Xi_{b}^{0}\to\Xi^{0}\pi^{0}$  & $(D_{3}^{T}-B_{6}^{T}+2E_{6}^{T}+D_{6}^{T}-B_{15}^{T}-4E_{15}^{T}-D_{15}^{T})/{\sqrt{2}}$ & $-(2\bar{a}_{2}-\bar{a}_{8}-\bar{a}_{11}+\bar{a}_{16}+\bar{a}_{17}+\bar{a}_{19}+\bar{b}_{6}+\bar{b}_{7})/{\sqrt{2}}$\tabularnewline
\hline 
$\Xi_{b}^{0}\to\Xi^{0}\eta_{q}$  & \tabincell{c}{$(-2A_{3}^{T}-D_{3}^{T}+2A_{6}^{T}+B_{6}^{T}-D_{6}^{T}$\\$+2A_{15}^{T}+B_{15}^{T}-2E_{15}^{T}+D_{15}^{T})/{\sqrt{2}}$} &  \tabincell{c}{$-(2\bar{a}_{2}-2\bar{a}_{3}-\bar{a}_{8}+2\bar{a}_{9}-\bar{a}_{11}+\bar{a}_{16}+\bar{a}_{17}$\\$+\bar{a}_{19}+4\bar{b}_{1}+2\bar{b}_{3}+\bar{b}_{6}+\bar{b}_{7})/{\sqrt{2}}$}\tabularnewline
\hline 
$\Xi_{b}^{0}\to\Xi^{0}\eta_{s}$  & $-A_{3}^{T}-B_{3}^{T}+A_{6}^{T}+C_{6}^{T}+A_{15}^{T}+C_{15}^{T}+2E_{15}^{T}$ & \tabincell{c}{$\bar{a}_{3}+\bar{a}_{4}-\bar{a}_{9}-\bar{a}_{10}-\bar{a}_{12}-\bar{a}_{13}-2\bar{b}_{1}$\\$-2\bar{b}_{2}-\bar{b}_{3}-\bar{b}_{4}-\bar{b}_{5}-\bar{b}_{6}$}\tabularnewline
\hline 
$\Xi_{b}^{-}\to\Lambda^{0}K^{-}$  & \tabincell{c}{$(B_{3}^{T}-2D_{3}^{T}-2B_{6}^{T}+C_{6}^{T}+E_{6}^{T}$\\$-6B_{15}^{T}+3C_{15}^{T}+3E_{15}^{T}+6D_{15}^{T})/{\sqrt{6}}$} &  \tabincell{c}{$(2\bar{a}_{1}+\bar{a}_{15}+2\bar{a}_{17}+\bar{a}_{18}+\bar{a}_{19}$\\$+2\bar{b}_{2}+\bar{b}_{4}+\bar{b}_{5}+2\bar{b}_{6}+\bar{b}_{7})/{\sqrt{6}}$}\tabularnewline
\hline 
$\Xi_{b}^{-}\to\Sigma^{0}K^{-}$  & $(B_{3}^{T}+C_{6}^{T}+E_{6}^{T}+2D_{6}^{T}+3C_{15}^{T}+3E_{15}^{T}+4D_{15}^{T})/{\sqrt{2}}$ & $(2\bar{a}_{1}+\bar{a}_{15}+\bar{a}_{18}-\bar{a}_{19}+2\bar{b}_{2}+\bar{b}_{4}+\bar{b}_{5}-\bar{b}_{7})/{\sqrt{2}}$\tabularnewline
\hline 
$\Xi_{b}^{-}\to\Sigma^{-}\overline{K}^{0}$  & $B_{3}^{T}+C_{6}^{T}-E_{6}^{T}+3C_{15}^{T}-E_{15}^{T}$ & $2\bar{b}_{2}+\bar{b}_{4}+\bar{b}_{5}-\bar{b}_{7}$\tabularnewline
\hline 
$\Xi_{b}^{-}\to\Xi^{-}\pi^{0}$  & $(D_{3}^{T}+B_{6}^{T}-2E_{6}^{T}-D_{6}^{T}+3B_{15}^{T}+4E_{15}^{T}+3D_{15}^{T})/{\sqrt{2}}$ & $(2\bar{a}_{2}+\bar{a}_{16}-\bar{b}_{6}-\bar{b}_{7})/{\sqrt{2}}$\tabularnewline
\hline 
$\Xi_{b}^{-}\to\Xi^{-}\eta_{q}$  & \tabincell{c}{$(2A_{3}^{T}+D_{3}^{T}+2A_{6}^{T}+B_{6}^{T}-D_{6}^{T}$\\$+6A_{15}^{T}+3B_{15}^{T}+2E_{15}^{T}+3D_{15}^{T})/{\sqrt{2}}$} & $(2\bar{a}_{2}+\bar{a}_{16}+4\bar{b}_{1}+2\bar{b}_{3}+\bar{b}_{6}+\bar{b}_{7})/{\sqrt{2}}$\tabularnewline
\hline 
$\Xi_{b}^{-}\to\Xi^{-}\eta_{s}$  & $A_{3}^{T}+B_{3}^{T}+A_{6}^{T}+C_{6}^{T}+3A_{15}^{T}+3C_{15}^{T}-2E_{15}^{T}$ & $2\bar{b}_{1}+2\bar{b}_{2}+\bar{b}_{3}+\bar{b}_{4}+\bar{b}_{5}+\bar{b}_{6}$\tabularnewline
\hline 
$\Xi_{b}^{-}\to\Xi^{0}\pi^{-}$  & $D_{3}^{T}+B_{6}^{T}+D_{6}^{T}+3B_{15}^{T}-D_{15}^{T}$ & $-\bar{a}_{17}-\bar{a}_{19}-\bar{b}_{6}-\bar{b}_{7}$\tabularnewline
\hline 
\hline 
\end{tabular}
\end{table}

\begin{figure}
\begin{center}
\includegraphics[scale=0.4]{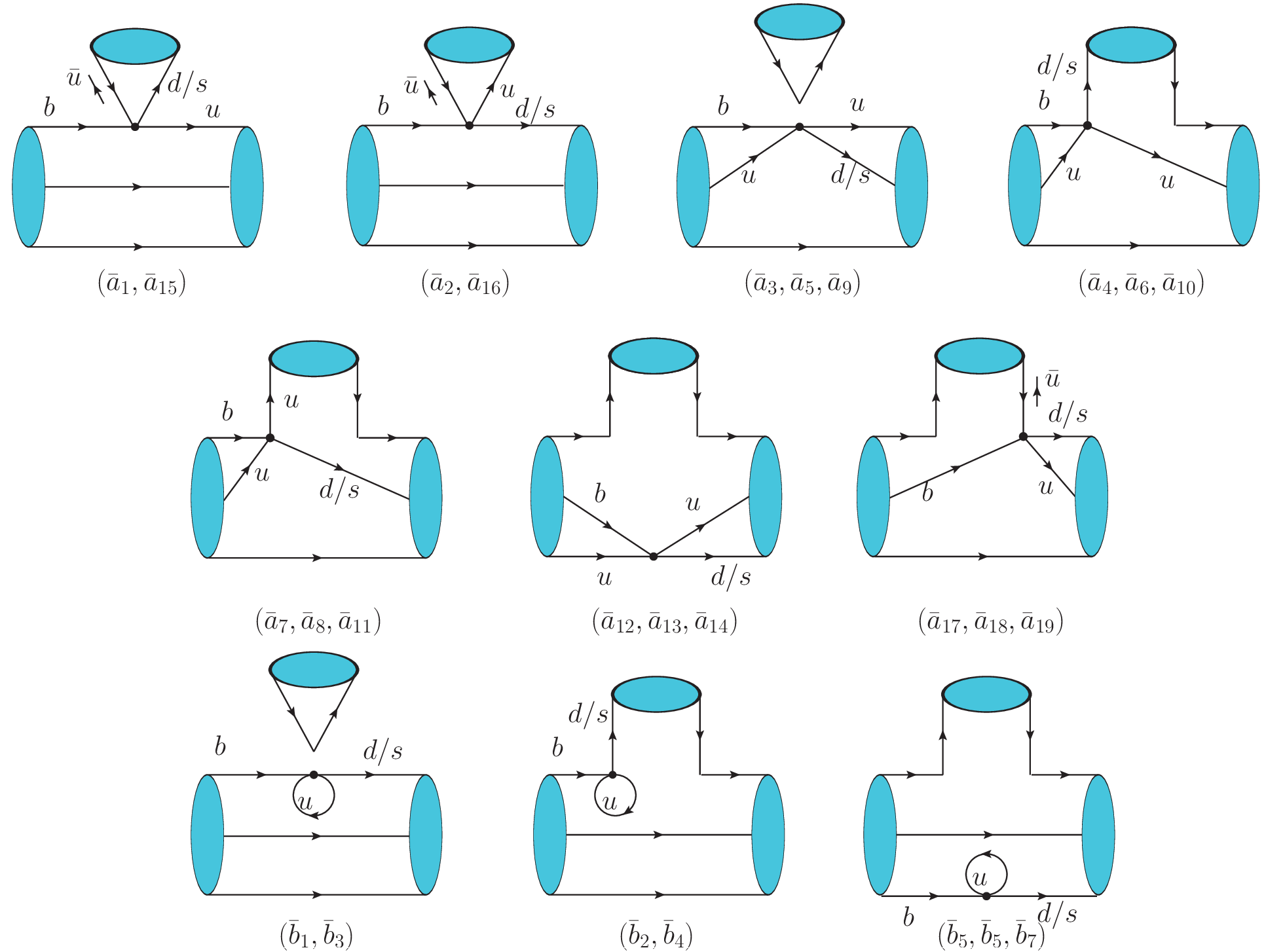}
\end{center}
\caption{Topology diagrams for the bottom baryon decays into an octet baryon and a light meson. Since the octet baryon is not fully symmetric or antisymmetric in flavor space, there are more than one amplitudes corresponding to one topological diagram. Actually the 10 topological diagrams correspond to 26 amplitudes shown in Eq.\eqref{Lb26TDAamps}. As in Fig.~\ref{fig:Feynman_Lambdab_T10}, the unspecified quark flavors can be $u,d,s$.  }\label{fig:Lambdab_B8P_TDA}
\end{figure}

If the final state contains a baryon octet,  the topological diagrams are shown in Fig.~\ref{fig:Lambdab_B8P_TDA} where ten diagrams can be found. However unlike the decuplet baryon, the octet baryon is not fully symmetric or antisymmetric in flavor space. Thus each of the diagrams can provide more than one amplitudes. 
In total, one can have 26 independent TDA amplitudes:
\begin{eqnarray}
{\cal A}_{u}^{TDA} & =&{\bar a}_{1}T_{b\bar 3}^{[ij]}H_{m}^{kl}(\overline T_{8})_{ijk}P_{l}^{m}+{\bar a}_{2}T_{b\bar 3}^{[ij]}H_{m}^{kl}(\overline T_{8})_{ijl}P_{k}^{m}+{\bar a}_{3}T_{b\bar 3}^{[ij]}H_{i}^{kl}(\overline T_{8})_{jkl}P_{m}^{m}+{\bar a}_{4}T_{b\bar 3}^{[ij]}H_{i}^{kl}(\overline T_{8})_{jkm}P_{l}^{m}\nonumber\\
 & &+{\bar a}_{5}T_{b\bar 3}^{[ij]}H_{i}^{kl}(\overline T_{8})_{jlk}P_{m}^{m}+{\bar a}_{6}T_{b\bar 3}^{[ij]}H_{i}^{kl}(\overline T_{8})_{jmk}P_{l}^{m}+{\bar a}_{7}T_{b\bar 3}^{[ij]}H_{i}^{kl}(\overline T_{8})_{jlm}P_{k}^{m}+{\bar a}_{8}T_{b\bar 3}^{[ij]}H_{i}^{kl}(\overline T_{8})_{jml}P_{k}^{m}\nonumber\\
 & &+{\bar a}_{9}T_{b\bar 3}^{[ij]}H_{i}^{kl}(\overline T_{8})_{klj}P_{m}^{m}+{\bar a}_{10}T_{b\bar 3}^{[ij]}H_{i}^{kl}(\overline T_{8})_{kmj}P_{l}^{m}+{\bar a}_{11}T_{b\bar 3}^{[ij]}H_{i}^{kl}(\overline T_{8})_{lmj}P_{k}^{m}+{\bar a}_{12}T_{b\bar 3}^{[ij]}H_{i}^{kl}(\overline T_{8})_{klm}P_{j}^{m}\nonumber\\
 & &+{\bar a}_{13}T_{b\bar 3}^{[ij]}H_{i}^{kl}(\overline T_{8})_{kml}P_{j}^{m}+{\bar a}_{14}T_{b\bar 3}^{[ij]}H_{i}^{kl}(\overline T_{8})_{lmk}P_{j}^{m}+{\bar a}_{15}T_{b\bar 3}^{[ij]}H_{m}^{kl}(\overline T_{8})_{ikj}P_{l}^{m}+{\bar a}_{16}T_{b\bar 3}^{[ij]}H_{m}^{kl}(\overline T_{8})_{ilj}P_{k}^{m}\nonumber\\
 & &+{\bar a}_{17}T_{b\bar 3}^{[ij]}H_{m}^{kl}(\overline T_{8})_{ikl}P_{j}^{m}+{\bar a}_{18}T_{b\bar 3}^{[ij]}H_{m}^{kl}(\overline T_{8})_{ilk}P_{j}^{m}+{\bar a}_{19}T_{b\bar 3}^{[ij]}H_{m}^{kl}(\overline T_{8})_{klj}P_{i}^{m}\nonumber\\
 & &+{\bar b}_{1}T_{b\bar 3}^{[ij]}H_{l}^{lk}(\overline T_{8})_{ijk}P_{m}^{m}+{\bar b}_{2}T_{b\bar 3}^{[ij]}H_{l}^{lk}(\overline T_{8})_{ijm}P_{k}^{m}+{\bar b}_{3}T_{b\bar 3}^{[ij]}H_{l}^{lk}(\overline T_{8})_{ikj}P_{m}^{m}+{\bar b}_{4}T_{b\bar 3}^{[ij]}H_{l}^{lk}(\overline T_{8})_{imj}P_{k}^{m}\nonumber\\
 & &+{\bar b}_{5}T_{b\bar 3}^{[ij]}H_{l}^{lk}(\overline T_{8})_{ikm}P_{j}^{m}+{\bar b}_{6}T_{b\bar 3}^{[ij]}H_{l}^{lk}(\overline T_{8})_{imk}P_{j}^{m}+{\bar b}_{7}T_{b\bar 3}^{[ij]}H_{l}^{lk}(\overline T_{8})_{kmi}P_{j}^{m}. \label{Lb26TDAamps}
\end{eqnarray}
In the IRA approach, one can construct 14 amplitudes:
\begin{eqnarray}
{\cal A}_{u}^{IRA} &=&A_{3}^{T}(T_{b\bar{3}})_{i}H_{\bar{3}}^{j}(\overline T_{8})_{j}^{i}P_{k}^{k}+B_{3}^{T}(T_{b\bar{3}})_{i}H_{\bar{3}}^{j}(\overline T_{8})_{k}^{i}P_{j}^{k}+C_{3}^{T}(T_{b\bar{3}})_{i}H_{\bar{3}}^{i}(\overline T_{8})_{l}^{k}P_{k}^{l}+D_{3}^{T}(T_{b\bar{3}})_{i}H_{\bar{3}}^{j}(\overline T_{8})_{j}^{k}P_{k}^{i}\nonumber \\
 & &+A_{6}^{T}(T_{b\bar{3}})_{i}(H_{6})_{j}^{[ik]}(\overline T_{8})_{k}^{j}P_{l}^{l}+B_{6}^{T}(T_{b\bar{3}})_{i}(H_{6})_{j}^{[ik]}(\overline T_{8})_{k}^{l}P_{l}^{j}+C_{6}^{T}(T_{b\bar{3}})_{i}(H_{6})_{j}^{[ik]}(\overline T_{8})_{l}^{j}P_{k}^{l}\nonumber \\
 & &+E_{6}^{T}(T_{b\bar{3}})_{i}(H_{6})_{l}^{[jk]}(\overline T_{8})_{j}^{i}P_{k}^{l}+D_{6}^{T}(T_{b\bar{3}})_{i}(H_{6})_{l}^{[jk]}(\overline T_{8})_{j}^{l}P_{k}^{i}+A_{15}^{T}(T_{b\bar{3}})_{i}(H_{\overline {15}})_{j}^{\{ik\}}(\overline T_{8})_{k}^{j}P_{l}^{l}\nonumber \\
 & &+B_{15}^{T}(T_{b\bar{3}})_{i}(H_{\overline {15}})_{j}^{\{ik\}}(\overline T_{8})_{k}^{l}P_{l}^{j}+C_{15}^{T}(T_{b\bar{3}})_{i}(H_{\overline {15}})_{j}^{\{ik\}}(\overline T_{8})_{l}^{j}P_{k}^{l}+E_{15}^{T}(T_{b\bar{3}})_{i}(H_{\overline {15}})_{l}^{\{jk\}}(\overline T_{8})_{j}^{i}P_{k}^{l}\nonumber \\
 & &+D_{15}^{T}(T_{b\bar{3}})_{i}(H_{\overline {15}})_{l}^{\{jk\}}(\overline T_{8})_{j}^{l}P_{k}^{i}.\label{Lb14IRAamps}
\end{eqnarray}
It should be noticed that in the above hadrons  have been written in different forms for the same  SU(3) multiplet. For illustration, we use the heavy bottom anti-triplet as the example.   Since it is an anti-triplet,  it is most straightforward to use the $(T_{b\bar 3})_i$  to represent this particle multiplet, as in IRA approach. The advantage of this form is its compactness, however, with this form it is not easy to understand the quark flows in the decays.  Instead there are two anti-symmetric light quarks in the anti-triplet heavy bottom baryon, and thus it is viable  to use $(T_{b\bar 3})^{ij}$ (with $ij$ anti-symmetric) to denote this multiplet.  The second form contains two SU(3) indices,  less compact, but it can reflect the quark flows in the decays. Thus the second form is suitable for drawing Feynman diagrams, and thus adopted in the above TDA amplitude. 
These two forms are equal, and  one can establish relations between the SU(3) parameters corresponding to these two forms. The explicit discussions in IRA approach are given in Appendix A.

The 14 IRA amplitudes and 26 TDA amplitudes are related as follows: 
\begin{eqnarray}
A_{3}^{T}&=&\frac{1}{8} \left(-2 \bar{a}_1+6 \bar{a}_{2}-5
   \bar{a}_{3}-\bar{a}_{5}+\bar{a}_{6}-3 \bar{a}_{8}+4
   \bar{a}_{9}+\bar{a}_{10}-3 \bar{a}_{11}+2 \bar{a}_{13}+2
   \bar{a}_{14}-\bar{a}_{15}+3 \bar{a}_{16}+3
   \bar{a}_{17}-\bar{a}_{18}+4 \bar{a}_{19} 
  \right) \nonumber\\
   && +2 \bar{b}_{1}+\bar{b}_{3}+\bar{b}_{6}+\bar{b}_{7},\nonumber\\
   B_{3}^{T}&=&\frac{1}{8} \left(6 \bar{a}_1-2 \bar{a}_{2}-5 \bar{a}_{4}-3
   \bar{a}_{6}-\bar{a}_{7}+\bar{a}_{8}+2 \bar{a}_{10}+2 \bar{a}_{11}+4
   \bar{a}_{12}+\bar{a}_{13}-3 \bar{a}_{14}+3
   \bar{a}_{15}-\bar{a}_{16}-\bar{a}_{17}+3 \bar{a}_{18}-4 \bar{a}_{19}\right) \nonumber\\
   &&+  \left(2 \bar{b}_{2}+\bar{b}_{4}+\bar{b}_{5}-\bar{b}_{7}\right),\nonumber\\
   C_{3}^{T}&=&\frac{1}{8} \left(-\bar{a}_{4}+3 \bar{a}_{7}+\bar{a}_{10}-3 \bar{a}_{11}-4
   \bar{a}_{12}-\bar{a}_{13}+3 \bar{a}_{14}+\bar{a}_{17}-3 \bar{a}_{18}+4
   \bar{a}_{19}-8 \bar{b}_{5}\right)+\bar{b}_{7},\nonumber\\
   D_{3}^{T}&=&\frac{1}{8} \left(-4 \left(\bar{a}_{19}+2
   \left(\bar{b}_{6}+\bar{b}_{7}\right)\right)-\bar{a}_{6}+3
   \bar{a}_{8}-\bar{a}_{10}+3 \bar{a}_{11}-2 \bar{a}_{13}-2 \bar{a}_{14}-3
   \bar{a}_{17}+\bar{a}_{18}\right),\nonumber\\
   A_{6}^{T}&=&\frac{1}{4} \left(\bar{a}_{3}-\bar{a}_{5}-2
   \bar{a}_{9}-\bar{a}_{13}+\bar{a}_{14}\right), \;\;\; B_{6}^{T}=\frac{1}{4} \left(\bar{a}_{13}-\bar{a}_{14}-\bar{a}_{17}+\bar{a}_{18}-2 \bar{a}_{19}\right),\nonumber\\
   C_{6}^{T}&=&\frac{1}{4} \left(\bar{a}_{4}-\bar{a}_{7}-\bar{a}_{10}+\bar{a}_{11}-2 \bar{a}_{12}\right), \;\;\; D_{6}^{T}=\frac{1}{4} \left(\bar{a}_{6}-\bar{a}_{8}+\bar{a}_{10}-\bar{a}_{11}-\bar{a}_{13}+\bar{a}_{14}\right), \nonumber\\
    E_{6}^{T}&=&\frac{1}{4} \left(2 \bar{a}_1-2 \bar{a}_{2}-\bar{a}_{6}+\bar{a}_{8}-\bar{a}_{10}+\bar{a}_{11}+\bar{a}_{13}-\bar{a}_{14}+\bar{a}_{15}-\bar{a}_{16}-\bar{a}_{17}+\bar{a}_{18}-2 \bar{a}_{19}\right),\nonumber\\
    A_{15}^{T}&=& \frac{1}{8} \left(\bar{a}_{3}+\bar{a}_{5}-\bar{a}_{13}-\bar{a}_{14}\right), \;\;\; B_{15}^{T}= \frac{1}{8} \left(\bar{a}_{13}+\bar{a}_{14}-\bar{a}_{17}-\bar{a}_{18}\right),\nonumber\\
    C_{15}^{T}&=& \frac{1}{8} \left(\bar{a}_{4}+\bar{a}_{7}-\bar{a}_{10}-\bar{a}_{11}\right), \;\;\; D_{15}^{T}= \frac{1}{8} \left(\bar{a}_{6}+\bar{a}_{8}+\bar{a}_{10}+\bar{a}_{11}+\bar{a}_{13}+\bar{a}_{14}\right),\nonumber\\
    E_{15}^{T}&=& \frac{1}{8} \left(2 \bar{a}_1+2 \bar{a}_{2}-\bar{a}_{6}-\bar{a}_{8}-\bar{a}_{10}-\bar{a}_{11}-\bar{a}_{13}-\bar{a}_{14}+\bar{a}_{15}+\bar{a}_{16}+\bar{a}_{17}+\bar{a}_{18}\right). \label{IRATDA}
\end{eqnarray}

However, even after such reduction, there still exists one independent degree of freedom among the 14 IRA amplitudes. The redundant amplitude can be made explicit with the redefinitions:
\begin{eqnarray}
A_{6}^{T\prime}= A_{6}^{T}+B_{6}^{T}, \;\;\; B_{6}^{T\prime}=B_{6}^{T}-C_{6}^{T},\;\;\; C_{6}^{T\prime}= C_{6}^{T}-E_{6}^{T},\;\;\; D_{6}^{T\prime}= C_{6}^{T}+D_{6}^{T}. \label{redundant14to13}
\end{eqnarray}

In addition, this redundancy can be understood more explicitly. In this work as well as the previous work Ref.~\cite{He:2018php} we use the irreducible representation operators for IRA as $(H_{6/ \overline{15}})_k^{ij}$. Actually there exists a simpler $H_6$ representation introduced by Ref \cite{Geng:2018plk}, where $H_6$ has only two lower indexes $(H_{6})_{ij}$. With the use of $(H_{6})_{ij}$ we do have only 13 IRA amplitudes. However, Since the IRA operators $(H_{6/ \overline{15}})_k^{ij}$ have the same index structure as the TDA operators. They make the derivation of IRA/TDA correspondence more directly so we will keep the use of them.


The expanded amplitudes can be found in Tab.~\ref{tab:Lambdab_B8P_bd} for the $b\to d$ transition  and \ref{tab:Lambdab_B8P_bs} for the $b\to s$ transition, respectively. Again if the final state is a vector meson, the amplitudes can be derived similarly.     

A few remarks are given in order. 
\begin{itemize}
\item  At a first sight, the diagrammatic approach, as depicted  in Fig.~\ref{fig:Lambdab_B8P_TDA},  is more intuitive, however there are more TDA amplitudes than the corresponding diagrams.  For an octet baryon in the final state,  there are three light quarks. In the same diagram, the symmetry of the quarks in flavor space could be different.   For example, in the third diagram of Fig.~\ref{fig:Lambdab_B8P_TDA},   the $u$ quark and another light quark could be  flavor anti-symmetric, or   the two unspecified quarks could be flavor anti-symmetric. These different combinations  will lead to different TDA amplitudes.  
Thus it is very hard to determine  the independent amplitudes in this approach, which  will introduce  subtleties to the global fit in the diagrammatic approach. The mismatch between Feynman diagrams and TDA amplitudes will not happen for the decays into decuplet baryons, since all three quarks are symmetric in flavor space. 

\item Without including  the polarization, one can see from the IRA approach, there exist 13 independent complex amplitudes  with CKM factor $V_{ub}V_{uq}^*$ and another 13 amplitudes accompanied by $V_{tb}V_{tq}^*$.

\item Two polarization configurations exist for decays into a pseudoscalar meson, while  there are four possibilities   for decays into a vector meson.

\item The U-spin related decay pairs are given in Tab.~\ref{tab:UspinLbB8P}, which completely fits with the results given by Ref.~\cite{He:2015fsa}. Here only the case for $T_b \to B_{8} P$ is listed. Since no unphysical states $\eta_q$ and $\eta_s$ exist in Tab.~\ref{tab:UspinLbB8P}. The U-spin pairs for  $T_b \to B_{8} V$ are similar by  replacing pseudoscalar octets by vector octets.

\item Some theoretical  analyses of nonleptonic bottom baryon decays based on either explicit modes or the flavor symmetry can be found in Refs.~\cite{Zhu:2016bra,Lu:2009cm,Hsiao:2017tif,Hsiao:2014mua,Wei:2009np,Wang:2015ndk}, while the experimental measurements can be found in Refs.~\cite{Aaij:2012as,Aaltonen:2014vra,Aaij:2018tlk}.  To date, the available measurements of two-body $\Lambda_b$  branching fractions are~\cite{Patrignani:2016xqp,Tanabashi:2018oca}:
\begin{eqnarray}
{\cal B}(\Lambda_b\to p\pi^-) = (4.2\pm0.8)\times 10^{-6},\nonumber\\
{\cal B}(\Lambda_b\to pK^-) = (5.1\pm0.9)\times 10^{-6},\nonumber\\
{\cal B}(\Lambda_b\to \Lambda\eta) = (9^{+7}_{-5})\times 10^{-6},\nonumber\\
{\cal B}(\Lambda_b\to \Lambda\eta') < 3.1\times 10^{-6},\nonumber\\
{\cal B}(\Lambda_b\to p\phi) = (9.2\pm2.5)\times 10^{-6}. 
\end{eqnarray}

\item The CP asymmetries for $\Lambda_b \to p\pi^-/pK^-$~\cite{Aaij:2018tlk} have been measured: 
\begin{eqnarray}
A_{CP}^{p\pi^-} &=&-0.020\pm0.013\pm0.019, \;\;
A_{CP}^{p K^-} =-0.035\pm0.017\pm0.020.  
\end{eqnarray}
Thus measuring the branching fractions and CP asymmetries for $\Xi_b^0\to \pi^-\Sigma^+$ and $\Xi_{b}^0\to K^-\Sigma^+$ will help us to understand the U-spin in baryonic decays.


\end{itemize}

\begin{table}
\caption{U-spin relations for $T_b \to T_{8} P$. }\label{tab:UspinLbB8P}
\begin{tabular}{cccccc}
\hline 
\hline 
$b\to d$ & $b\to s$ &$r$ & $b\to d$ & $b\to s$ & $r$ \tabularnewline
\hline 
${\cal A}(\Xi_{b}^{-}\to K^{-}n)$ & ${\cal A}(\Xi_{b}^{-}\to\pi^{-}\Xi^{0})$ & $+1$ & ${\cal A}(\Xi_{b}^{0}\to\bar{K}^{0}n)$ & ${\cal A}(\Lambda_{b}^{0}\to K^{0}\Xi^{0})$ & $-1$\tabularnewline
\hline 
${\cal A}(\Xi_{b}^{-}\to K^{0}\Xi^{-})$ & ${\cal A}(\Xi_{b}^{-}\to\bar{K}^{0}\Sigma^{-})$ & $+1$ & ${\cal A}(\Xi_{b}^{0}\to K^{0}\Xi^{0})$ & ${\cal A}(\Lambda_{b}^{0}\to\bar{K}^{0}n)$ & $-1$\tabularnewline
\hline 
${\cal A}(\Xi_{b}^{0}\to\pi^{-}\Sigma^{+})$ & ${\cal A}(\Lambda_{b}^{0}\to K^{-}p)$ & $-1$ & ${\cal A}(\Lambda_{b}^{0}\to\pi^{-}p)$ & ${\cal A}(\Xi_{b}^{0}\to K^{-}\Sigma^{+})$ & $-1$\tabularnewline
\hline 
${\cal A}(\Xi_{b}^{0}\to\pi^{+}\Sigma^{-})$ & ${\cal A}(\Lambda_{b}^{0}\to K^{+}\Xi^{-})$ & $-1$ & ${\cal A}(\Lambda_{b}^{0}\to K^{+}\Sigma^{-})$ & ${\cal A}(\Xi_{b}^{0}\to\pi^{+}\Xi^{-})$ & $-1$\tabularnewline
\hline 
${\cal A}(\Xi_{b}^{0}\to K^{-}p)$ & ${\cal A}(\Lambda_{b}^{0}\to\pi^{-}\Sigma^{+})$ & $-1$ & ${\cal A}(\Xi_{b}^{0}\to K^{+}\Xi^{-})$ & ${\cal A}(\Lambda_{b}^{0}\to\pi^{+}\Sigma^{-})$ & $+1$\tabularnewline
\hline 
\hline 
\end{tabular}
\end{table}


\section{ Antitriplet Charmed  Baryon $T_c(\Lambda_c,\Xi_{c}^+, \Xi_c^0)$ Decays }

\begin{figure}
\begin{center}
\includegraphics[scale=0.45]{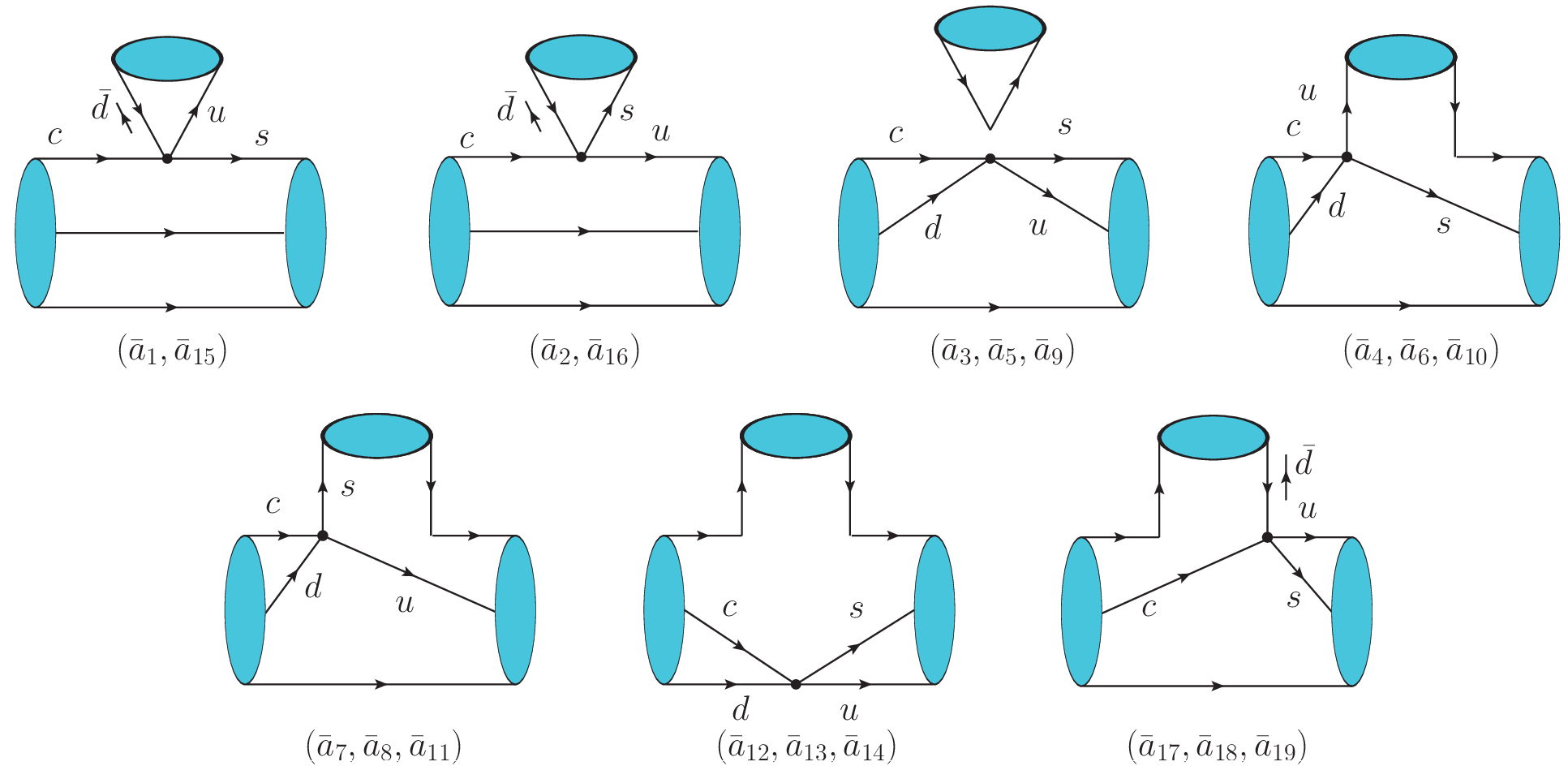}
\end{center}
\caption{Topology diagrams for the charmed baryon decays into an octet baryon and a light meson.  Due to the same reason as that of bottom baryon decays, one topological diagrams correspond to more than one TDA amplitudes.  Here the 7 topological diagrams correspond to 19 TDA amplitudes as given in Eq.~\eqref{eq:TDA_charm_19}. }\label{fig:Lambdac_B8P_TDA}
\end{figure}

\begin{table}
\global\long\def\arraystretch{1.4}
 \newcommand{\tabincell}[2]{\begin{tabular}{@{}#1@{}}#2\end{tabular}}
 \caption{Decay amplitudes for two-body Cabibblo-Allowed charmed baryon decays.}
\label{tab:Two_body_Cabibblo_Allowed}%
\begin{tabular}{cccccc}
\hline 
\hline 
{channel}  & {IRA}  & {TDA}  &  &  & \tabularnewline
\hline 
$\Lambda_{c}^{+}\to\Lambda^{0}\pi^{+}$  & $(B_{6}^{T}+C_{6}^{T}-2E_{6}^{T}+B_{15}^{T}+C_{15}^{T}-2E_{15}^{T})/{\sqrt{6}}$ & \tabincell{c}{$(-4\bar{a}_{1}+\bar{a}_{4}+2\bar{a}_{6}+\bar{a}_{10}-\bar{a}_{12}+\bar{a}_{13}+2\bar{a}_{14}$\\$-2\bar{a}_{15}-\bar{a}_{17}-2\bar{a}_{18}+\bar{a}_{19})/{\sqrt{6}}$}  &  &  & \tabularnewline
\hline 
$\Lambda_{c}^{+}\to\Sigma^{+}\pi^{0}$  & $(-B_{6}^{T}+C_{6}^{T}-B_{15}^{T}+C_{15}^{T})/{\sqrt{2}}$ & $(\bar{a}_{4}-\bar{a}_{10}-\bar{a}_{12}-\bar{a}_{13}+\bar{a}_{17}+\bar{a}_{19})/{\sqrt{2}}$  &  &  & \tabularnewline
\hline 
$\Lambda_{c}^{+}\to\Sigma^{+}\eta_{q}$  & $(2A_{6}^{T}+B_{6}^{T}+C_{6}^{T}+2A_{15}^{T}+B_{15}^{T}+C_{15}^{T})/{\sqrt{2}}$ & $(2\bar{a}_{3}+\bar{a}_{4}-2\bar{a}_{9}-\bar{a}_{10}-\bar{a}_{12}-\bar{a}_{13}-\bar{a}_{17}-\bar{a}_{19})/{\sqrt{2}}$  &  &  & \tabularnewline
\hline 
$\Lambda_{c}^{+}\to\Sigma^{+}\eta_{s}$  & $A_{6}^{T}-D_{6}^{T}+A_{15}^{T}+D_{15}^{T}$ & $\bar{a}_{3}+\bar{a}_{8}-\bar{a}_{9}+\bar{a}_{11}$  &  &  & \tabularnewline
\hline 
$\Lambda_{c}^{+}\to\Sigma^{0}\pi^{+}$  & $(B_{6}^{T}-C_{6}^{T}+B_{15}^{T}-C_{15}^{T})/{\sqrt{2}}$ & $(-\bar{a}_{4}+\bar{a}_{10}+\bar{a}_{12}+\bar{a}_{13}-\bar{a}_{17}-\bar{a}_{19})/{\sqrt{2}}$  &  &  & \tabularnewline
\hline 
$\Lambda_{c}^{+}\to{p}\overline{K}^{0}$  & $B_{6}^{T}-E_{6}^{T}+B_{15}^{T}+E_{15}^{T}$ & $2\bar{a}_{2}-\bar{a}_{8}-\bar{a}_{11}+\bar{a}_{16}$  &  &  & \tabularnewline
\hline 
$\Lambda_{c}^{+}\to\Xi^{0}K^{+}$  & $C_{6}^{T}+D_{6}^{T}+C_{15}^{T}+D_{15}^{T}$ & $\bar{a}_{4}+\bar{a}_{6}-\bar{a}_{12}+\bar{a}_{14}$  &  &  & \tabularnewline
\hline 
$\Xi_{c}^{+}\to\Sigma^{+}\overline{K}^{0}$  & $E_{6}^{T}+D_{6}^{T}-E_{15}^{T}-D_{15}^{T}$ & $-2\bar{a}_{2}-\bar{a}_{16}-\bar{a}_{17}-\bar{a}_{19}$  &  &  & \tabularnewline
\hline 
$\Xi_{c}^{+}\to\Xi^{0}\pi^{+}$  & $-E_{6}^{T}-D_{6}^{T}-E_{15}^{T}-D_{15}^{T}$ & $-2\bar{a}_{1}-\bar{a}_{15}-\bar{a}_{18}+\bar{a}_{19}$  &  &  & \tabularnewline
\hline 
$\Xi_{c}^{0}\to\Lambda^{0}\overline{K}^{0}$  & $(2B_{6}^{T}-C_{6}^{T}-E_{6}^{T}-2B_{15}^{T}+C_{15}^{T}+E_{15}^{T})/{\sqrt{6}}$ & \tabincell{c}{$(2\bar{a}_{2}+\bar{a}_{7}-\bar{a}_{8}-2\bar{a}_{11}+\bar{a}_{12}-\bar{a}_{13}-2\bar{a}_{14}$\\$+\bar{a}_{16}+\bar{a}_{17}+2\bar{a}_{18}-\bar{a}_{19})/{\sqrt{6}}$}  &  &  & \tabularnewline
\hline 
$\Xi_{c}^{0}\to\Sigma^{+}K^{-}$  & $-C_{6}^{T}-D_{6}^{T}+C_{15}^{T}+D_{15}^{T}$ & $\bar{a}_{7}+\bar{a}_{8}+\bar{a}_{12}+\bar{a}_{13}$  &  &  & \tabularnewline
\hline 
$\Xi_{c}^{0}\to\Sigma^{0}\overline{K}^{0}$  & $(C_{6}^{T}-E_{6}^{T}-C_{15}^{T}+E_{15}^{T})/{\sqrt{2}}$ & $(2\bar{a}_{2}-\bar{a}_{7}-\bar{a}_{8}-\bar{a}_{12}-\bar{a}_{13}+\bar{a}_{16}+\bar{a}_{17}+\bar{a}_{19})/{\sqrt{2}}$  &  &  & \tabularnewline
\hline 
$\Xi_{c}^{0}\to\Xi^{-}\pi^{+}$  & $-B_{6}^{T}+E_{6}^{T}+B_{15}^{T}+E_{15}^{T}$ & $2\bar{a}_{1}-\bar{a}_{6}-\bar{a}_{10}+\bar{a}_{15}$  &  &  & \tabularnewline
\hline 
$\Xi_{c}^{0}\to\Xi^{0}\pi^{0}$  & $(B_{6}^{T}+D_{6}^{T}-B_{15}^{T}+D_{15}^{T})/{\sqrt{2}}$ & $(\bar{a}_{6}+\bar{a}_{10}+\bar{a}_{18}-\bar{a}_{19})/{\sqrt{2}}$  &  &  & \tabularnewline
\hline 
$\Xi_{c}^{0}\to\Xi^{0}\eta_{q}$  & $(-2A_{6}^{T}-B_{6}^{T}+D_{6}^{T}+2A_{15}^{T}+B_{15}^{T}+D_{15}^{T})/{\sqrt{2}}$ & $(2\bar{a}_{5}+\bar{a}_{6}+2\bar{a}_{9}+\bar{a}_{10}-\bar{a}_{18}+\bar{a}_{19})/{\sqrt{2}}$  &  &  & \tabularnewline
\hline 
$\Xi_{c}^{0}\to\Xi^{0}\eta_{s}$  & $-A_{6}^{T}-C_{6}^{T}+A_{15}^{T}+C_{15}^{T}$ & $\bar{a}_{5}+\bar{a}_{7}+\bar{a}_{9}-\bar{a}_{11}+\bar{a}_{12}-\bar{a}_{14}$  &  &  & \tabularnewline
\hline 
\hline 
\end{tabular}
\end{table}
\begin{table}
 \renewcommand\arraystretch{1.2}
 \newcommand{\tabincell}[2]{\begin{tabular}{@{}#1@{}}#2\end{tabular}}
\caption{Decay amplitudes for two-body Singly Cabibblo-Suppressed charmed baryon decays.}\label{tab:Two_body_Singly_Cabibblo_Suppressed}\begin{tabular}{cccccc}\hline\hline
channel  & IRA  & TDA \tabularnewline
\hline 
$\Lambda_{c}^{+}\to\Lambda^{0}K^{+}$  & \tabincell{c}{$(B_{6}^{T}-2C_{6}^{T}-2E_{6}^{T}-3D_{6}^{T}+B_{15}^{T}$\\$-2C_{15}^{T}-2E_{15}^{T}-3D_{15}^{T})/{\sqrt{6}}$} & \tabincell{c}{$-(4\bar{a}_{1}+2\bar{a}_{4}+\bar{a}_{6}-\bar{a}_{10}-2\bar{a}_{12}-\bar{a}_{13}$\\$+\bar{a}_{14}+2\bar{a}_{15}+\bar{a}_{17}+2\bar{a}_{18}-\bar{a}_{19})/{\sqrt{6}}$}\tabularnewline
\hline 
$\Lambda_{c}^{+}\to\Sigma^{+}K^{0}$  & $B_{6}^{T}+D_{6}^{T}+B_{15}^{T}-D_{15}^{T}$ & $-\bar{a}_{8}-\bar{a}_{11}-\bar{a}_{17}-\bar{a}_{19}$\tabularnewline
\hline 
$\Lambda_{c}^{+}\to\Sigma^{0}K^{+}$  & $(B_{6}^{T}+D_{6}^{T}+B_{15}^{T}+D_{15}^{T})/{\sqrt{2}}$ & $(\bar{a}_{6}+\bar{a}_{10}+\bar{a}_{13}+\bar{a}_{14}-\bar{a}_{17}-\bar{a}_{19})/{\sqrt{2}}$\tabularnewline
\hline 
$\Lambda_{c}^{+}\to{p}\pi^{0}$  & $(C_{6}^{T}-E_{6}^{T}+C_{15}^{T}+E_{15}^{T})/{\sqrt{2}}$ & \tabincell{c}{$(2\bar{a}_{2}+\bar{a}_{4}-\bar{a}_{8}-\bar{a}_{10}-\bar{a}_{11}$\\$-\bar{a}_{12}-\bar{a}_{13}+\bar{a}_{16}+\bar{a}_{17}+\bar{a}_{19})/{\sqrt{2}}$}\tabularnewline
\hline 
$\Lambda_{c}^{+}\to{p}\eta_{q}$  & $(2A_{6}^{T}+C_{6}^{T}+E_{6}^{T}+2A_{15}^{T}+C_{15}^{T}-E_{15}^{T})/{\sqrt{2}}$ & \tabincell{c}{$-(2\bar{a}_{2}-2\bar{a}_{3}-\bar{a}_{4}-\bar{a}_{8}+2\bar{a}_{9}+\bar{a}_{10}-\bar{a}_{11}$\\$+\bar{a}_{12}+\bar{a}_{13}+\bar{a}_{16}+\bar{a}_{17}+\bar{a}_{19})/{\sqrt{2}}$}\tabularnewline
\hline 
$\Lambda_{c}^{+}\to{p}\eta_{s}$  & $A_{6}^{T}+B_{6}^{T}-E_{6}^{T}-D_{6}^{T}+A_{15}^{T}+B_{15}^{T}+E_{15}^{T}+D_{15}^{T}$ & $2\bar{a}_{2}+\bar{a}_{3}-\bar{a}_{9}+\bar{a}_{16}$\tabularnewline
\hline 
$\Lambda_{c}^{+}\to{n}\pi^{+}$  & $C_{6}^{T}-E_{6}^{T}+C_{15}^{T}-E_{15}^{T}$ & $-2\bar{a}_{1}+\bar{a}_{4}+\bar{a}_{6}-\bar{a}_{12}+\bar{a}_{14}-\bar{a}_{15}-\bar{a}_{18}+\bar{a}_{19}$\tabularnewline
\hline 
$\Xi_{c}^{+}\to\Lambda^{0}\pi^{+}$  & $(B_{6}^{T}+C_{6}^{T}+E_{6}^{T}+3D_{6}^{T}+B_{15}^{T}+C_{15}^{T}+E_{15}^{T}+3D_{15}^{T})/{\sqrt{6}}$ & \tabincell{c}{$(2\bar{a}_{1}+\bar{a}_{4}+2\bar{a}_{6}+\bar{a}_{10}-\bar{a}_{12}+\bar{a}_{13}+2\bar{a}_{14}$\\$+\bar{a}_{15}-\bar{a}_{17}+\bar{a}_{18}-2\bar{a}_{19})/{\sqrt{6}}$}\tabularnewline
\hline 
$\Xi_{c}^{+}\to\Sigma^{+}\pi^{0}$  & $(-B_{6}^{T}+C_{6}^{T}+E_{6}^{T}+D_{6}^{T}-B_{15}^{T}+C_{15}^{T}-E_{15}^{T}-D_{15}^{T})/{\sqrt{2}}$ & $-(2\bar{a}_{2}-\bar{a}_{4}+\bar{a}_{10}+\bar{a}_{12}+\bar{a}_{13}+\bar{a}_{16})/{\sqrt{2}}$\tabularnewline
\hline 
$\Xi_{c}^{+}\to\Sigma^{+}\eta_{q}$  & \tabincell{c}{$(2A_{6}^{T}+B_{6}^{T}+C_{6}^{T}-E_{6}^{T}-D_{6}^{T}+2A_{15}^{T}$\\$+B_{15}^{T}+C_{15}^{T}+E_{15}^{T}+D_{15}^{T})/{\sqrt{2}}$} & $(2\bar{a}_{2}+2\bar{a}_{3}+\bar{a}_{4}-2\bar{a}_{9}-\bar{a}_{10}-\bar{a}_{12}-\bar{a}_{13}+\bar{a}_{16})/{\sqrt{2}}$\tabularnewline
\hline 
$\Xi_{c}^{+}\to\Sigma^{+}\eta_{s}$  & $A_{6}^{T}+E_{6}^{T}+A_{15}^{T}-E_{15}^{T}$ & $-2\bar{a}_{2}+\bar{a}_{3}+\bar{a}_{8}-\bar{a}_{9}+\bar{a}_{11}-\bar{a}_{16}-\bar{a}_{17}-\bar{a}_{19}$\tabularnewline
\hline 
$\Xi_{c}^{+}\to\Sigma^{0}\pi^{+}$  & $(B_{6}^{T}-C_{6}^{T}-E_{6}^{T}-D_{6}^{T}+B_{15}^{T}-C_{15}^{T}-E_{15}^{T}-D_{15}^{T})/{\sqrt{2}}$ & $-(2\bar{a}_{1}+\bar{a}_{4}-\bar{a}_{10}-\bar{a}_{12}-\bar{a}_{13}+\bar{a}_{15}+\bar{a}_{17}+\bar{a}_{18})/{\sqrt{2}}$\tabularnewline
\hline 
$\Xi_{c}^{+}\to{p}\overline{K}^{0}$  & $B_{6}^{T}+D_{6}^{T}+B_{15}^{T}-D_{15}^{T}$ & $-\bar{a}_{8}-\bar{a}_{11}-\bar{a}_{17}-\bar{a}_{19}$\tabularnewline
\hline 
$\Xi_{c}^{+}\to\Xi^{0}K^{+}$  & $C_{6}^{T}-E_{6}^{T}+C_{15}^{T}-E_{15}^{T}$ & $-2\bar{a}_{1}+\bar{a}_{4}+\bar{a}_{6}-\bar{a}_{12}+\bar{a}_{14}-\bar{a}_{15}-\bar{a}_{18}+\bar{a}_{19}$\tabularnewline
\hline 
$\Xi_{c}^{0}\to\Lambda^{0}\pi^{0}$  & $-(B_{6}^{T}+C_{6}^{T}+E_{6}^{T}+3D_{6}^{T}-B_{15}^{T}-C_{15}^{T}-E_{15}^{T}+3D_{15}^{T})/{2\sqrt{3}}$ & \tabincell{c}{$(2\bar{a}_{2}-3\bar{a}_{6}+\bar{a}_{7}-\bar{a}_{8}-3\bar{a}_{10}-2\bar{a}_{11}+\bar{a}_{12}$\\$-\bar{a}_{13}-2\bar{a}_{14}+\bar{a}_{16}+\bar{a}_{17}-\bar{a}_{18}+2\bar{a}_{19})/{2\sqrt{3}}$}\tabularnewline
\hline 
$\Xi_{c}^{0}\to\Lambda^{0}\eta_{q}$  & \tabincell{c}{$(6A_{6}^{T}+B_{6}^{T}+C_{6}^{T}+E_{6}^{T}-3D_{6}^{T}-6A_{15}^{T}$\\$-B_{15}^{T}-C_{15}^{T}-E_{15}^{T}-3D_{15}^{T})/{2\sqrt{3}}$} & \tabincell{c}{$-(1/{2\sqrt{3}})(2\bar{a}_{2}+6\bar{a}_{5}+3\bar{a}_{6}+\bar{a}_{7}-\bar{a}_{8}+6\bar{a}_{9}+3\bar{a}_{10}$\\$-2\bar{a}_{11}+\bar{a}_{12}-\bar{a}_{13}-2\bar{a}_{14}+\bar{a}_{16}+\bar{a}_{17}-\bar{a}_{18}+2\bar{a}_{19})$}\tabularnewline
\hline 
$\Xi_{c}^{0}\to\Lambda^{0}\eta_{s}$  & \tabincell{c}{$(3A_{6}^{T}+2B_{6}^{T}+2C_{6}^{T}-E_{6}^{T}-3A_{15}^{T}$\\$-2B_{15}^{T}-2C_{15}^{T}+E_{15}^{T})/{\sqrt{6}}$} & \tabincell{c}{$(2\bar{a}_{2}-3\bar{a}_{5}-2\bar{a}_{7}-\bar{a}_{8}-3\bar{a}_{9}+\bar{a}_{11}-2\bar{a}_{12}$\\$-\bar{a}_{13}+\bar{a}_{14}+\bar{a}_{16}+\bar{a}_{17}+2\bar{a}_{18}-\bar{a}_{19})/{\sqrt{6}}$}\tabularnewline
\hline 
$\Xi_{c}^{0}\to\Sigma^{+}\pi^{-}$  & $C_{6}^{T}+D_{6}^{T}-C_{15}^{T}-D_{15}^{T}$ & $-\bar{a}_{7}-\bar{a}_{8}-\bar{a}_{12}-\bar{a}_{13}$\tabularnewline
\hline 
$\Xi_{c}^{0}\to\Sigma^{0}\pi^{0}$  & $1/{2}(B_{6}^{T}+C_{6}^{T}-E_{6}^{T}+D_{6}^{T}-B_{15}^{T}-C_{15}^{T}+E_{15}^{T}+D_{15}^{T})$ & \tabincell{c}{$(1/{2})(2\bar{a}_{2}+\bar{a}_{6}-\bar{a}_{7}-\bar{a}_{8}+\bar{a}_{10}-\bar{a}_{12}$\\$-\bar{a}_{13}+\bar{a}_{16}+\bar{a}_{17}+\bar{a}_{18})$}\tabularnewline
\hline 
$\Xi_{c}^{0}\to\Sigma^{0}\eta_{q}$  &  \tabincell{c}{$1/{2}(-2A_{6}^{T}-B_{6}^{T}-C_{6}^{T}+E_{6}^{T}+D_{6}^{T}$\\$+2A_{15}^{T}+B_{15}^{T}+C_{15}^{T}-E_{15}^{T}+D_{15}^{T})$} &  \tabincell{c}{$(1/{2})(-2\bar{a}_{2}+2\bar{a}_{5}+\bar{a}_{6}+\bar{a}_{7}+\bar{a}_{8}+2\bar{a}_{9}$\\$+\bar{a}_{10}+\bar{a}_{12}+\bar{a}_{13}-\bar{a}_{16}-\bar{a}_{17}-\bar{a}_{18})$}\tabularnewline
\hline 
$\Xi_{c}^{0}\to\Sigma^{0}\eta_{s}$  & $(-A_{6}^{T}-E_{6}^{T}+A_{15}^{T}+E_{15}^{T})/{\sqrt{2}}$ & \tabincell{c}{$(2\bar{a}_{2}+\bar{a}_{5}-\bar{a}_{8}+\bar{a}_{9}-\bar{a}_{11}-\bar{a}_{13}-$\\$\bar{a}_{14}+\bar{a}_{16}+\bar{a}_{17}+\bar{a}_{19})/{\sqrt{2}}$}\tabularnewline
\hline 
$\Xi_{c}^{0}\to\Sigma^{-}\pi^{+}$  & $B_{6}^{T}-E_{6}^{T}-B_{15}^{T}-E_{15}^{T}$ & $-2\bar{a}_{1}+\bar{a}_{6}+\bar{a}_{10}-\bar{a}_{15}$\tabularnewline
\hline 
$\Xi_{c}^{0}\to{p}K^{-}$  & $-C_{6}^{T}-D_{6}^{T}+C_{15}^{T}+D_{15}^{T}$ & $\bar{a}_{7}+\bar{a}_{8}+\bar{a}_{12}+\bar{a}_{13}$\tabularnewline
\hline 
$\Xi_{c}^{0}\to{n}\overline{K}^{0}$  & $B_{6}^{T}-C_{6}^{T}-B_{15}^{T}+C_{15}^{T}$ & $\bar{a}_{7}-\bar{a}_{11}+\bar{a}_{12}-\bar{a}_{14}+\bar{a}_{18}-\bar{a}_{19}$\tabularnewline
\hline 
$\Xi_{c}^{0}\to\Xi^{-}K^{+}$  & $-B_{6}^{T}+E_{6}^{T}+B_{15}^{T}+E_{15}^{T}$ & $2\bar{a}_{1}-\bar{a}_{6}-\bar{a}_{10}+\bar{a}_{15}$\tabularnewline
\hline 
$\Xi_{c}^{0}\to\Xi^{0}K^{0}$  & $-B_{6}^{T}+C_{6}^{T}+B_{15}^{T}-C_{15}^{T}$ & $-\bar{a}_{7}+\bar{a}_{11}-\bar{a}_{12}+\bar{a}_{14}-\bar{a}_{18}+\bar{a}_{19}$\tabularnewline
\hline 
\hline 
\end{tabular}
\end{table}

\begin{table}
\renewcommand\arraystretch{1.4}
 \newcommand{\tabincell}[2]{\begin{tabular}{@{}#1@{}}#2\end{tabular}}
\caption{Decay amplitudes for two-body Doubly Cabibblo-Suppressed charmed baryon decays.}\label{tab:Two_body_Doubly_Cabibblo_Suppressed}\begin{tabular}{cccccc}\hline\hline
{channel}  &{IRA}   & {TDA}  \\\hline
$\Lambda_{c}^{+}\to{p}K^{0}$  & $-E_{6}^{T}-D_{6}^{T}+E_{15}^{T}+D_{15}^{T}$ & $2\bar{a}_{2}+\bar{a}_{16}+\bar{a}_{17}+\bar{a}_{19}$ &  &  & \tabularnewline
\hline 
$\Lambda_{c}^{+}\to{n}K^{+}$  & $E_{6}^{T}+D_{6}^{T}+E_{15}^{T}+D_{15}^{T}$ & $2\bar{a}_{1}+\bar{a}_{15}+\bar{a}_{18}-\bar{a}_{19}$ &  &  & \tabularnewline
\hline 
$\Xi_{c}^{+}\to\Lambda^{0}K^{+}$  & $-(B_{6}^{T}-2C_{6}^{T}+E_{6}^{T}+B_{15}^{T}-2C_{15}^{T}+E_{15}^{T})/{\sqrt{6}}$ & \tabincell{c}{$-(2\bar{a}_{1}-2\bar{a}_{4}-\bar{a}_{6}+\bar{a}_{10}+2\bar{a}_{12}+\bar{a}_{13}-\bar{a}_{14}$\\$+\bar{a}_{15}-\bar{a}_{17}+\bar{a}_{18}-2\bar{a}_{19})/{\sqrt{6}}$} &  &  & \tabularnewline
\hline 
$\Xi_{c}^{+}\to\Sigma^{+}K^{0}$  & $-B_{6}^{T}+E_{6}^{T}-B_{15}^{T}-E_{15}^{T}$ & $-2\bar{a}_{2}+\bar{a}_{8}+\bar{a}_{11}-\bar{a}_{16}$ &  &  & \tabularnewline
\hline 
$\Xi_{c}^{+}\to\Sigma^{0}K^{+}$  & $(-B_{6}^{T}+E_{6}^{T}-B_{15}^{T}+E_{15}^{T})/{\sqrt{2}}$ & $(2\bar{a}_{1}-\bar{a}_{6}-\bar{a}_{10}-\bar{a}_{13}-\bar{a}_{14}+\bar{a}_{15}+\bar{a}_{17}+\bar{a}_{18})/{\sqrt{2}}$ &  &  & \tabularnewline
\hline 
$\Xi_{c}^{+}\to{p}\pi^{0}$  & $-(C_{6}^{T}+D_{6}^{T}+C_{15}^{T}-D_{15}^{T})/{\sqrt{2}}$ & $(-\bar{a}_{4}+\bar{a}_{8}+\bar{a}_{10}+\bar{a}_{11}+\bar{a}_{12}+\bar{a}_{13})/{\sqrt{2}}$ &  &  & \tabularnewline
\hline 
$\Xi_{c}^{+}\to{p}\eta_{q}$  & $-(2A_{6}^{T}+C_{6}^{T}-D_{6}^{T}+2A_{15}^{T}+C_{15}^{T}+D_{15}^{T})/{\sqrt{2}}$ & $(-2\bar{a}_{3}-\bar{a}_{4}-\bar{a}_{8}+2\bar{a}_{9}+\bar{a}_{10}-\bar{a}_{11}+\bar{a}_{12}+\bar{a}_{13})/{\sqrt{2}}$ &  &  & \tabularnewline
\hline 
$\Xi_{c}^{+}\to{p}\eta_{s}$  & $-A_{6}^{T}-B_{6}^{T}-A_{15}^{T}-B_{15}^{T}$ & $-\bar{a}_{3}+\bar{a}_{9}+\bar{a}_{17}+\bar{a}_{19}$ &  &  & \tabularnewline
\hline 
$\Xi_{c}^{+}\to{n}\pi^{+}$  & $-C_{6}^{T}-D_{6}^{T}-C_{15}^{T}-D_{15}^{T}$ & $-\bar{a}_{4}-\bar{a}_{6}+\bar{a}_{12}-\bar{a}_{14}$ &  &  & \tabularnewline
\hline 
$\Xi_{c}^{0}\to\Lambda^{0}K^{0}$  & $(-B_{6}^{T}+2C_{6}^{T}-E_{6}^{T}+B_{15}^{T}-2C_{15}^{T}+E_{15}^{T})/{\sqrt{6}}$ & \tabincell{c}{$(2\bar{a}_{2}-2\bar{a}_{7}-\bar{a}_{8}+\bar{a}_{11}-2\bar{a}_{12}-\bar{a}_{13}+\bar{a}_{14}$\\$+\bar{a}_{16}+\bar{a}_{17}-\bar{a}_{18}+2\bar{a}_{19})/{\sqrt{6}}$} &  &  & \tabularnewline
\hline 
$\Xi_{c}^{0}\to\Sigma^{0}K^{0}$  & $(B_{6}^{T}-E_{6}^{T}-B_{15}^{T}+E_{15}^{T})/{\sqrt{2}}$ & $(2\bar{a}_{2}-\bar{a}_{8}-\bar{a}_{11}-\bar{a}_{13}-\bar{a}_{14}+\bar{a}_{16}+\bar{a}_{17}+\bar{a}_{18})/{\sqrt{2}}$ &  &  & \tabularnewline
\hline 
$\Xi_{c}^{0}\to\Sigma^{-}K^{+}$  & $-B_{6}^{T}+E_{6}^{T}+B_{15}^{T}+E_{15}^{T}$ & $2\bar{a}_{1}-\bar{a}_{6}-\bar{a}_{10}+\bar{a}_{15}$ &  &  & \tabularnewline
\hline 
$\Xi_{c}^{0}\to{p}\pi^{-}$  & $-C_{6}^{T}-D_{6}^{T}+C_{15}^{T}+D_{15}^{T}$ & $\bar{a}_{7}+\bar{a}_{8}+\bar{a}_{12}+\bar{a}_{13}$ &  &  & \tabularnewline
\hline 
$\Xi_{c}^{0}\to{n}\pi^{0}$  & $(C_{6}^{T}+D_{6}^{T}-C_{15}^{T}+D_{15}^{T})/{\sqrt{2}}$ & $(\bar{a}_{6}-\bar{a}_{7}+\bar{a}_{10}+\bar{a}_{11}-\bar{a}_{12}+\bar{a}_{14})/{\sqrt{2}}$ &  &  & \tabularnewline
\hline 
$\Xi_{c}^{0}\to{n}\eta_{q}$  & $(-2A_{6}^{T}-C_{6}^{T}+D_{6}^{T}+2A_{15}^{T}+C_{15}^{T}+D_{15}^{T})/{\sqrt{2}}$ & $(2\bar{a}_{5}+\bar{a}_{6}+\bar{a}_{7}+2\bar{a}_{9}+\bar{a}_{10}-\bar{a}_{11}+\bar{a}_{12}-\bar{a}_{14})/{\sqrt{2}}$ &  &  & \tabularnewline
\hline 
$\Xi_{c}^{0}\to{n}\eta_{s}$  & $-A_{6}^{T}-B_{6}^{T}+A_{15}^{T}+B_{15}^{T}$ & $\bar{a}_{5}+\bar{a}_{9}-\bar{a}_{18}+\bar{a}_{19}$ &  &  & \tabularnewline
\hline 
\hline 
\end{tabular}
\end{table}

For the charmed baryon decays, the $H_3$ contributions are vanishingly small, and thus we have 19 amplitudes in TDA:  
\begin{eqnarray}
{\cal A}_{u}^{TDA} & =&{\bar a}_{1}T_{c\bar 3}^{[ij]}H_{m}^{kl}(\overline T_{8})_{ijk}P_{l}^{m}+{\bar a}_{2}T_{c\bar 3}^{[ij]}H_{m}^{kl}(\overline T_{8})_{ijl}P_{k}^{m}+{\bar a}_{3}T_{c\bar 3}^{[ij]}H_{i}^{kl}(\overline T_{8})_{jkl}P_{m}^{m}+{\bar a}_{4}T_{c\bar 3}^{[ij]}H_{i}^{kl}(\overline T_{8})_{jkm}P_{l}^{m}\nonumber\\
 & &+{\bar a}_{5}T_{c\bar 3}^{[ij]}H_{i}^{kl}(\overline T_{8})_{jlk}P_{m}^{m}+{\bar a}_{6}T_{c\bar 3}^{[ij]}H_{i}^{kl}(\overline T_{8})_{jmk}P_{l}^{m}+{\bar a}_{7}T_{c\bar 3}^{[ij]}H_{i}^{kl}(\overline T_{8})_{ilm}P_{k}^{m}+{\bar a}_{8}T_{c\bar 3}^{[ij]}H_{i}^{kl}(\overline T_{8})_{jml}P_{k}^{m}\nonumber\\
 & &+{\bar a}_{9}T_{c\bar 3}^{[ij]}H_{i}^{kl}(\overline T_{8})_{klj}P_{m}^{m}+{\bar a}_{10}T_{c\bar 3}^{[ij]}H_{i}^{kl}(\overline T_{8})_{kmj}P_{l}^{m}+{\bar a}_{11}T_{c\bar 3}^{[ij]}H_{i}^{kl}(\overline T_{8})_{lmj}P_{k}^{m}+{\bar a}_{12}T_{c\bar 3}^{[ij]}H_{i}^{kl}(\overline T_{8})_{klm}P_{j}^{m}\nonumber\\
 & &+{\bar a}_{13}T_{c\bar 3}^{[ij]}H_{i}^{kl}(\overline T_{8})_{kml}P_{j}^{m}+{\bar a}_{14}T_{c\bar 3}^{[ij]}H_{i}^{kl}(\overline T_{8})_{lmk}P_{j}^{m}+{\bar a}_{15}T_{c\bar 3}^{[ij]}H_{m}^{kl}(\overline T_{8})_{ikj}P_{l}^{m}+{\bar a}_{16}T_{c\bar 3}^{[ij]}H_{m}^{kl}(\overline T_{8})_{ilj}P_{m}^{k}\nonumber\\
 & &+{\bar a}_{17}T_{c\bar 3}^{[ij]}H_{m}^{kl}(\overline T_{8})_{ikl}P_{j}^{m}+{\bar a}_{18}T_{c\bar 3}^{[ij]}H_{m}^{kl}(\overline T_{8})_{ilk}P_{j}^{m}+{\bar a}_{19}T_{c\bar 3}^{[ij]}H_{m}^{kl}(\overline T_{8})_{klj}P_{i}^{m}. \label{eq:TDA_charm_19}
\end{eqnarray}
The corresponding Feynman diagrams are given in Fig.~\ref{fig:Lambdac_B8P_TDA}, in which 7 Feynman diagrams can be found. The analysis for independent amplitudes are almost the same as that of bottom baryon decays. On the other side, 
10 IRA amplitudes can be constructed as:
\begin{eqnarray}
{\cal A}_{u}^{IRA} & =&A_{6}^{T}(T_{c\bar{3}})_{i}(H_{6})_{j}^{[ik]}(\overline T_{8})_{k}^{j}P_{l}^{l}+B_{6}^{T}(T_{c\bar{3}})_{i}(H_{6})_{j}^{[ik]}(\overline T_{8})_{k}^{l}P_{l}^{j}+C_{6}^{T}(T_{c\bar{3}})_{i}(H_{6})_{j}^{[ik]}(\overline T_{8})_{l}^{j}P_{k}^{l}\nonumber\\
 & &+E_{6}^{T}(T_{c\bar{3}})_{i}(H_{6})_{l}^{[jk]}(\overline T_{8})_{j}^{i}P_{k}^{l}+D_{6}^{T}(T_{c\bar{3}})_{i}(H_{6})_{l}^{[jk]}(\overline T_{8})_{j}^{l}P_{k}^{i}+A_{15}^{T}(T_{c\bar{3}})_{i}(H_{\overline {15}})_{j}^{\{ik\}}(\overline T_{8})_{k}^{j}P_{l}^{l}\nonumber\\
 & &+B_{15}^{T}(T_{c\bar{3}})_{i}(H_{\overline {15}})_{j}^{\{ik\}}(\overline T_{8})_{k}^{l}P_{l}^{j}+C_{15}^{T}(T_{c\bar{3}})_{i}(H_{\overline {15}})_{j}^{\{ik\}}(\overline T_{8})_{l}^{j}P_{k}^{l}+E_{15}^{T}(T_{c\bar{3}})_{i}(H_{\overline {15}})_{l}^{\{jk\}}(\overline T_{8})_{j}^{i}P_{k}^{l}\nonumber\\
 & &+D_{15}^{T}(T_{c\bar{3}})_{i}(H_{\overline {15}})_{l}^{\{ik\}}(\overline T_{8})_{j}^{l}P_{k}^{i}. \label{LcIRA}
\end{eqnarray}
Only  9  of them are independent, and  one redundant amplitude can be made explicit with the redefinitions: 
\begin{eqnarray}
A_{6}^{T\prime}= A_{6}^{T}+B_{6}^{T}, \;\;\; B_{6}^{T\prime}=B_{6}^{T}-C_{6}^{T},\;\;\; C_{6}^{T\prime}= C_{6}^{T}-E_{6}^{T},\;\;\; D_{6}^{T\prime}= C_{6}^{T}+D_{6}^{T}, \label{LcIndepend}
\end{eqnarray}
which is exactly the same as Eq.~(\ref{redundant14to13}).

After a careful examination, one can also find the relations:
\begin{eqnarray}
A_{6}^{T}&=& \frac{1}{2} \left(\bar{a}_{3}-\bar{a}_{5}-2 \bar{a}_{9}-\bar{a}_{13}+\bar{a}_{14}\right), \;\;\; B_{6}^{T}= \frac{1}{2} \left(\bar{a}_{13}-\bar{a}_{14}-\bar{a}_{17}+\bar{a}_{18}-2 \bar{a}_{19}\right), \nonumber\\
C_{6}^{T}&=& \frac{1}{2} \left(\bar{a}_{4}-\bar{a}_{7}-\bar{a}_{10}+\bar{a}_{11}-2 \bar{a}_{12}\right), \;\;\; D_{6}^{T}= \frac{1}{2} \left(\bar{a}_{6}-\bar{a}_{8}+\bar{a}_{10}-\bar{a}_{11}-\bar{a}_{13}+\bar{a}_{14}\right), \nonumber\\
E_{6}^{T}&=& \frac{1}{2} \left(2 \bar{a}_{1}-2 \bar{a}_{2}-\bar{a}_{6}+\bar{a}_{8}-\bar{a}_{10}+\bar{a}_{11}+\bar{a}_{13}-\bar{a}_{14}+\bar{a}_{15}-\bar{a}_{16}-\bar{a}_{17}+\bar{a}_{18}-2 \bar{a}_{19}\right), \nonumber\\
A_{15}^{T}&=& \frac{1}{2} \left(\bar{a}_{3}+\bar{a}_{5}-\bar{a}_{13}-\bar{a}_{14}\right), \;\;\; B_{15}^{T}= \frac{1}{2} \left(\bar{a}_{13}+\bar{a}_{14}-\bar{a}_{17}-\bar{a}_{18}\right), \nonumber\\
C_{15}^{T}&=& \frac{1}{2} \left(\bar{a}_{4}+\bar{a}_{7}-\bar{a}_{10}-\bar{a}_{11}\right), \;\;\; D_{15}^{T}= \frac{1}{2} \left(\bar{a}_{6}+\bar{a}_{8}+\bar{a}_{10}+\bar{a}_{11}+\bar{a}_{13}+\bar{a}_{14}\right), \nonumber\\
E_{15}^{T}&=& \frac{1}{2} \left(2 \bar{a}_{1}+2 \bar{a}_{2}-\bar{a}_{6}-\bar{a}_{8}-\bar{a}_{10}-\bar{a}_{11}-\bar{a}_{13}-\bar{a}_{14}+\bar{a}_{15}+\bar{a}_{16}+\bar{a}_{17}+\bar{a}_{18}\right). 
\end{eqnarray} 

Some further remarks are given in order. 
\begin{itemize}
\item   The flavor SU(3) symmetry in charmed baryon decays and the symmetry breaking effects have been extensively explored  in Refs.~\cite{Lu:2016ogy,Geng:2017esc,Geng:2017mxn,Wang:2017gxe,Geng:2018plk,Cheng:2018hwl,Zhao:2018zcb,Geng:2018bow,Geng:2018upx}, and we refer the reader to these references for detailed discussions.   

\item 
On the experimental  side, BESIII collaboration has given  the first measurement of decay branching fractions for the $W$-exchange induced decays~\cite{Ablikim:2018bir}: 
\begin{eqnarray}
{\cal B}(\Lambda_c\to \Xi^0K^+)&=&(5.90\pm0.86\pm0.39)\times 10^{-3}, \\
{\cal B}(\Lambda_c\to \Xi(1530)^0K^+)&=&(5.02\pm0.99\pm0.31)\times 10^{-3}. 
\end{eqnarray}
It indicates that the decays into a decuplet baryon might  not be power suppressed compared to those decays into an octet baryon.  This introduces a theoretical  difficulty to understand the charmed baryon decays. 
\item One can find  some relations between the different channels listed in Table~\ref{tab:Two_body_Cabibblo_Allowed}, Table~\ref{tab:Two_body_Singly_Cabibblo_Suppressed} and Table~\ref{tab:Two_body_Doubly_Cabibblo_Suppressed}. For the charmed baryon two-body decay, there is only one   relation for decay width:
 \begin{eqnarray}
    \Gamma(\Lambda_c^+\to \Sigma^+\pi^0 )= \Gamma(\Lambda_c^+\to \Sigma^0\pi^+ ). 
 \end{eqnarray}
This relation  fits well with the data in Ref.~\cite{Patrignani:2016xqp}:
     \begin{eqnarray}
    {\cal B}(\Lambda_c^+\to \Sigma^+\pi^0 )=1.24 \pm 0.10\%, \;\;\;
    {\cal B}(\Lambda_c^+\to \Sigma^0\pi^+ )=1.28 \pm 0.07\%. 
 \end{eqnarray}
 
\item  In   Ref \cite{Geng:2018plk}, a global fit was conducted for charmed baryon decays, particularly inspired by the recent BESIII data~\cite{Ablikim:2018bir,Ablikim:2015flg,Ablikim:2018czr}. In that work the sextet contribution  was expressed in a different representation.  Relating  the four coefficients  in \cite{Geng:2018plk} with our notations, we have~\footnote{In Ref.~\cite{Geng:2018plk},  these parameters are denoted as $a_1,\ a_2,\ a_3,\ h$. Here we add  primes in order to distinguish them with the parameters used in this work. }: 
\begin{eqnarray}
-A_{6}^{T}+D_{6}^{T}=h=(0.105 \pm 0.073)\ {\rm GeV^3}, \;\;\; -B_{6}^{T}+E_{6}^{T}=a_{1}=(0.244 \pm 0.006)\ {\rm GeV^3},\\
-C_{6}^{T}-D_{6}^{T}=a_{2}=(0.115 \pm 0.014)\ {\rm GeV^3},\;\;\; E_{6}^{T}+D_{6}^{T}=a_{3}=(0.088 \pm 0.019)\ {\rm GeV^3}.
\end{eqnarray}
Such a fit was conducted with the neglect of the  $H_{\overline {15}}$ terms, which might be challenged in interpreting the  $\Lambda_c\to p\pi^0$~\cite{Cheng:2018hwl,Geng:2017esc,Geng:2018bow}. 
\end{itemize}

 \section{Discussions and Conclusions}
\label{sec:conclusion}

In this work, we have carried out a comprehensive  analysis comparing two different  realizations of   the flavor SU(3) symmetry, the irreducible operator representation amplitude and topological diagram amplitude, to study various bottom/charm meson and baryon decays.

We find that previous analyses in the literature using these two methods  do not match consistently in several ways. 
The TDA approach provides a more intuitive understanding of the decays, however it also suffers from a few subtleties.  Using  two-body $B/D$ meson decays, we have  demonstrated that  a few SU(3) independent amplitudes (the last 6 diagrams in Fig.1) are sometimes overlooked in TDA (for instance Refs.~\cite{Gronau:1994rj,Cheng:2014rfa}).  Most of these amplitudes arises from higher order loop corrections, but they are irreducible in the flavor SU(3) space, and thus can not be neglected in principle.  
Taking these new amplitudes into account, we find a consistent description in both approaches.  In addition, using $B$ and $D$ decays we have found that   one of the $A,C,E,T$ amplitudes should be absorbed into others, which has been pointed out in Ref.~\cite{Muller:2015lua}. 
For heavy baryon decays,  we pointed out though the TDA approach is very intuitive, it suffers the difficulty in providing the independent amplitudes.  On this point, the IRA approach is more helpful.

All results derived in this paper can be used to study the heavy meson and baryon decays in the future when sufficient data become available. Then one can have a better understanding of the role of flavor SU(3) symmetry in heavy meson and baryon decays.

For charm quark decays, we did not include the penguin contributions, which can also be studied in a similar manner. 
It is also necessary to notice that the flavor SU(3) symmetry has been applied to study weak decays of doubly heavy baryons~\cite{Wang:2017azm,Shi:2017dto,Wang:2018utj}, and multi-body $\Lambda_c$ decays~\cite{Geng:2018upx}. The equivalence between the TDA and IRA approaches in these decay modes can be studied similarly.

\section*{Acknowledgement}
 
The authors are   grateful to Hai-Yang Cheng, Chao-Qiang Geng, Di Wang,  Fu-Sheng Yu and Ruilin Zhu  for useful discussions.   This work is supported  in part   by National  Natural
Science Foundation of China under Grant
 No.11575110, 11575111, 11735010, and   Natural  Science Foundation of Shanghai under Grant No.~15DZ2272100.   X.G. He was also supported in part by  MOST (Grant No. MOST104-2112-M-002-015-MY3 and 106-2112-M-002-003-MY3).

\appendix

\section{Relations for  bottom antitriplet  baryon decay amplitudes}

In Section V, we have argued that due to different forms to represent the hadron SU(3) multiplet, the  IRA amplitudes can be constructed in different ways. In Eq.~\ref{Lb14IRAamps}, the antitriplet baryons and octet baryons are used as $(T_{b\bar 3})_i$ and $(T_{8})^{ij}$.  Instead  it is possible to use another set of forms, $(T_{b\bar 3})^{ij}$ and $(T_{8})_{ijk}$, which gives the IRA amplitudes with 26 terms:
\begin{eqnarray}
{\cal A}_{u}^{IRA} & = & \bar{A}_{1}T_{b\bar 3}^{[ij]} H_{\bar{3}}^{k}\epsilon_{ijn}(T_{8})_{k}^{n}P_{l}^{l}+\bar{A}_{2}T_{b\bar 3}^{[ij]}H_{\bar{3}}^{k}\epsilon_{ijn}(T_{8})_{l}^{n}P_{k}^{l}+\bar{A}_{3}T_{b\bar 3}^{[ij]}H_{\bar{3}}^{k}\epsilon_{ikn}(T_{8})_{j}^{n}P_{k}^{l}+\bar{A}_{4}T_{b\bar 3}^{[ij]}H_{\bar{3}}^{k}\epsilon_{iln}(T_{8})_{j}^{n}P_{k}^{l} \nonumber \\
 & &+\bar{A}_{5}T_{b\bar 3}^{[ij]}H_{\bar{3}}^{k}\epsilon_{ikn}(T_{8})_{l}^{n}P_{j}^{l}+\bar{A}_{6}T_{b\bar 3}^{[ij]}H_{\bar{3}}^{k}\epsilon_{iln}(T_{8})_{k}^{n}P_{j}^{l}+\bar{A}_{7}T_{b\bar 3}^{[ij]}H_{\bar{3}}^{k}\epsilon_{kln}(T_{8})_{i}^{n}P_{j}^{l}\nonumber \\
 & &+\bar{B}_{1}T_{b\bar 3}^{[ij]}(H_{6})_{m}^{[kl]}\epsilon_{ijn}(T_{8})_{k}^{n}P_{l}^{m}+\bar{B}_{2}T_{b\bar 3}^{[ij]}(H_{6})_{m}^{[kl]}\epsilon_{ikn}(T_{8})_{j}^{n}P_{l}^{m}+\bar{B}_{3}T_{b\bar 3}^{[ij]}(H_{6})_{m}^{[kl]}\epsilon_{ikn}(T_{8})_{l}^{n}P_{j}^{m}\nonumber \\
 & &+\bar{B}_{4}T_{b\bar 3}^{[ij]}(H_{6})_{m}^{[kl]}\epsilon_{kln}(T_{8})_{i}^{n}P_{j}^{m}+\bar{B}_{5}T_{b\bar 3}^{[ij]}(H_{6})_{i}^{[kl]}\epsilon_{jkn}(T_{8})_{m}^{n}P_{l}^{m}+\bar{B}_{6}T_{b\bar 3}^{[ij]}(H_{6})_{i}^{[kl]}\epsilon_{kln}(T_{8})_{m}^{n}P_{j}^{m}\nonumber \\
 & &+\bar{B}_{7}T_{b\bar 3}^{[ij]}(H_{6})_{i}^{[kl]}\epsilon_{jkn}(T_{8})_{l}^{n}P_{m}^{m}+\bar{B}_{8}T_{b\bar 3}^{[ij]}(H_{6})_{i}^{[kl]}\epsilon_{kln}(T_{8})_{j}^{n}P_{m}^{m}+\bar{B}_{9}T_{b\bar 3}^{[ij]}(H_{6})_{i}^{[kl]}\epsilon_{mjn}(T_{8})_{k}^{n}P_{l}^{m}\nonumber \\
 & &+\bar{B}_{10}T_{b\bar 3}^{[ij]}(H_{6})_{i}^{[kl]}\epsilon_{mkn}(T_{8})_{j}^{n}P_{l}^{m}+\bar{B}_{11}T_{b\bar 3}^{[ij]}(H_{6})_{i}^{[kl]}\epsilon_{mkn}(T_{8})_{l}^{n}P_{j}^{m}\nonumber \\
 & &+\bar{C}_{1}T_{b\bar 3}^{[ij]}(H_{\overline {15}})_{m}^{\{kl\}}\epsilon_{ijn}(T_{8})_{k}^{n}P_{l}^{m}+\bar{C}_{2}T_{b\bar 3}^{[ij]}(H_{\overline {15}})_{m}^{\{kl\}}\epsilon_{ikn}(T_{8})_{j}^{n}P_{l}^{m}+\bar{C}_{3}T_{b\bar 3}^{[ij]}(H_{\overline {15}})_{m}^{\{kl\}}\epsilon_{ikn}(T_{8})_{l}^{n}P_{j}^{m}\nonumber \\
 & &+\bar{C}_{4}T_{b\bar 3}^{[ij]}(H_{\overline {15}})_{i}^{\{kl\}}\epsilon_{jkn}(T_{8})_{m}^{n}P_{l}^{m}+\bar{C}_{5}T_{b\bar 3}^{[ij]}(H_{\overline {15}})_{i}^{\{kl\}}\epsilon_{jkn}(T_{8})_{l}^{n}P_{m}^{m}+\bar{C}_{6}T_{b\bar 3}^{[ij]}(H_{\overline {15}})_{i}^{\{kl\}}\epsilon_{mjn}(T_{8})_{k}^{n}P_{l}^{m}\nonumber \\
 & &+\bar{C}_{7}T_{b\bar 3}^{[ij]}(H_{\overline {15}})_{i}^{\{kl\}}\epsilon_{mkn}(T_{8})_{j}^{n}P_{l}^{m}+\bar{C}_{8}T_{b\bar 3}^{[ij]}(H_{\overline {15}})_{i}^{\{kl\}}\epsilon_{mkn}(T_{8})_{l}^{n}P_{j}^{m}.
\end{eqnarray}
The relation between the two sets of baryon SU(3) representation is:
\begin{eqnarray}
T_{b\bar 3}^{[ij]}=\frac{1}{2}\epsilon^{ijk}(T_{b\bar 3})_{k},\ \ (T_8)_{ijk}=\epsilon_{ijl}(T_8)^{l}_{k}.
\end{eqnarray}
These two sets of IRA are related to each other by:
\begin{eqnarray}
A_{3}^{T} & =&2\bar{A}_{1}+\bar{A}_{3}+\bar{A}_{6}+\bar{A}_{7},\;\;\;
B_{3}^{T} =2\bar{A}_{2}+\bar{A}_{4}+\bar{A}_{5}-\bar{A}_{7},\;\;\; C_{3}^{T}  =\bar{A}_{7}-\bar{A}_{5},\;\;\;
D_{3}^{T}  =-\bar{A}_{6}-\bar{A}_{7}\nonumber \\
A_{6}^{T} & =&\bar{B}_{7}+\bar{B}_{11}-2\bar{B}_{8},\;\;\;
B_{6}^{T}  =-(\bar{B}_{3}+\bar{B}_{11})+2\bar{B}_{4},\;\;
C_{6}^{T} =\bar{B}_{5}+\bar{B}_{10}-2\bar{B}_{6},\;\;\;
D_{6}^{T} =-(\bar{B}_{9}+\bar{B}_{10}-\bar{B}_{11}),
\nonumber \\
A_{15}^{T} & =&\bar{C}_{5}+\bar{C}_{8},\;\;\;
B_{15}^{T}  =-(\bar{C}_{3}+\bar{C}_{8}),\;\;\; 
C_{15}^{T} = \bar{C}_{4}+\bar{C}_{7},\;\;\;
D_{15}^{T}  =-(\bar{C}_{6}+\bar{C}_{8}+\bar{C}_{7}),\nonumber \\
E_{6}^{T} &=&2\bar{B}_{1}+2\bar{B}_{4}+\bar{B}_{2}-\bar{B}_{3}+\bar{B}_{9}+\bar{B}_{10}-\bar{B}_{11},\;\;\; E_{15}^{T} =2\bar{C}_{1}+\bar{C}_{2}+\bar{C}_{3}+\bar{C}_{6}+\bar{C}_{7}+\bar{C}_{8}. \label{IRA1426}
\end{eqnarray}

In addition, the relation between the coefficients of 26 amplitudes in  ${\cal A}_{T_{b}\rightarrow PT_{8}(u)}^{IRA}$ and ${\cal A}_{T_{b}\rightarrow PT_{8}(u)}^{TDA}$ is:
\begin{eqnarray}
{\bar A}_{1} & =&\frac{1}{8}(-{\bar a}_{1}+3{\bar a}_{2}-3{\bar a}_{3}+{\bar a}_{5})+{\bar b}_{1},\;\;\;
{\bar A}_{2}  =\frac{1}{8}(3{\bar a}_{1}-{\bar a}_{2}-3{\bar a}_{4}+{\bar a}_{7})+{\bar b}_{2},\nonumber\\
{\bar A}_{3} & =&\frac{1}{8}({\bar a}_{3}-3{\bar a}_{5}+4{\bar a}_{9})+{\bar b}_{3},\;\;\;
{\bar A}_{4}  =\frac{1}{8}(-3{\bar a}_{6}+{\bar a}_{8}+3{\bar a}_{10}-{\bar a}_{11}+3{\bar a}_{15}-{\bar a}_{16})+{\bar b}_{4},\nonumber\\
{\bar A}_{5} & =&\frac{1}{8}(4{\bar a}_{12}-{\bar a}_{17}+3{\bar a}_{18}+{\bar a}_{4}-3{\bar a}_{7})+{\bar b}_{5},\;\;\;
{\bar A}_{6}  =\frac{1}{8}(3{\bar a}_{13}-{\bar a}_{14}+3{\bar a}_{17}-{\bar a}_{18}+{\bar a}_{6}-3{\bar a}_{8})+{\bar b}_{6},\nonumber\\
{\bar A}_{7} & =&\frac{1}{8}({\bar a}_{10}-3{\bar a}_{11}-{\bar a}_{13}+3{\bar a}_{14}+4{\bar a}_{19})+{\bar b}_{7},\nonumber\\
{\bar B}_{1} & =&\frac{1}{4}({\bar a}_{1}-{\bar a}_{2}),\ \ \ \ {\bar C}_{1}=\frac{1}{8}({\bar a}_{1}+{\bar a}_{2}),\;\;\;
{\bar B}_{2}  =\frac{1}{4}({\bar a}_{15}-{\bar a}_{16}),\ \ \ \ {\bar C}_{2}=\frac{1}{8}({\bar a}_{15}+{\bar a}_{16}),\nonumber\\
{\bar B}_{3} & =&\frac{1}{4}({\bar a}_{17}-{\bar a}_{18}),\ \ \ \ {\bar C}_{3}=\frac{1}{8}({\bar a}_{17}+{\bar a}_{18}),\;\;\;
{\bar B}_{4} =-\frac{1}{4}{\bar a}_{19},\;\;\;
{\bar B}_{5}  =\frac{1}{4}({\bar a}_{4}-{\bar a}_{7}),\ \ \ \ {\bar C}_{4}=\frac{1}{8}({\bar a}_{4}+{\bar a}_{7}),\;\;
{\bar B}_{6}  =\frac{1}{4}{\bar a}_{12},\nonumber\\
{\bar B}_{7} & =&\frac{1}{4}({\bar a}_{3}-{\bar a}_{5}),\ \ \ \ {\bar C}_{5}=\frac{1}{8}({\bar a}_{3}+{\bar a}_{5}),\;\;\;
{\bar B}_{8} =\frac{1}{4}{\bar a}_{9},\;\;\;
{\bar B}_{9} =\frac{1}{4}({\bar a}_{8}-{\bar a}_{6}),\ \ \ \ {\bar C}_{6}=-\frac{1}{8}({\bar a}_{6}+{\bar a}_{8}),\nonumber\\
{\bar B}_{10} & =&\frac{1}{4}({\bar a}_{11}-{\bar a}_{10}),\ \ \ \ {\bar C}_{7}=-\frac{1}{8}({\bar a}_{10}+{\bar a}_{11}),\;\;\;
{\bar B}_{11} =\frac{1}{4}({\bar a}_{14}-{\bar a}_{13}),\ \ \ \ {\bar C}_{8}=-\frac{1}{8}({\bar a}_{13}+{\bar a}_{14}). \label{IRATDA26}
\end{eqnarray}
Combination of Eq.(\ref{IRA1426}) and Eq.(\ref{IRATDA26}) leads to Eq.(\ref{IRATDA}).






The authors of Ref.~\cite{He:2015fwa} have given another parametrization of IRA amplitudes, in which they have focused on the flavor non-singlet. Comparing with their results, one finds the following relations: 
\begin{eqnarray}
B_{3}^{T} & =&2b(\bar{3})_{2}+d(\bar{3})_{1}-e(\bar{3})_{2}+c(\bar{3}),\nonumber \\
C_{3}^{T} & =&2a(\bar{3})-c(\bar{3}),\nonumber \\
D_{3}^{T} & =&2b(\bar{3})_{1}+d(\bar{3})_{2}-e(\bar{3})_{1}+c(\bar{3}), \nonumber\\
B_{6}^{T} & =&2(a(6)_{2}-g(6)-n(6)_{1})+d(6)_{2}-e(6)_{1}-e(6)_{2},\nonumber \\
C_{6}^{T} & =&2(a(6)_{1}+f(6)-n(6)_{2})+d(6)_{1},\nonumber \\
E_{6}^{T} & =&2(b(6)_{2}-g(6))-c(6)+d(6)_{1}+d(6)_{2}-e(6)_{1}-e(6)_{2}-g(6),\nonumber \\
D_{6}^{T} & =&-2(b(6)_{1}+f(6))+c(6)-d(6)_{1}-d(6)_{2}+e(6)_{1}+e(6)_{2}, \nonumber\\
B_{15}^{T} & =&2a(\overline{15})_{2}+d(\overline{15})_{2}-e(\overline{15})_{1},\nonumber \\
C_{15}^{T} & =&2a(\overline{15})_{1}+d(\overline{15})_{1}-e(\overline{15})_{2},\nonumber \\
E_{15}^{T} & =&2b(\overline{15})_{2}+c(\overline{15})+d(\overline{15})_{1}-d(\overline{15})_{2}+e(\overline{15})_{1}-e(\overline{15})_{2},\nonumber \\
D_{15}^{T} & =&2b(\overline{15})_{1}-c(\overline{15})-d(\overline{15})_{1}+d(\overline{15})_{2}-e(\overline{15})_{1}+e(\overline{15})_{2}.
\end{eqnarray}

\section{Relations for charmed baryon decays}

Similar with the previous section, one can construct another set of IRA for  charmed antitriplet baryon decays with 19 amplitudes:    
\begin{eqnarray}
{\cal A}_{u}^{IRA} & =&{\bar B}_{1}T_{c\bar 3}^{[ij]}(H_{6})_{m}^{kl}\epsilon_{ijn}(T_{8})_{k}^{n}P_{l}^{m}+{\bar B}_{2}T_{c\bar 3}^{[ij]}(H_{6})_{m}^{kl}\epsilon_{ikn}(T_{8})_{j}^{n}P_{l}^{m}+{\bar B}_{3}T_{c\bar 3}^{[ij]}(H_{6})_{m}^{kl}\epsilon_{ikn}(T_{8})_{l}^{n}P_{j}^{m}\nonumber\\
 & &+{\bar B}_{4}T_{c\bar 3}^{[ij]}(H_{6})_{m}^{kl}\epsilon_{kln}(T_{8})_{i}^{n}P_{j}^{m}+{\bar B}_{5}T_{c\bar 3}^{[ij]}(H_{6})_{i}^{kl}\epsilon_{jkn}(T_{8})_{m}^{n}P_{l}^{m}+{\bar B}_{6}T_{c\bar 3}^{[ij]}(H_{6})_{i}^{kl}\epsilon_{kln}(T_{8})_{m}^{n}P_{j}^{m}\nonumber\\
 & &+{\bar B}_{7}T_{c\bar 3}^{[ij]}(H_{6})_{i}^{kl}\epsilon_{jkn}(T_{8})_{l}^{n}P_{m}^{m}+{\bar B}_{8}T_{c\bar 3}^{[ij]}(H_{6})_{i}^{kl}\epsilon_{kln}(T_{8})_{l}^{n}P_{m}^{m}+{\bar B}_{9}T_{c\bar 3}^{[ij]}(H_{6})_{i}^{kl}\epsilon_{mjn}(T_{8})_{k}^{n}P_{l}^{m}\nonumber\\
 & &+{\bar B}_{10}T_{c\bar 3}^{[ij]}(H_{6})_{i}^{kl}\epsilon_{mkn}(T_{8})_{j}^{n}P_{l}^{m}+{\bar B}_{11}T_{c\bar 3}^{[ij]}(H_{6})_{i}^{kl}\epsilon_{mkn}(T_{8})_{l}^{n}P_{j}^{m}\nonumber\\
 & &+{\bar C}_{1}T_{c\bar 3}^{[ij]}(H_{\overline {15}})_{m}^{kl}\epsilon_{ijn}(T_{8})_{k}^{n}P_{l}^{m}+{\bar C}_{2}T_{c\bar 3}^{[ij]}(H_{\overline {15}})_{m}^{kl}\epsilon_{ikn}(T_{8})_{j}^{n}P_{l}^{m}+{\bar C}_{3}T_{c\bar 3}^{[ij]}(H_{\overline {15}})_{m}^{kl}\epsilon_{ikn}(T_{8})_{l}^{n}P_{j}^{m}\nonumber\\
 & &+{\bar C}_{4}T_{c\bar 3}^{[ij]}(H_{\overline {15}})_{i}^{kl}\epsilon_{jkn}(T_{8})_{m}^{n}P_{l}^{m}+{\bar C}_{5}T_{c\bar 3}^{[ij]}(H_{\overline {15}})_{i}^{kl}\epsilon_{jkn}(T_{8})_{l}^{n}P_{m}^{m}+{\bar C}_{6}T_{c\bar 3}^{[ij]}(H_{\overline {15}})_{i}^{kl}\epsilon_{mjn}(T_{8})_{k}^{n}P_{l}^{m}\nonumber\\
 & &+{\bar C}_{7}T_{c\bar 3}^{[ij]}(H_{\overline {15}})_{i}^{kl}\epsilon_{mkn}(T_{8})_{j}^{n}P_{l}^{m}+{\bar C}_{8}T_{c\bar 3}^{[ij]}(H_{\overline {15}})_{i}^{kl}\epsilon_{mkn}(T_{8})_{l}^{n}P_{j}^{m}. 
\end{eqnarray}

The relation between  the two sets of IRA is given as:
\begin{eqnarray}
A_{6}^{T} & =&\frac{1}{2}(\bar{B}_{7}+\bar{B}_{11})-\bar{B}_{8},\;\;\;
B_{6}^{T}  =-\frac{1}{2}(\bar{B}_{3}+\bar{B}_{11})+\bar{B}_{4},\;\;\;
C_{6}^{T}  =\frac{1}{2}(\bar{B}_{5}+\bar{B}_{10})-\bar{B}_{6},\;\;\;
D_{6}^{T} =-\frac{1}{2}(\bar{B}_{9}+\bar{B}_{10}-\bar{B}_{11})\nonumber \\
A_{15}^{T} & =&\frac{1}{2}(\bar{C}_{5}+\bar{C}_{8}),\;\;\;
B_{15}^{T}  =-\frac{1}{2}(\bar{C}_{3}+\bar{C}_{8}),\;\;\;
C_{15}^{T} =\frac{1}{2}(\bar{C}_{4}+\bar{C}_{7}),\;\;\;
D_{15}^{T} =-\frac{1}{2}(\bar{C}_{6}+\bar{C}_{8}+\bar{C}_{7})\nonumber \\
E_{6}^{T} & =&\bar{B}_{1}+\bar{B}_{4}+\frac{1}{2}(\bar{B}_{2}-\bar{B}_{3}+\bar{B}_{9}+\bar{B}_{10}-\bar{B}_{11}),\;\;\;
E_{15}^{T} =\bar{C}_{1}+\frac{1}{2}(\bar{C}_{2}+\bar{C}_{3}+\bar{C}_{6}+\bar{C}_{7}+\bar{C}_{8}). 
\end{eqnarray}

The relation between the  new ${\cal A}_{T_{c}\rightarrow PT_{8}(u)}^{IRA}$ and ${\cal A}_{T_{c}\rightarrow PT_{8}(u)}^{TDA}$ can be obtained as: 
\begin{eqnarray}
\bar B_{1} & =&\frac{1}{2}(\bar a_{1}-\bar a_{2}),\ \ \ \ \bar C_{1}=\frac{1}{2}(\bar a_{1}+\bar a_{2}),\;\;\; \bar B_{2} =\frac{1}{2}(\bar a_{15}-\bar a_{16}),\ \ \ \ \bar C_{2}=\frac{1}{2}(\bar a_{15}+\bar a_{16})\nonumber\\
\bar B_{3} & =&\frac{1}{2}(\bar a_{17}-\bar a_{18}),\ \ \ \ \bar C_{3}=\frac{1}{2}(\bar a_{17}+\bar a_{18}),\;\;\;
\bar B_{4}  =-\frac{1}{2}\bar a_{19},\;\;\;
\bar B_{5} =\frac{1}{2}(\bar a_{4}-\bar a_{7}),\ \ \ \ \bar C_{4}=\frac{1}{2}(\bar a_{4}+\bar a_{7}),\nonumber\\
\bar B_{6} & =&\frac{1}{2}\bar a_{12},\;\;\;
\bar B_{7} =\frac{1}{2}(\bar a_{3}-\bar a_{5}),\ \ \ \ \bar C_{5}=\frac{1}{2}(\bar a_{3}+\bar a_{5}),\;\;\;
\bar B_{8}  =\frac{1}{2}\bar a_{9},\;\;\;
\bar B_{9} =\frac{1}{2}(\bar a_{8}-\bar a_{6}),\ \ \ \ \bar C_{6}=-\frac{1}{2}(\bar a_{6}+\bar a_{8}),\nonumber\\
\bar B_{10} & =&\frac{1}{2}(\bar a_{11}-\bar a_{10}),\ \ \ \ \bar C_{7}=-\frac{1}{2}(\bar a_{10}+\bar a_{11}),\;\;\;
\bar B_{11} =\frac{1}{2}(\bar a_{14}-\bar a_{13}),\ \ \ \ \bar C_{8}=-\frac{1}{2}(\bar a_{13}+\bar a_{14}). 
\end{eqnarray}


\end{document}